\documentclass[english,smallextended]{svjour3}
\pdfoutput=1 
\usepackage[T2A,T1]{fontenc}
\usepackage[latin9]{inputenc}
\usepackage{color}
\usepackage{babel}
\usepackage{url}
\usepackage{amsmath}
\usepackage{amssymb}
\usepackage{graphicx}
\usepackage{esint}
\usepackage[unicode=true,
bookmarks=false,
breaklinks=false,pdfborder={0 0 1},colorlinks=true]
{hyperref}
\hypersetup{
	pdfstartview=FitH,linkcolor=linkcolor,urlcolor=urlcolor,citecolor=linkcolor}

\makeatletter
\newenvironment{svmultproof}{\begin{proof}}{\qed\end{proof}}

\usepackage{bbm}

\@ifundefined{textcolor}{}{%
	\definecolor{BLACK}{gray}{0}
	\definecolor{WHITE}{gray}{1}
	\definecolor{RED}{rgb}{1,0,0}
	\definecolor{GREEN}{rgb}{0,1,0}
	\definecolor{BLUE}{rgb}{0,0,1}
	\definecolor{CYAN}{cmyk}{1,0,0,0}
	\definecolor{MAGENTA}{cmyk}{0,1,0,0}
	\definecolor{YELLOW}{cmyk}{0,0,1,0}
}
\DeclareMathAlphabet{\mathbbold}{U}{bbold}{m}{n}

\usepackage{xcolor}

\definecolor{linkcolor}{HTML}{000000}
\definecolor{urlcolor}{HTML}{000000}

\makeatother

\begin{document}

\title{Nonstationary generalized TASEP in KPZ and jamming regimes.}

\author{A.E. Derbyshev \and  A.M. Povolotsky}

\institute{A.E. Derbyshev \and A.M. Povolotsky \at Bogoliubov laboratory of
theoretical physics, Joint Institute for Nuclear Research, Dubna,
Russia\\
\email{andreyderbishev@yandex.com}\\ \email{alexander.povolotsky@gmail.com}\\
 \and A.M. Povolotsky \at  Center  for  Advanced  Studies,  Skolkovo  Institute  of  Science  and  Technology,  Nobel  Street  1,  121205 Moscow, Russia\\
}
\maketitle
\begin{abstract}
We study the model of the totally asymmetric exclusion process with generalized update, which compared to the usual totally asymmetric 
exclusion process, has an additional parameter enhancing clustering of particles. We derive the exact multiparticle distributions of distances travelled by particles on the infinite lattice for two types of initial conditions: step and alternating once.  Two different  scaling limits of the exact formulas are studied. 
Under the first scaling associated to Kardar-Parisi-Zhang (KPZ) universality class we prove  convergence of  joint distributions of the scaled particle positions to finite-dimensional distributions of  the universal   Airy$_2$ and Airy$_1$ processes. 
Under the second scaling we prove  convergence of the  same  position distributions to  finite-dimensional distributions of two new random processes, which   describe the transition between the KPZ regime and the deterministic aggregation regime, in which the particles stick together into a single giant cluster moving as one particle. It is shown that the transitional distributions have the Airy processes and fully correlated Gaussian fluctuations as limiting cases. We also give  the heuristic arguments explaining how the non-universal scaling constants  appearing from the asymptotic analysis in the KPZ regime are related to the properties of   translationally invariant  stationary states in the infinite system and how the parameters of the model should scale in the transitional regime.

\end{abstract}

\tableofcontents{}

\section{Introduction}

The Kardar-Parisi-Zhang (KPZ) universality class was first introduced
in context of interface growth in 1986 \cite{KPZ}. It unifies a large
class of models of growing interfaces, interacting particle systems
and other systems with many degrees of freedom and local interactions
subject to an uncorrelated random forcing \cite{Corwin2012}. The landmark of KPZ class
is two critical exponents responsible for the time scaling of fluctuations
and correlations. Values of the exponents are exactly known only in
1+1 space-time dimensions, where they are $1/3$ and $2/3$ respectively. 

The totally asymmetric simple exclusion process (TASEP) is probably
the most renowned model from this class.  It is an interacting particle system,  in which   particles move in the same direction through the one-dimensional lattice with nearest neighbor  stochastic jumps obeying exclusion interaction that prevents a particle from jumping to an already occupied site (see  the next section for  rigorous definition). Despite very simple formulation,
it has a rich mathematical structure encapsulated in the term ``quantum
integrability'' \cite{KorepinBogoliubovIzergin}, which practically suggests that the model is exactly
solvable, at least potentially \cite{Derrida1998}. Indeed, its first Bethe ansatz solution announced in 
\cite{dhar1987exactly} was presented in  \cite{gwa1992six,gwa1992bethe}, where it was used to 
obtain the dynamical exponent $z=3/2$, which is inverse of the above mentioned correlation exponent $2/3$,    serving as a manifestation of the KPZ scaling behavior.    
Later the Bethe ansatz was also used to obtain exact current large deviation function of  TASEP on a periodic lattice \cite{Derrida_Lebowitz}. 
These results were  extended to partially asymmetric generalization of TASEP referred to as  ASEP \cite{kim1995bethe,lee1999large}  and to the system with open boundary conditions \cite{de2005bethe,de2006exact,de2011large}. Also, one  has to mention many important results on  TASEP and its generalizations  obtained with the matrix product method \cite{derrida1993exact}, see \cite{blythe2007nonequilibrium} for review and references therein. 

The early  studies   addressed mainly the stationary and large time  behavior of the  TASEP in  finite systems. Another line of research that finally lead to a breakthrough in the description of finite time KPZ behavior concerned  with TASEP on the infinite lattice.   Remarkably the necessary tools had been developing independently by physical and mathematical communities until their efforts merged  in the beginning of two thousands to give a cumulative effect on the development of the subject. Sch\"utz first used the Bethe ansatz to obtain a determinantal formula for the finite time transition probabilities between two arbitrary finite particle configurations in the continuous time TASEP  \cite{SchutzG.1997}. Exploiting connections 
with the problem of last passage percolation, which in turn is related to  the  statistics of random  Young diagrams, random permutations and random matrices, Johansson \cite{Johansson} obtained  particle current in the discrete time TASEP  on the infinite lattice with the step initial conditions. Nagao, Sasamoto \cite{nagao2004asymmetric} and R{\'a}kos,  Sch{\"u}tz   
\cite{rakos2005current} shown that similar results can be obtained for continuous time TASEP and discrete time TASEP with backward sequential update using the Sch\"utz's determinantal  formula as a starting point, which is also  applicable  to various initial conditions.    
Later, the Bethe ansatz approach was shown efficient for  a generalization of these results to  ASEP, i.e. the partially asymmetric version of TASEP   \cite{tracy2008integral,tracy2008fredholm}.

What makes the TASEP special compared
to many other related integrable models of interacting particles is
the structure of determinantal point process \cite{BorodinOkounkovOlshaski,Borodin2015} hidden behind its transition
probabilities. This fact,  allowed
an exact calculation of all spacial and space-like  finite-dimensional
distributions of particle positions and particle currents \cite{Sasamoto2005,borodin2007fluctuation,Borodin2,BFS,imamura2007dynamics,PoghosyanS.2010,exit} . The calculations
made for the TASEP on infinite integer lattice  for several special
types of initial conditions (IC)  \cite{Ferrari2008},  finally led to a recipe applicable for general IC \cite{MQR}. Recently these results were extended for the TASEP on a finite periodic lattice \cite{BL_1,BL_2}.

Of special interest are functional forms of the distributions in the
so called ``scaling limit''. They are believed to be  universal
scaling functions insensitive to details of microscopic dynamics and
characterizing the KPZ fixed point in one dimension. The limiting
distributions still depend on global geometry of IC.
Explicit expressions were obtained for three main types of initial
conditions, flat  \cite{Sasamoto2005}, step  \cite{Johansson} and stationary  \cite{Baik2000} , the basic ones, which survive
the scaling limit owing to their self-similarity property. They led
to discovery of three basic universal random processes, Airy$_{1}$,
Airy$_{2}$ and Airy$_{\mathrm{stat}}$, respectively \cite{Sasamoto2005,BFPS_2007,Spohn2002,Baik2010}. Their finite-dimensional
distributions  can be represented in the form of the Fredholm determinants
of trace-class operators with explicitly defined kernels, some of
which were previously known from the theory of random matrices  \cite{Mehta} and
some were new. The universality of these processes was confirmed by
results on several other models also possessing the structure of the
determinatal process, such as ensembles of non-intersecting paths \cite{WeissFerrariSpohn2017} or
non-colliding Brownian motions, domino and lozenge tilings \cite{Johansson2005}, Schur processes \cite{Okounkov2003},
e.t.c. Also the universal one-point distributions were obtained in
a number of non-determinantal models, see e.g. \cite{tracy2009asymptotics,ferrari2015tracy,vetHo2015tracy,imamura2019fluctuations}.

The three types of IC play the
role of building blocks. Logically, the next step was to identify
the transitional kernels connecting different basic subclasses within the
KPZ class. This task was also completed for several
models by obtaining transitional distributions in the form of Fredholm
determinants with explicit kernels \cite{Borodin2008},\cite{Imamura2004}.

Another interesting problem is to describe the crossover between
the KPZ and non-KPZ scaling behaviors in the cases when the KPZ universality
breaks down. An example of such a crossover is the transition between
KPZ and Edwards-Wilkinson universality classes \cite{krug1997origins} . The recent groundbreaking
derivation of the one-point distribution of the interface height governed
by KPZ equation resulted in the function transforming from Gaussian
to Tracy-Widom distribution as the nonlinearity coefficient varies
from zero to infinity \cite{sasamoto2010one,dotsenko2010bethe,Amir2011ProbabilityDO,calabrese2010free}. 
Another important example of crossover  from the KPZ  to  the equilibrium behavior is given by 
 the TASEP on a periodic lattice \cite{BL_1}.
Though there are other examples, the list of
the crossover functions studied so far is far from being complete. 

In the present paper we study the one-parameter generalization of
the TASEP, TASEP with generalized update (GTASEP). This model was
first proposed in \cite{Woelki} as an example of TASEP-like model,
which can be mapped to a system with factorized steady state. It was
later rediscovered in \cite{gtasep} as an integrable generalization
of the TASEP and also was found to be the $q=0$ degeneration of three-parametric
family of chipping models \cite{P2013} also known as q-Hahn model \cite{BCPS2013}.
It reappeared again in \cite{KnizelPetrovSaenz} from the studies of the  Schur measures related to deformations 
of the Robinson-Schensted-Knuth dynamics. The  stationary state of GTASEP was studied on the ring \cite{DPP2015,aneva2016matrix} and on the interval \cite{bunzarova2017one,brankov2018model,bunzarova2019one,bunzarova2021aggregation}.

In addition to the usual discrete time dynamics and exclusion interaction
the GTASEP has an extra tuning parameter responsible for an attractive-like
interaction that affects clustering of particles. As the parameter
varies in  its range, the model transforms from the discrete time
TASEP with parallel update to what we call the deterministic aggregation
(DA), the regime where all particles tend to stick together to a giant
cluster moving as a single particle. Therefore, if we look at the
large scale statistics of particle flow, e.g. dependence of particle
position on time, it is naturally to expect typical KPZ-like fluctuations
on one end and purely diffusive behavior on the other. The effect
of this transition on the structure of the stationary state and on
the current large deviation function (LDF) was  studied in
the GTASEP on the ring \cite{DPP2015}. The main conclusion derived
was that at moderate interaction strength far enough from the DA regime
the current LDF is of KPZ type being characterized by the universal
scaling function obtained by Derrida and Lebowitz \cite{Derrida_Lebowitz}.
On the other hand, the scaling behavior starts changing, when the
interaction strength is scaled with the system size, so that the stationary
state correlation length or, equivalently, the typical cluster size become
extensive, i.e. of order of the system size. In such defined transitional
regime all particles are typically distributed among finitely many
clusters moving diffusively, and the current LDF obtained under the
diffusive scaling interpolates between the Derrida-Lebowitz and Gaussian
LDF. 

The next question to ask is how this change of behavior shows
up in the non-stationary setting.  Qualitatively it is natural to expect that  the  typical
KPZ behaviour at moderate interaction strength  crosses over to the diffusive motion of giant 
clusters  close to DA limit. Once we identify the appropriate transition scale, we may hope to observe the transition between 
the Tracy-Widom and normal distributions. One example of such a transition was described by Baik, Ben Arous and P{\'e}ch{\'e} (BBP),  \cite{baik2005phase},
in limiting distributions of the largest eigenvalue of a complex Gaussian sample covariance matrix with all but finitely many coinciding eigenvalues.
It was later noticed that this transition is common in interacting particle systems like TASEP with finitely many slow particles, see e.g. \cite{baik2006painleve,imamura2007dynamics,barraquand2015phase}. These particles play the role of an  obstacle for faster particles behind leading to formation of large particle clusters. Then, the  particles far behind the slow particles suffer the BBP transition in an appropriate time scale.
It would be interesting to understand, whether the transition to the DA limit is of the same type.    

The aim of this paper is twofold. First, we test the KPZ universality
in the GTASEP on infinite lattice with step and alternating IC and
moderate interaction strength. Second, we identify the transitional
regime and obtain the crossover distributions interpolating between
KPZ and diffusive fully correlated particle motion.

We start with deriving the exact formulas for finite dimensional distributions
of particle positions obtained as usual in the form of Fredholm determinants
of functional operators with explicit kernels. This is done with the
use of the determinantal structure of the Green function obtained
in \cite{gtasep}. Up to technical complications this part mostly
follows  the line of the previous works on the TASEP, especially
the TASEP with parallel update \cite{BFS}.

Then, we turn to asymptotic analysis of the kernels obtained. 
The KPZ part of the analysis is also
similar  to the those for particular cases of TASEP studied previously.
The main result is the statement of convergence of the joint distributions  
of the distances travelled by particles to the multipoint distributions  
 of the Airy$_{1}$ and Airy$_{2}$ processes, which takes
place under a proper scaling. 

In the second part of the asymptotic analysis we consider the scaling
limit corresponding to the transitional regime. It suggests that the
parameter controlling the interaction and the space-time window is
scaled together with time. Similarly to what was observed in the periodic
lattice in the limit of large time the new scaling regime corresponds
to the situation, when the finite number of giant clusters are under
consideration. In the transitional regime we prove the convergence 
of the exact multi-particle distance distributions  to the multipoint 
distributions of two new random processes. The whole one-parameter families of 
limiting processes  are obtained depending  on a single crossover
parameter, which controls the transition from the KPZ regime  to the
fully correlated particle motion.  
We also demonstrate that the kernels obtained converge to Airy kernels
and the kernel describing the fully correlated Gaussian fluctuations in the two extreme limits of the crossover parameter. 

Note,  when this article was being   prepared for publication, the results on the exact multiparticle distributions for step initial condition as well as their KPZ asymptotics  were independently published in  \cite{KnizelPetrovSaenz}. Unlike us having departed from the Bethe ansatz solution, the starting point for the analysis of those  authors was the representation of the GTASEP via the Schur process. Hence  the step IC were considered there and the results equivalent to the ones presented here were obtained for this case. To our knowledge the results on exact  distributions for the alternating IC and their KPZ limit  as well as on the transitional distributions for both cases are new.

The article is organized as follows. The main section of the article is  section \ref{sec:model}. We start it by  formulating  the model and describing its relations with other models  studied before.  Then, we state the results.  The first set of results contains  exact multiparticle distributions for step and alternating IC. Next, we move to the results of the asymptotic analysis of this formulas, which is   performed under two different scalings, KPZ and transitional ones. Preceding  the rigorous 
limiting statements we first shortly discuss the qualitative picture to get an idea about  the anticipated form of the scaling limits on a heuristic level. Specifically, we discuss the form of the model-dependent parameters involved into the KPZ  scaling 
basing on  the hydrodynamic description of the particle flow and the hypothesis of quasi-equilibrium that relates the local behavior of non-stationary KPZ systems with  the properties of the  translation invariant  stationary states in such systems.  This consideration predicts the deterministic large scale part of the distance traveled by particles. Using additional scaling hypothesis about the KPZ dimensionful invariants we also predict the form of the model-dependent  constants  describing the correlation and fluctuation scales   characterizing the  fluctuating part of the distance. In the section we sketch only the final  formulas of the heuristic consideration, while all the related calculations are brought into appendices \ref{app: Hydrodynamics}-\ref{app: KPZ dimensional invariants }. These formulas then find confirmation in the rigorous limiting statements based on the asymptotic analysis of the exact formulas. We also discuss the anticipated scaling corresponding to  the transitional regime based on assumption  of the diffusive  particle motion and the finite number of particle clusters in the space-time window under consideration. Finally we formulate the main theorems about the limiting random processes obtained from  KPZ and transitional limits of our model. We conclude this section by discussing the limiting behavior of the transitional processes, showing that they interpolate between the KPZ-specific Airy processes and the fully correlated Brownian motion.  

Thus, section \ref{sec:model} together with appendices \ref{app: Hydrodynamics}-\ref{app: KPZ dimensional invariants } contain  the information necessary for understanding 
 the results of the article by the physically oriented Reader.  The Reader interested in mathematical details of the proofs then may proceed to the following sections.    
 In sections \ref{sec: Determinantal point process}-\ref{sec: Asymptotic analysis: transitional regime}  we  prove  the results stated in Section \ref{sec:model}. Specifically, in section \ref{sec: Determinantal point process} starting from the formula for the Green function, obtained earlier from the Bethe ansatz solution, and using the machinery of the determinantal point processes we obtain the exact multiparticle distributions for the GTASEP with step and alternating IC in the form of Fredholm determinants with explicit kernels. Sections \ref{sec: Asymptotic analysis: KPZ regime} and \ref{sec: Asymptotic analysis: transitional regime} are devoted to asymptotic analysis of the exact distributions under the KPZ and transitional scalings respectively. The general line is the same in all cases. We prove  convergence of the kernels on bounded sets and obtain the large deviation estimates for them. In the KPZ regime (section \ref{sec: Asymptotic analysis: KPZ regime}), this, together with Hadamard inequality, guarantees  a uniform convergence of the series representing the Fredholm determinant, which  allows passing to the limit in the kernel inside the determinant. In the transitional regime (section \ref{sec: Asymptotic analysis: transitional regime}) the situation is more tricky, because of  the unusual form  of the transitional kernels that prevents us from the direct use of the Hadamard bound.  Therefore, we  analyse every term of the Fredholm determinant series more accurately to ensure the uniform convergence of the series. In the  last subsection of  section \ref{sec: Asymptotic analysis: transitional regime} we discuss the extreme limits of the transitional kernels and of the corresponding Fredholm determinants at  large and small values of the  crossover parameter. We show that in the former case, though being in the diffusive scale, we return back to the universal KPZ processes.  In the opposite limit we arrive at the regime where all particles move synchronously as a single particle performing a simple Brownian motion. In particular we obtain an unusual representation of the joint distribution of several identical random variables in the form of  Fredholm determinant, which to our knowledge did not appear in the literature before. 

\begin{acknowledgements}
	This project started  from discussions with our colleague and teacher  Vyacheslav Priezzhev, who passed away in 2017. We would like to appreciate his support and  inspiration he gave us on  early stages of the work. We are grateful to Leonid Petrov for stimulating discussions and for keeping us informed of his work on the subject. Part of this work was completed and first presented during the visit of AP to the program   ``New approaches to non-equilibrium and random systems: KPZ integrability, universality, applications and experiments'' in KITP, Santa Barbara. AP would like to thank the organizers of the program for kind invitation.  
	The work of AP in part of the results of sections 3,4 has been  funded by the Russian Science Foundation under grant  19-11-00275 via  Skolkovo  Institute  of  Science  and  Technology.

	This version of the article has been accepted for publication, after peer
	review  but is not the Version of Record and does not reflect post-acceptance
	improvements, or any corrections. The Version of Record is available online at:
	http://dx.doi.org/10.1007/s10955-021-02840-z. Use of this Accepted Version is subject to the publisher?s Accepted
	Manuscript terms of use https://www.springernature.com/gp/open-research/policies/acceptedmanuscript-
	terms.
\end{acknowledgements}  

\section{Model definition, results and discussion}

\label{sec:model}

\subsection{Model}

GTASEP is a model formulated in terms of particle configurations on
an integer lattice evolving stochastically in discrete time. In the present
paper we deal with the infinite lattice $\mathbb{Z}$. A particle
configuration at a time step $t$ can be recorded as an infinite binary
string $\eta(t)\in\{0,1\}^{\mathbb{Z}},$ where $\eta_{i}(t)=1\,\,\,(\eta_{i}(t)=0)$
means that the site at a position $i$ is occupied with a particle
(empty). The fact that configurations consist only of zeroes and ones,
i.e. at most one particle at a site is allowed, is referred to as
an exclusion interaction. The update of particle configuration at
each time step is most convenient to formulate in terms of \textit{clusterwise}
backward sequential update. Here by cluster we mean a compact group
of particles between two empty sites without empty sites inside, i.e.
a subconfiguration of the form $(\eta_{i}(t),\dots,\eta_{i+k+1}(t))=(0,1^{(k)},0)$.
At every time step all clusters are updated simultaneously and independently, particle
by particle, from right to left\footnote{One can see that our   clusterwise update coincides with usual backward sequential update, when a  configuration  bounded from the right is considered on the infinite lattice.  We, however, keep to this terminology as it remains well defined also on the ring or for unbounded   configurations.}. From a cluster with the rightmost
particle at site $x$ 
\begin{enumerate}
\item the first particle of the cluster decides to jump to $x+1$ with probability $p$
or to stay in $x$ with probability $1-p$;
\item if the first particle has jumped, the second particle of the  cluster follows it with probability
$\mu$ or stays with probability $1-\mu$, and so does every next
particle of  the  same cluster if  the previous particle has jumped;
\item if some particle has decided to stay, all the other particles to the
left of it within the same cluster stay with probability $1$. 
\end{enumerate}
\begin{figure}[h]
	\center{\includegraphics[width=0.7\textwidth]{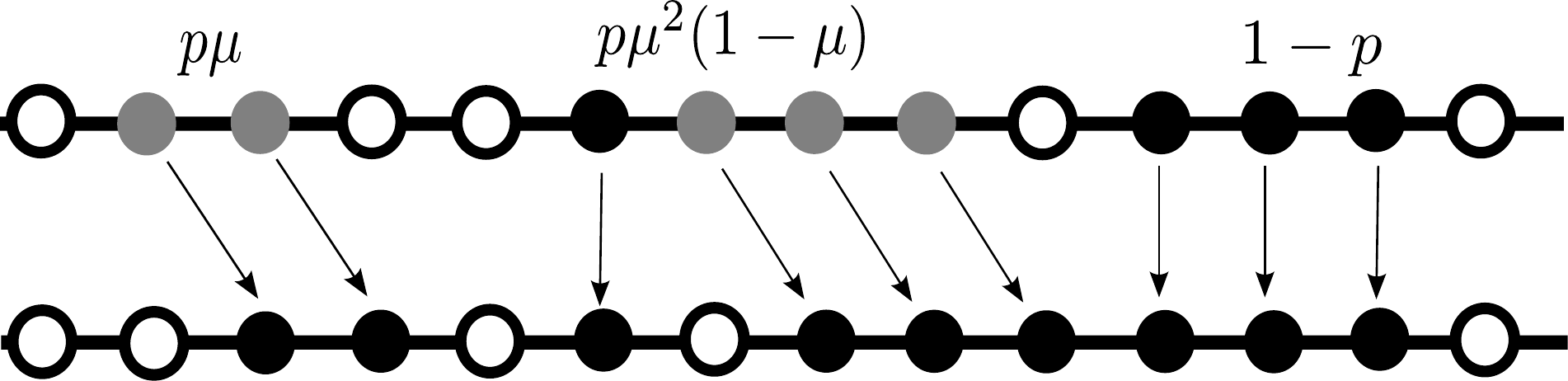}}
	\caption{Example of cluster updates in  GTASEP. The gray particles in the starting configuration are those, which   have decided to jump, and the black ones will stay. In the rightmost cluster all particles stay on their places with probability $1-p$, which is the probability for the first particle to stay, while the others stay with probability one. Three particles from the middle cluster  jump with probabilities $p,\mu$ and $\mu$ and the last one stays with probability $1-\mu$. Both particles in the leftmost cluster jump with probabilities $p$ and $\mu$.    }
	\label{fig: GTASEP}
\end{figure}
This dynamics is illustrated in fig. \ref{fig: GTASEP}.
Particular limiting cases of this dynamics corresponding to $\mu=0$ and $\mu=p$ are known as  TASEP with  parallel and backward sequential updates respectively \cite{rajewsky1998asymmetric}. In the former, at each time step  all  particles are updated simultaneously, having the same probability $p$ of jump to the right, provided that  the right neighboring site in the current configuration is empty.     In the latter they also jump to the right with the same probability $p$ being updated sequentially from right to left. The fact that GTASEP interpolates between the two updates is responsible for the coined  term generalized update.

\begin{figure}[h]
	\center{\includegraphics[width=0.7\textwidth]{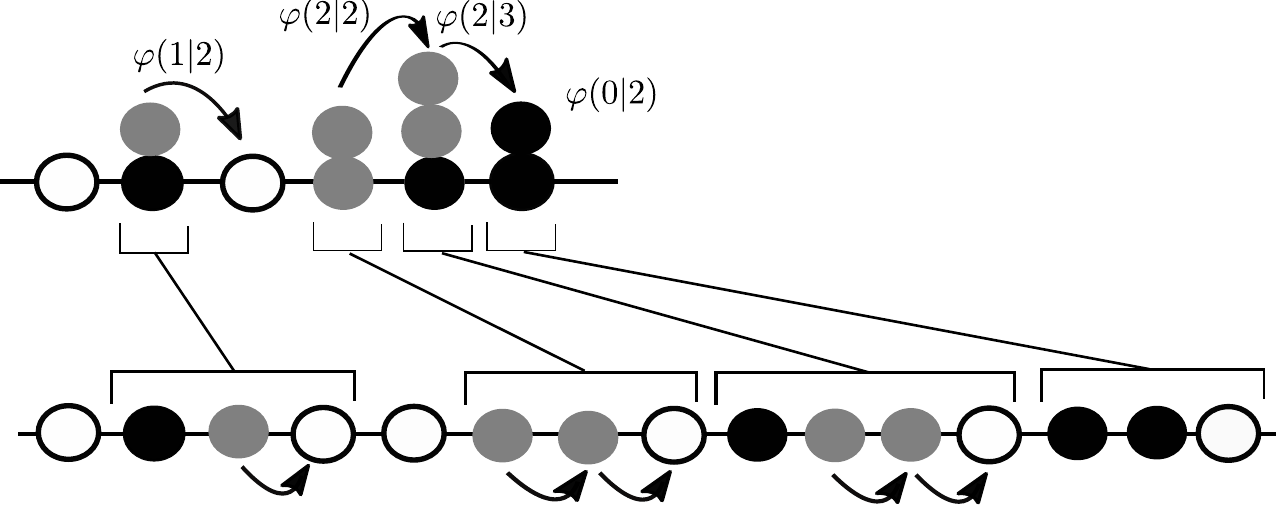}}
	\caption{An update of the q-Hahn boson model and of the associated model with exclusion interaction obtained via the ZRP-ASEP mapping. In the q-Hahn boson model $m$ particles jump to the next site on the right from a site with $n\geq m$ particles with probability $\varphi(m|n)$. Particles having decided to jump  are shown in gray color. Under the  ZRP-ASEP mapping a site with $n$ particles is replaced by a cluster of $n$ one-particle sites  plus one empty site on the right. A jump of $m$ particles from a site in the q-Hahn boson model corresponds to the  shift of  the rightmost $m$-particle piece of  the corresponding cluster one step  to the right.}
	\label{fig: q-Hahn & ZRP-ASEP}
	\end{figure}

One can  summarize the update of each cluster saying that the transition 
\[
(0,1^{(k)},0)\to(0,1^{(k-l)},0,1^{(l)})
\]
 occurs with probability $\varphi(l|k)$ defined by
\begin{eqnarray}
\varphi(l|k) &=&p\mu^{l-1}(1-\mu),\,\,\,\mathrm{for\,\,\,}0<l<k\nonumber \\
\varphi(0|k) &=&(1-p),\,\,\,\,\,\,\,\,\,\,\,\,\,\,\,\,\,\,\,\,\,\,\,\,\,\,\,\,0<k\label{eq: phi(m|n)}\\
\varphi(k|k) &=&p\mu^{k-1},\,\,\,\,\,\,\,\,\,\,\,\,\,\,\,\,\,\,\,\,\,\,\,\,\,\,\,\,\,\,0<k\nonumber 
\end{eqnarray}
 for every cluster independently. One may recognize in these formulas
the $q=0$ limit of the jumping probabilities of three-parametric
integrable chipping model \cite{P2013} also known as q-Hahn boson model
\cite{BCPS2013}. That model is  defined as a system of particles 
on a one-dimensional lattice with zero-range type interaction. Configurations $\mathbf{n}=\{n_i\}_{i\in \mathbb{Z}}$ of the model are also defined in terms of the numbers of particles at each site,  which now can take any value $n_i\geq 0$. The system evolves in  discrete time. Within an update, given current configuration $\mathbf{n}$,     we draw the number $0 \leq m_i\leq n_i $ of particles to jump from every site $i\in\mathbb{Z}$  from the probability distribution $\varphi(m_i|n_i)$ independently of the others. The exmples of updates are shown in  fig. \ref{fig: q-Hahn & ZRP-ASEP}.      The  three-parametric family of distributions  $\varphi(m|n)$   ensuring the Bethe ansatz solvability of the model was obtained in \cite{P2013} in the form  
\begin{equation}
\varphi(m|n)=\mu^m\frac{(\nu/\mu;q)_m(\mu;q)_{n-m}}{(\nu;q)_n}\frac{(q;q)_n}{(q;q)_m(q;q)_{n-m}},\label{eq: q-Hahn phi(m|n)}
\end{equation} 
where   $\mu,\nu$ and $q$ are real parameters to be chosen  such that   that $\varphi(m|n)$ is a probability distribution,  and $(a;q)_n=(1-a)\dots(1-aq^{n-1})$ is the q-Pochhammer symbol.
 One can check that the $q=0$ limit of (\ref{eq: q-Hahn phi(m|n)})  coincides with (\ref{eq: phi(m|n)}) if one sets 
\[
\nu=\frac{\mu-p}{1-p}.
\]
To see that  the two models are  tightly related we  establish a connection between the  zero-range process (ZRP), where the occupation numbers are unbounded,  and the ASEP-like system with at most one particle in a site referred to as ZRP-ASEP mapping in \cite{P2013}.  To this end, we   replace a site with  $n$ particles in the ZRP-like system by the sequence of $n$ sites with one particle in each plus one empty site on the right, see fig. \ref{fig: q-Hahn & ZRP-ASEP}.  The correspondence between  evolutions of these two systems is then one-to-one. 

Taking the $q=0$ limit leads to a crucial simplification
of the model. In this case it acquires the structure of a determinantal
process, which makes the calculation of all finite-dimensional distributions
of particle positions possible. The results on the distribution for
two cases of IC are given in the next subsection. 

\subsection{Finite-dimensional distributions: exact formulas}

Here we present exact formulas for joint distributions of positions
of tagged particles at fixed time $t$, given initially the particles
either densely occupied the negative half of the lattice 
\[
\eta(0)=\eta_{\mathrm{step}}:=\left\{ \eta_{i}=\mathbbm{1}_{i<0}\right\} _{i\in\mathbb{Z}}
\]
or every second site of the whole lattice 
\[
\eta(0)=\eta_{\text{alt}}:=\left\{ \eta_{i}=\mathbbm{1}_{i\in2\mathbb{Z}}\right\} _{i\in\mathbb{Z}}.
\]
As usual, these two configurations are referred to as step and alternating
IC, respectively. To keep track of particles' history
we assign every particle with an integer index $n$ and denote the
position of particles at time $t$ by $x_{n}(t)$, assuming that initially
\begin{align*}
x_{n}(0) & =-n,\,\,n\in\mathbb{Z}_{>0}
\end{align*}
for step initial conditions, and 

\[
x_{n}(0)=-2n,\,\,n\in\mathbb{Z}
\]
for alternating IC.

A simpler problem of description of the GTASEP evolution of a finite
particle configuration was solved in \cite{gtasep}. There, the Green
function, i.e. the joint distribution of positions of all particles
given an arbitrary initial $N$-particle configuration was obtained
in form of the determinant of $N\times N$ matrix. The following result
based on that result, however, addresses the evolution of infinite
particle configurations. An infinite-dimensional random process can
be described by specifying the complete set of finite dimensional
distributions, associated in general with arbitrary sets of time points.
Below we provide the formulas for purely spacial multipoint distributions,
aka the distributions of positions of finite subsets of particles
associated with the fixed moment of time. 

The finite dimensional  distributions will have   the form  similar to 
that obtained earlier for several other models of this type. Specifically, the $m$-point distribution will be given by the  Fredholm determinant of an  operator acting on functions on the set   $\mathbb{Z}\times(1,\dots,m)$, i.e. the the disjoint union of $m$ copies of the set $\mathbb{Z}$, where particle coordinates live. Let the action of the operator, say denoted by $K$, on functions  
\begin{equation}
(Kf)(n,x)=\sum_{i=1}^m \sum_{y\in\mathbb{Z}} K(n,x;i,y)f(i,y)\label{eq: K action}
\end{equation}
be defined in terms of a  matrix  kernel\footnote{The term ``matrix kernel'' refers to the fact that the operator $K$ can be interpreted as acting on  the   $m$-dimensional vector functions $(f(1,x_1),\dots,f(m,x_m))$.  Then the arguments $n$  and $i$ in $K(n,x;i,y)$ are interpreted as matrix indices.} $K(n,x ;i,y)$.  Then, the Fredhom determinant can be represented as  series 
\begin{eqnarray}
&&\det(1-K)_{\ell^2\left(\{1,\dots,m\}\times\mathbb{Z}\right)}\label{eq: Fredholm series}\\&&:=\sum_{k=0}^\infty\frac{(-1)^k}{k!}\sum_{i_1,\dots,i_k=1}^m \sum_{x_{i_1},\dots,x_{i_k}\in \mathbb{Z}}\det\{K(i_l,x_{i_l},i_s,x_{i_s})\}_{1\leq s,l\leq k}.\notag
\end{eqnarray}
Though,  other more abstract  definitions of the Fredhom determinant exist,  it is the  definition  (\ref{eq: Fredholm series}) referring only to the form of the kernel that  will be exploited in the present paper. 
\begin{theorem}
\label{K} Consider $m$ particles with indices $n_{1}<\dots<n_{m}$
evolving under  GTASEP  conditioned to an (infinite) initial
configuration $\eta(0)$, which can be either $\eta_{step}$ or $\eta_{alt}$.
The joint probability for  their positions $x_{n_{1}}(t)>\dots>x_{n_{m}}(t)$
to take values in half-axes $(x_{n_1}\geq a_1,\dots,x_{n_m}\geq a_m)$ is given by Fredholm
determinant 
\begin{equation}
\mathbbm{P}\Big(\bigcap_{k=1}^{m}\big\{ x_{n_{k}}(t)\geq a_{k}\big\}|\eta(0)\Big)=\det(\boldsymbol{1}-\bar{P}_{a}K_{t}\bar{P}_{a})_{\ell^{2}(\{1,\dots,m\}\times\mathbbm{Z})}\label{Prob-1}
\end{equation}
of the  operator $\bar{P}_{a}K_{t}\bar{P}_{a}$ with the $m\times m$ matrix kernel 
$$K(k,x;l,y)=\bar{P}_a(k,x)K_t(n_k,x;n_l,y)\bar{P}_a(l,y),\quad 1\leq k,l\leq m,$$
composed of  projectors
\begin{equation}
\bar{P}_{a}(n,x)=\mathbbm{1}_{x<a_{n}}\label{projector}
\end{equation}
restricting the internal summations in (\ref{eq: K action},\ref{eq: Fredholm series}) to the complementary half-axes and the kernel of the operator $K_t$ of the form
\begin{equation}
K_{t}(n_{k},x;n_{l},y)=-\phi^{*(n_{k},n_{l})}(x,y)+\widetilde{K}_{t}(n_{k},x;n_{l},y)\label{eq: K_t}
\end{equation}
where 
\begin{eqnarray}
\phi^{*(n_{k},n_{l})}(x,y)=\mathbbold{1}_{n_{l}>n_{k}}\oint_{\Gamma_{1}}\frac{dv}{2\pi\mathrm{i}}\frac{(\nu-1)(1-v)^{n_{l}+y-n_{k}-x-1}}{v^{n_{l}-n_{k}}(1-\nu v)^{n_{l}+y-n_{k}-x+1}}\label{Phi}
\end{eqnarray}
and $\widetilde{K}_{t}(n_{k},x;n_{l},y)$ is either
\begin{align}
\!\!\!\!\widetilde{K}_{t}^{step}(n_{k},x;n_{l},y)\nonumber \\
 & \!\!\!\!\!\!\!\!\!\!\!\!\!\!\!\!\!\!\!\!\!\!\!\!\!\!\!\!\!\!\!\!\!\!\!\!\!\!\!\!\!\!\!\!\!=\oint_{\Gamma_{1}}\frac{du}{2\pi\mathrm{i}}\oint_{\Gamma_{0}}\frac{dv}{2\pi\mathrm{i}}\frac{u^{n_{k}}(1-\mu u)^{t}(1-v)^{y+n_{l}}(1-\nu u)^{n_{k}+x-t-1}}{v^{n_{l}}(1-\mu v)^{t}(1-u)^{x +n_{k}+1}(1-\nu v)^{n_{l}+y-t}}\frac{(1-\nu)}{(v-u)}\label{eq:K_step}
\end{align}
or 
\begin{align}
\widetilde{K}_{t}^{alt}(n_{k},x ;n_{l},y) & =\oint_{\Gamma_{0}}\frac{dv}{2\pi\mathrm{i}}\frac{(1-v)^{y+n_{l}+n_{k}}(1-p+vp)^{t}}{v^{x+n_{k}+n_{l}+1}(1-\mu v)^{t}(1-\nu v)^{y+n_{l}+n_{k}-t}}\label{K_alt}
\end{align}
for step and alternating IC respectively. Here $\Gamma_{0}$ (resp
$\Gamma_{1}$) is any simple loop, anticlockwise oriented, which encloses
the only pole $v=0$ $(v=1)$ and no any other poles of the integrand. 
\end{theorem}
Formulas (\ref{Phi}) and (\ref{eq:K_step}) of the kernel for step IC were first obtained in \cite{KnizelPetrovSaenz}. 
\subsection{Scaling limits}

\begin{figure}
	\includegraphics[width=0.45\textwidth]{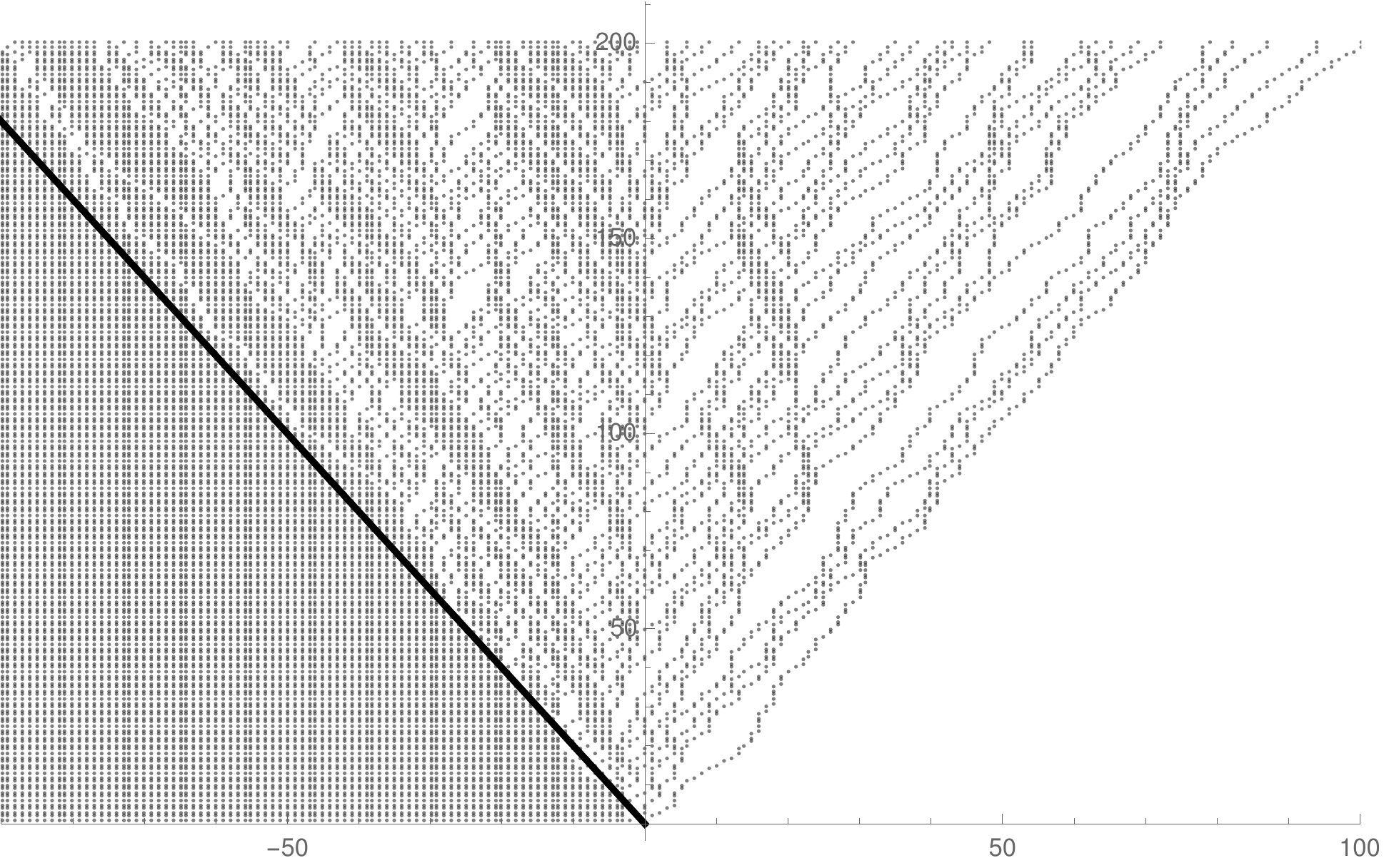}\vspace{0.2\textwidth}	\includegraphics[width=0.45\textwidth]{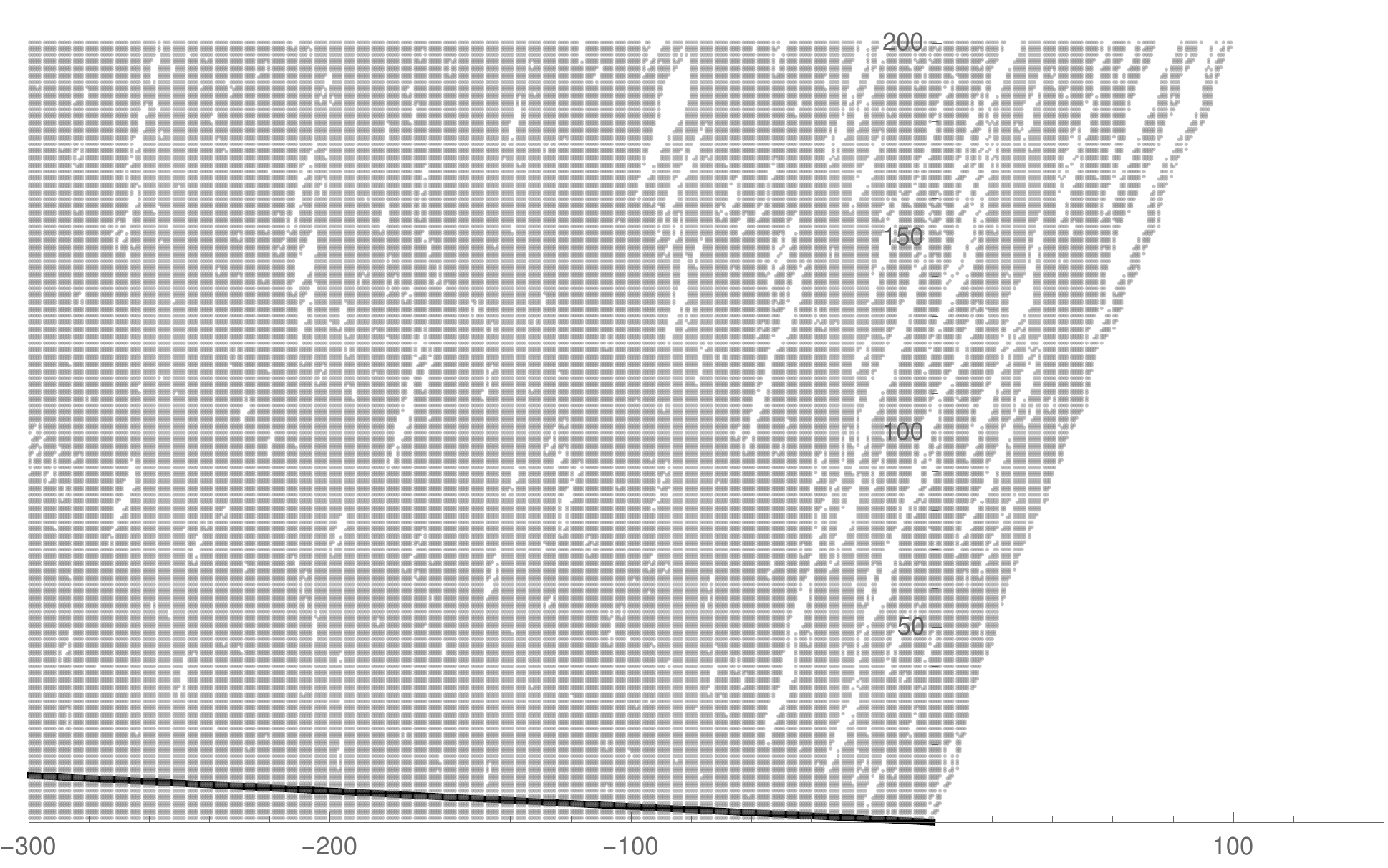}\vspace{-2.5cm}\\
\centerline{\small\hspace{-0.11\textwidth}(a)\hspace{0.4\textwidth}(b)}\vspace{0.2cm}\\
\includegraphics[width=0.45\textwidth]{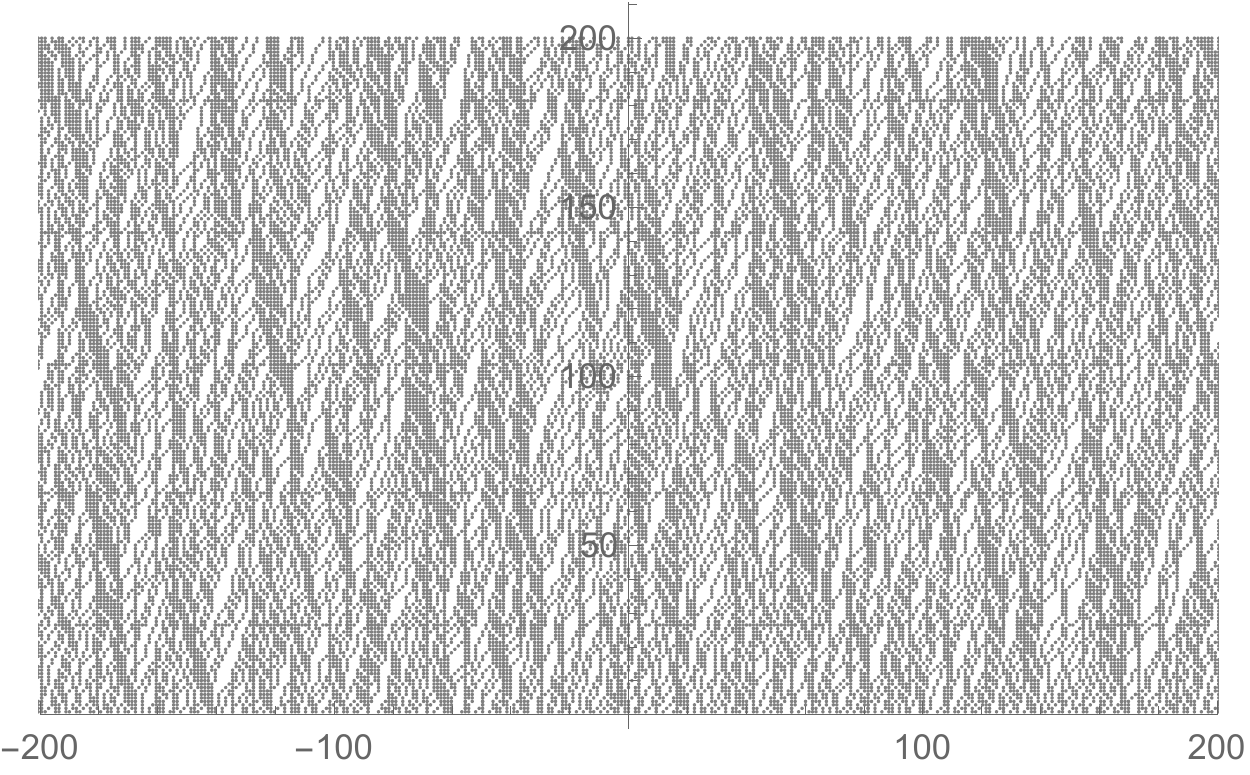}\vspace{0.2\textwidth}	\includegraphics[width=0.45\textwidth]{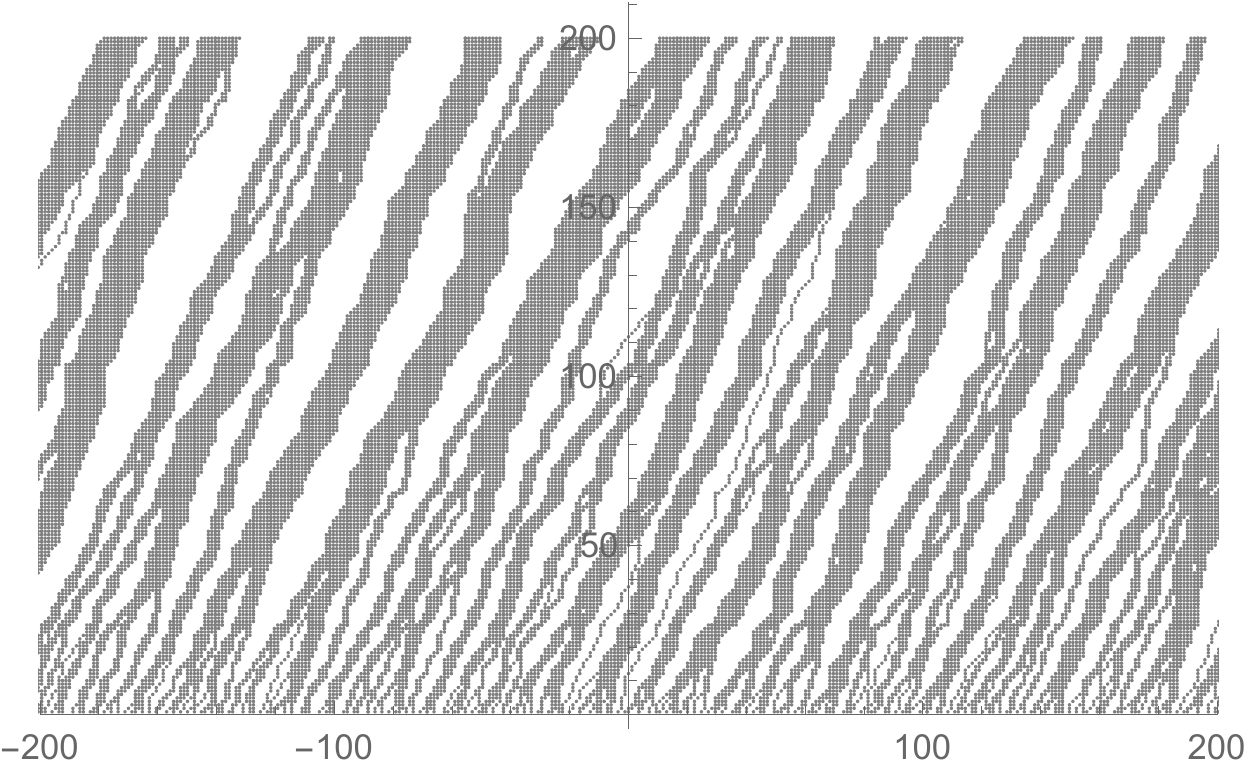}\vspace{-2.5cm}\\
		\centerline{\small\hspace{-0.11\textwidth}(c)\hspace{0.4\textwidth}(d)}\\
\caption{The 200 steps samples of evolution of GTASEP with step, (a) and (b), and alternating, (c) and (d), initial conditions  corresponding to KPZ and transitional regimes respectively. Space and time coordinates are displayed on the horizontal and vertical axes respectively.  
	The values of parameters are  $\mu=0$ in (a),(c) and $\mu=0.99$ in (b),(d) and $p=0.5$ in all cases. The black line in step IC plots (a) and (b) is the left border of the rarefaction fan. }	\label{fig: GTASEP samples}	
\end{figure}	

Though the finite dimensional distributions contain complete information
about the process, the above results are of limited scope, being of
complicated form, specific for the particular model only. They acquire
a general meaning in  scaling limits, when we zoom out the system
to time and space scales in which the microscopic details are not
important and the universal features specific for large classes of
systems remain. 

Below we consider two different scaling limits of the exact formulas  
corresponding to two types of behavior illustrated by  fig.\ref{fig: GTASEP samples}, which we refer to as KPZ and transitional regimes. 
The KPZ regime is observed in the large time limit $t\to \infty $   at  generic values of the parameters of the model, $0<p<1$ and $0\leq\mu<1$.
In this case  GTASEP  represents a  typical driven-diffusive system with short-range interactions 
subject to an uncorrelated random force, which is    generally believed to belong 
to the KPZ universality class.  The transitional regime  is associated with the simultaneous  limits $\mu \to 1$ and $t\to\infty$    competing 
 for the choice between the   KPZ regime and the DA limit, in which particles irreversibly merge into growing clusters moving diffusively.

Technically the limits will be obtained from the asymptotic analysis of the exact formulas. 
However, before going to exact  statements  let us discuss how the scaling limits are expected to look like on a heuristic level. 
In particular, in the KPZ part we introduce all the model-dependent constants that will then be used in the statement of the 
theorem \ref{thm: KPZ convergence} and explain  how they naturally appear from the quasi-equilibrium hypothesis as the characteristics 
of the translation invariant stationary state of the infinite system. Also, we  explain the  physical reasoning behind the 
choice of the mutual scaling of $t$ and $(1-\mu)$ that will be used in theorem \ref{thm: Transitional regim} for the transitional regime.

\subsubsection{KPZ scaling \label{subsec: KPZscaling}}

In the KPZ regime the distance traveled by a particle in the bulk of the system consists of the deterministic $O(t)$ part and of the random fluctuations  on the scale $O(t^{1/3})$.  
In a mathematically rigorous sense the convergence of random variables like $x_{n}(t)/t$  to non-random ones, known as the law of large
numbers (LLN), is widely believed and proved for some models to hold almost surely \cite{Rost}, \cite{Pablo_Ferrari}.
Heuristically, the value of the limit  follows from  the hypothesis of local
\textit{quasi-equilibrium} meaning that from the large scale
perspective particles are carried by the particle flow, which locally
behaves in the same way as the one in the stationary state of the
infinite translationally invariant system. The only function responsible
for everything taking place in the hydrodynamic (i.e. of order of
$t$) scale is the particle current $j_{\infty}(c)$ maintained by
such a system at particle density $c$. 

For example, in a  system with  constant particle density $c$ a particle moves with constant velocity, $x_n(t)-x_n(0)=v(c)t$, where the particle velocity $v(c)$ is related to the current by  $v(c) =j_{\infty}(c)/c $. It is the case for the alternating IC corresponding to $c=1/2$. In the non-stationary setting of step IC the particles move in the varying density profile described  
by the Euler hydrodynamics governed by the particle conservation law. We refer the reader to  appendix \ref{app: Hydrodynamics}, were we  explain that 
in the large time limit, $t\to\infty$, the coordinate of a particle with number  $n=\theta t$  is given by $x_n\simeq t\chi(\theta),$ where the function $\chi(\theta)$ is an  inversion of the function $\theta (\chi)$ obtained as a Legandre  transform
\begin{equation}
\theta=\inf_{c\in[0,1]}(j_{\infty}(c)-c\chi)\label{eq: thetavar}
\end{equation}
of  the function $j_\infty(c)$ assumed to be convex.

Thus,  having the function $j_\infty(c)$ as an input the hydrodynamics  fixes the linear in time deterministic part of $x_n(t)$ corresponding to LLN. 
The random part is the KPZ analogue of the central limit theorem. The KPZ universality  suggests   that the random fluctuations of order of $O(t^{1/3})$   are non-trivially correlated 
between the spacial positions within the scale  $O(t^{2/3})$.     Hence, choosing  the number $n$ of a particle  in  $O(t^{2/3})$ vicinity of some  macroscopic
reference value $t \theta $, 
\begin{equation}
n=t\theta+t^{2/3}\kappa_c u \label{eq: n}
\end{equation}
we expect that the distance traveled by a particle  will have the form
\begin{eqnarray}
x_{n}(t)\simeq x_n(0) +tj_\infty(c)/c+\kappa_f \mathcal{X}_{flat} (u)t^{1/3},\label{eq: x_n-x_0,flat}
\end{eqnarray}
for flat IC and
\begin{eqnarray}
x_{n}(t)\simeq t\chi(\theta)+\kappa_f \mathcal{X}_{step} (u)t^{1/3},\label{eq: x_n-x_0,step}
\end{eqnarray}
 for step IC with $\chi(\theta)$ being the inverse of   (\ref{eq: thetavar}).  The scaling factors  $\kappa_f=\kappa_f(c)$ and $\kappa_c=\kappa_c(c)$ are  model-dependent parameters to be defined below, aka units in the fluctuation and correlation scales respectively depending on  particle density $c$, the parameter $u$ is the scaled particle number relative to the macroscopic reference value $t \theta$ and $\mathcal{X}_{flat}(u),\mathcal{X}_{step}(u)$ are families of correlated random variables parameterized by $u$, i.e.  random processes expected to be universal within the KPZ universality class for the same large scale shapes of initial conditions.
In the spirit of local quasi-equilibrium $\kappa_f$ and $\kappa_c$   are functions of   $\theta$  in the case of step IC, since the density $c=c(\chi)$ depends on the scaled coordinate $\chi=\chi(\theta)$, which is in turn related with $\theta$ by (\ref{eq: thetavar}). That the model dependent scaling factors
have the same form in the two cases is another demonstration of the KPZ universality.   

The conjectures (\ref{eq: x_n-x_0,flat},\ref{eq: x_n-x_0,step}) for the distances traveled by particles are formulated in terms of yet unknown  functions $j_\infty(c),\chi(c),\kappa_f(c)$ and the function $\chi(\theta)$ related with $j_\infty(c)$ via (\ref{eq: thetavar}). 
They will be derived in course of  the asymptotic  analysis of the exact  distributions. In this way, the functions appear as a result 
of technical calculations, while their physical meaning remain beyond our scope. At the same time, these quantities can  be given a transparent interpretation in the spirit of the local quasi-equilibrium by expressing them in terms of observables of the   translation invariant  stationary state  in the  infinite system. We  sketch the argument here, while the details of calculations are moved to appendix sections    \ref{app: Stationary state and deterministic relations},\ref{app: KPZ dimensional invariants }.

Instead of considering  the GTASEP, it is simpler to deal with the equivalent ZRP-like system described above. There is a one-parametric family  of transitionally invariant stationary measures in such systems, which are product measures, see e.g. \cite{Woelki,DPP2015}. Specifically the numbers of particles  $n_i,i\in\mathbb{Z},$ \ are identically distributed  independent random variables with one-site distribution      
\begin{equation}   
\mathbb{P}(n_i=k)=\frac{f(k)z_c^k}{\mathfrak{z}(z_c)}	\label{eq: P(n_i)}
\end{equation}	
parameterized by a single parameter $0\leq z_c\leq 1$ called the particle fugacity.  Here $f(k)=\delta_{k,0}+(1-\delta_{k,0})(1-\nu)$ is the one site weight and $\mathfrak{z}(z_c)$ is the normalization factor or one-site partition function.
The particle  density and current in such a system can be calculated as corresponding averages
$$\rho=\mathbb{E}n_i, \quad j_\infty^{ZRP}=\mathbb{E}\sum_{m=0}^{n_i}m\varphi(m|n_i).$$
while for their counterparts in the ASEP-like system we obtain  $c=\rho/(1+\rho)$ and $j_\infty=j_\infty^{ZRP}(1-c)$. In terms of $z_c$ they take the following form
\begin{eqnarray}
c & =&\frac{(1-\nu)z_{c}}{\nu z_{c}^{2}-2\nu z_{c}+1},\label{eq: c(z)}\\
j_{\infty} & =&\frac{(\mu-\nu)\left(1-z_{c}\right)z_{c}}{\left(1-\mu z_{c}\right)\left(1-\nu\left(2-z_{c}\right)z_{c}\right)},\label{eq: j_infty} 
\end{eqnarray}
which used with (\ref{eq: j'=00003Dchi},\ref{eq: thetavar}) yeilds
\begin{eqnarray}
\chi & = & \frac{(\mu-\nu)(1-2z_{c}+z_{c}^{2}(\mu+\nu-2\mu\nu)-\mu\nu z_{c}^{4}+2\mu\nu z_{c}^{3})}{(1-\nu)(1-\mu z_{c})^{2}(1-\nu z_{c}^{2})},\label{eq: chi}\\
\theta & = & \frac{(\mu-\nu)(1-\mu)z_{c}^{2}}{(1-\mu z_{c})^{2}(1-\nu z_{c}^{2})}.\label{eq: theta}
\end{eqnarray}
\begin{figure}[t]
	\centerline{\includegraphics[width=0.45\textwidth]{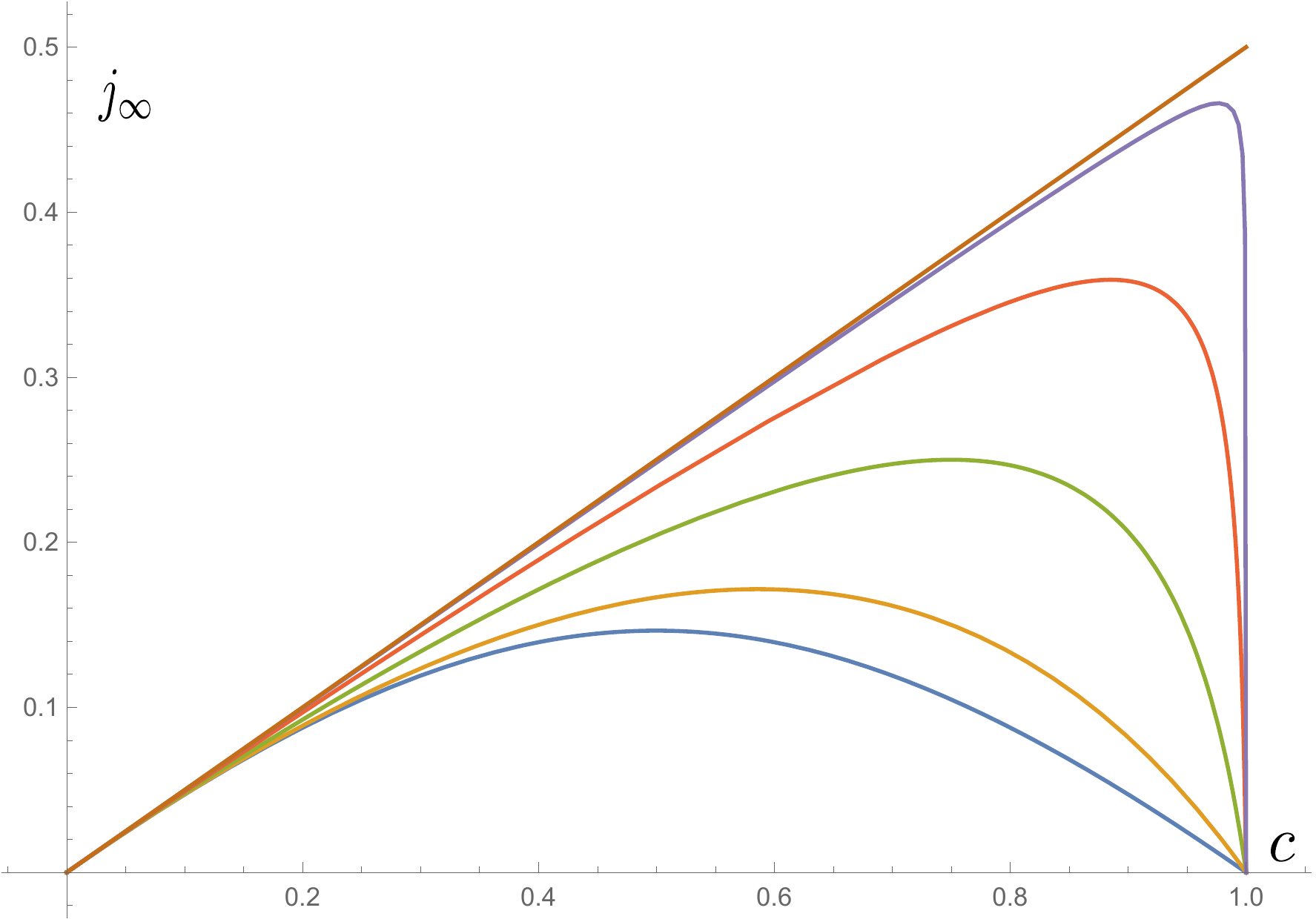} \hspace{0.05\textwidth}\includegraphics[width=0.45\textwidth]{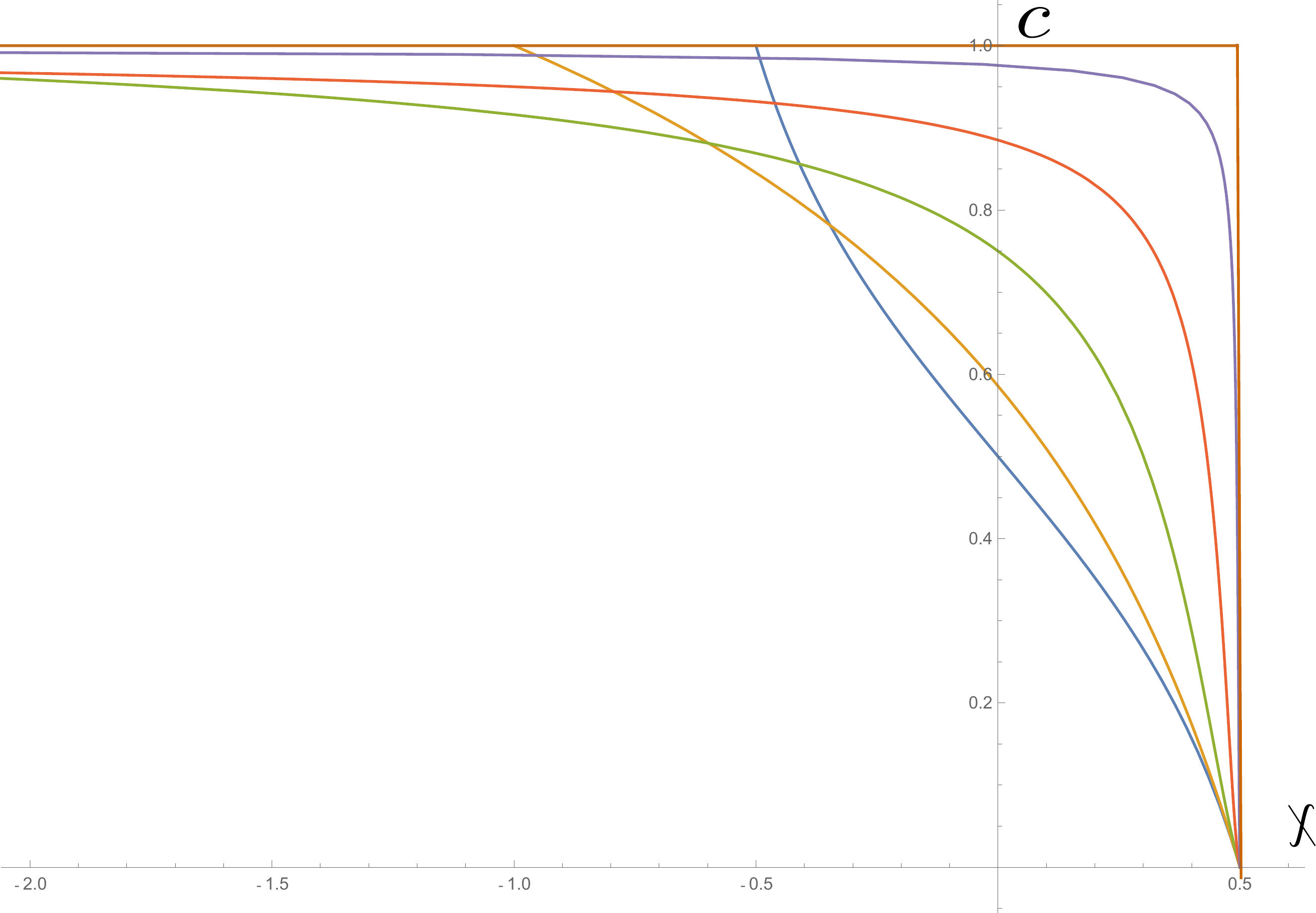}}
	
	\centerline{(a)\hspace{0.45\textwidth}(b)}
	\caption{Current-density diagrams $j_\infty(c)$  (a) and particle density profiles $c(\chi)$ (b) for $p=1/2$ and different values of $\mu$. Going from bottom to top the plots correspond to  $\mu=0,0.5,0.9,0.99,0.9999,1.$ \label{fig: current-density}}
\end{figure}
In  appendix section \ref{app: Stationary state and deterministic relations} we prepare  the stationary state of the GTASEP in the infinite system starting from that on the finite ring and then taking the thermodynamic limit. There, the parameter $z_c$ naturally appears as a saddle point of the integrand within the  integral represenation of the stationary state averages.
The latter approach also allows calculation of the universal  finite size correction to the current on the ring used in  construction of the scaling factors $\kappa_f,\kappa_c$.  In  section \ref{app: KPZ dimensional invariants }, these quantities are expressed in terms of two dimensionful scaling invariants of the KPZ theory  using the dimensional analysis and scaling hypothesis proposed in \cite{KM1990,amar1992universal,KMH1992}.  Connection of  the dimensionful invariants with  stationary state observables allows one to express $\kappa_f$ and $\kappa_c$ in terms of $z_c$.
\begin{eqnarray}
\kappa_{f} & =&\frac{\left[(1-\mu)(\mu-\nu)(1-\mu\nu z_{c}^{3})\right]^{1/3}\left[\left(1-\nu z_{c}\right)\left(1-z_{c}\right)\right]{}^{2/3}}{\left(1-\nu\right)\left(1-\mu z_{c}\right)\left[z_{c}(1-\nu z_{c}^{2})\right]^{1/3}}\label{eq: kappa_f}  \\
\kappa_{c} & =&\frac{z_c^{4/3}\left[\left(1-\nu z_{c}\right)\left(1-z_{c}\right)\right]^{1/3}\left[(1-\mu)(\mu-\nu)\left(1-\mu\nu z_{c}^{3}\right)\right]^{2/3}}{\left(1-\mu z_{c}\right){}^{2}\left(1-\nu z_{c}^{2}\right){}^{5/3}}\label{eq: kappa_c} 
\end{eqnarray}

The formulas (\ref{eq: c(z)}-\ref{eq: kappa_c}) obtained heuristically will appear below from the rigorous steepest descent analysis of the integral formulas (\ref{Phi}-\ref{K_alt}), where $z_c$ is the double saddle (critical) point of the integrand, hence the subscript.
To obtain one of the quantities as a function of another, one has to eliminate $z_c$ between the  pair of equations. An example is the function    $j_\infty(c)$ first appeared in \cite{Woelki}. Obtained  by resolving the quadratic equation (\ref{eq: c(z)}) with respect to $z_c$ and substituting the result into (\ref{eq: j_infty}) it has relatively simple form, while  the other functions, like e.g. $c(\chi)$, are  more complex involving the roots 
of polynomials of higher degrees. On the other hand the parametric form  (\ref{eq: c(z)}-\ref{eq: kappa_c})  turns out to be  more convenient to 
work with. In particular this is the form these quantities arising from the steepest descent analysis below.  

In fig. \ref{fig: current-density} we plot the current-density diagrams, $j_\infty(c)$, and  the density profiles, $c(\chi)$, for different values of $\mu$. One can see that as $\mu$ approaches one, the first graph  tends to the linear plot $j_\infty=pc$ and the second one becomes the step function $c=\mathbbm{1}(\chi<p)$. This limit is the  subject of the following discussion.

\subsubsection{Transitional scaling \label{subsec: Trans scaling}}

The closer $\mu$ is to one, the more effort particles make to take over the particles
ahead making the typical particle cluster in the system longer. In
the DA limit, which we define as the limit $\mu\to1$ as $p<1$ is kept
constant, the clusters irreversibly merge into bigger clusters, each
moving as a single particle.   In particular, they move diffusively,
i.e. in the limit $t\to\infty$ the fluctuations of traveled distance
$x_{n}(t)$ measured in the diffusive scale are Gaussian, $x_{n}(t)\simeq pt+\left(\Delta t\right)^{1/2}\mathcal{N},$
where $\mathcal{N}$ is the standard normal random variable.  The signature
of the diffusive behavior is a non-zero diffusion coefficient,
which is $\Delta=p(1-p)$ for a single particle. 

It was shown in \cite{DPP2015} for the case of periodic system that
these two types of universal behavior are connected by the transitional
regime. It takes place when the parameter 
\[
\lambda=\frac{1}{1-\nu},
\]
which diverges to infinity as $\mu\to1$ while $p$ is fixed,  scales as $L^{2}$, the
size of the system squared. When the crossover parameter   $\lambda/L^{2}$  varies from zero to infinity, 
the average number of particle clusters on the lattice gradually decreases from infinity to one and the 
statistics of particle current, specifically the  functional form of  LDF of the distance traveled by particle (measured
in the diffusive scale), changes from the KPZ to purely Gaussian
form.

To guess the scaling in the non-stationary setting we adopt the quasi-equilibrium 
hypothesis to the system of large particle clusters. Since  a cluster of particles in GTASEP corresponds
to an occupied site with occupation number distributed according to    (\ref{eq: P(n_i)}), the cluster length $l_{cl}$  is a geometric random variable with distribution 
$$\mathbb{P}(l_{cl}=k)=z_c^{k-1}(1-z_c),k\in \mathbb{N}.$$ 
In particular,   the mean cluster size is $\left\langle l_{cl}\right\rangle =(1-z_{c})^{-1}$. At large $\lambda $ and fixed density $c$, related with $z_c$ by  (\ref{eq: c(z)}), it is estimated to 
\[
\left\langle l_{cl}\right\rangle =(1-z_{c})^{-1}\simeq\sqrt{c\lambda/(1-c)}.
\]
We also estimate the average gap between clusters to be $\left\langle l_{cl}\right\rangle (1/c-1)$ by noting that the average  distance
per one cluster is  $\left\langle l_{cl}\right\rangle/c$ and the fraction of empty sites within this distance is 
$(1-c)$.

Then for the transitional regime we expect that  

-- the fluctuations of the distance traveled by a particle are diffusive,
\begin{equation}
\delta x_{n}(t)\varpropto\sqrt{p(1-p)t};\label{eq: delta x_n transition}
\end{equation}

-- for the  time under consideration  a cluster interacts with finitely many clusters, i.e. fluctuations of the traveled distance is of order of the typical gap between clusters
\begin{equation}
	\sqrt{tp(1-p)}\varpropto\sqrt{\lambda(1-c)/c};\label{eq: scales transition}
\end{equation}

-- particles are correlated, when they are  a finite number of clusters apart, i.e. the particle number $n$ should be measured  
in the units of  average cluster length, 
\begin{equation}
\delta n\varpropto\left\langle l_{cl}\right\rangle.\label{eq: delta n transition}
\end{equation}

Thus, when the particle density is fixed away from zero and one, i.e. the clusters and  gaps between them are of comparable size, both the correlation and fluctuation scales are   of order of $O(\sqrt{\lambda})$, while 
$\lambda \varpropto t$. This is  the case for the alternating IC. 

For step IC the density ranges from zero to one, and we  can also probe into the situations, in which the density is in the vicinity of the ends of its range. To this end, we   consider a family of scalings parameterized by a parameter $\beta\in(0,1)$ setting $z_c=1-O( \lambda^{ -\beta } )$, which  generalizes the above  $z_c=1-O(1/\sqrt{ \lambda} )$. Practically, this means that  the average cluster length is $\langle l_{cl}\rangle=O(\lambda^{\beta})$ and the gap length scales as $O(\lambda^{1-\beta})$.    
The length of the region containing finitely many rightmost clusters is dominated either by clusters or by gaps  corresponding to the densities  $c=1-O(\lambda^{1-2\beta})$ or $c=O(\lambda^{2\beta-1})$, when $1/2<\beta<1$ or  $0<\beta<1/2$ respectively. Then, the scaling  $\lambda^{1-\beta} \varpropto  \sqrt{t}$ as well as the correlation and fluctuation scales $\delta x_n(r)\varpropto \lambda^{1-\beta},\delta n\varpropto \lambda^{\beta}$ follow from the formulas (\ref{eq: delta x_n transition}-\ref{eq: delta n transition}).

To summarize, we expect that the transitional regime is associated with a double limit  $t,\lambda \to \infty$, such that the ratio $\lambda^{1-\beta}/\sqrt{t}$
is kept constant. Taking $n=[r \lambda^{\beta}]$ with $r$ fixed we expect that the distance traveled by the particle with number $n$ will be 
\begin{equation}
x_n(t)-x_n(0)\simeq pt+\sqrt{p(1-p)t}\mathcal{X}(r),\label{eq: transitional ansatz}
\end{equation}
where  $\mathcal{X}(r)$ is the random process expected be universal for a given global shape of initial conditions. In our cases, $\beta=1/2$ should be taken for the alternating IC, while for the step IC the whole range  $0<\beta<1$ is meaningful.

It will be convenient to introduce the crossover parameter 
\[
\tau_{\beta}=\frac{\sqrt{tp(1-p)}}{\lambda^{1-\beta}}.
\]
that controls the crossover between the KPZ regime and the DA limit. As $\tau_\beta$ increases the number of clusters involved  into the evolution grows to infinity, returning us to the KPZ regime. In the opposite  situation   $\tau_\beta\to 0$ the system approaches the DA limit, in which  $\mathcal{X}(r)$ is expected to be Gaussian and $r$-independent, since  all particles move synchronously as a single particle in this case. The coefficient of the random part  in (\ref{eq: transitional ansatz}) is chosen in such a way, that  $\mathcal{X}(r)$ becomes the standard normal variable in the DA  limit.

\subsubsection{Limiting processes}
\begin{figure}
\centerline{\includegraphics[width=0.8\textwidth]{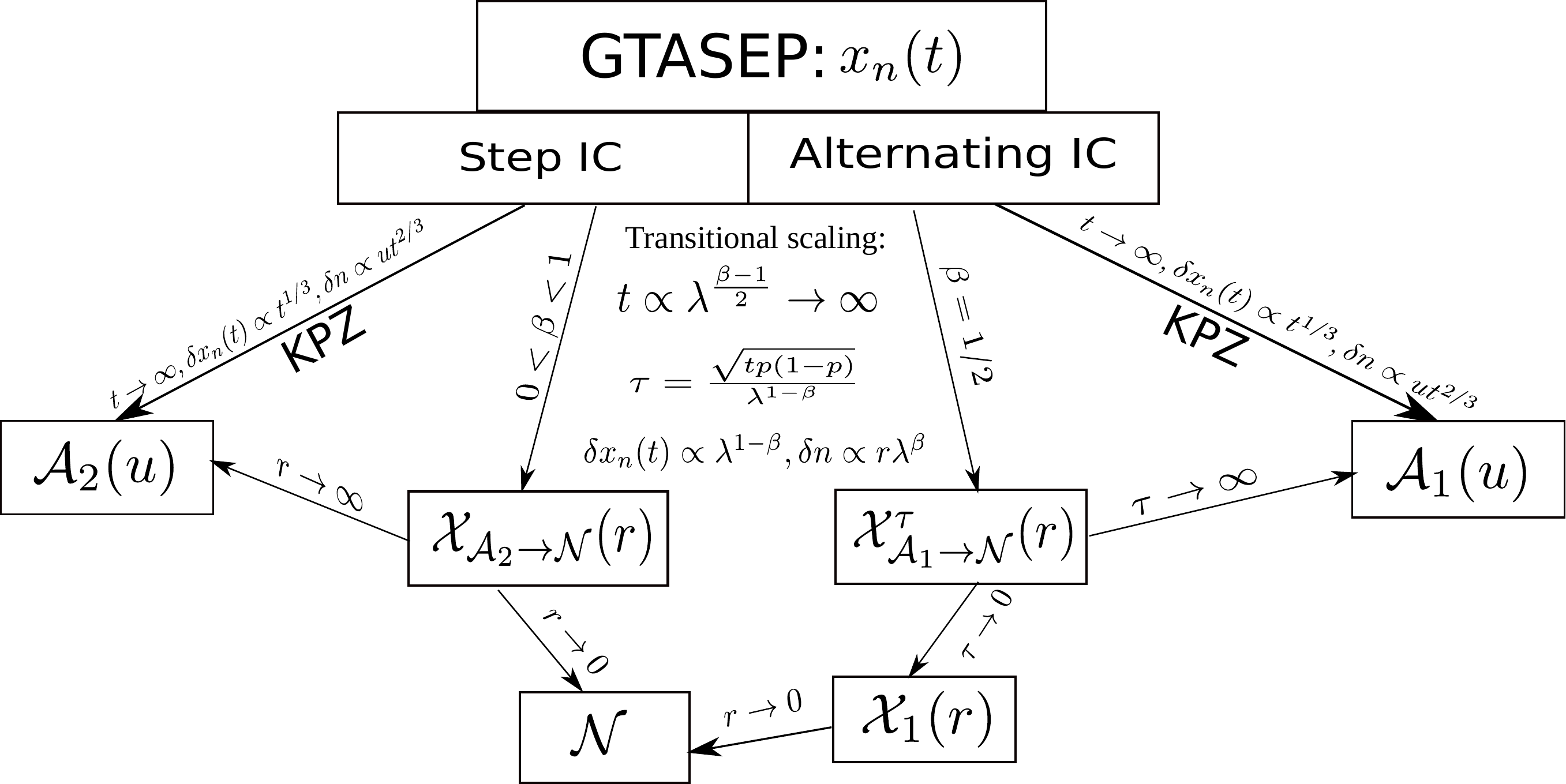}}
\caption{Limiting processes in GTASEP. The arrows show the directions  of limiting transitions. We explicitly indicate the form of the scalings corresponding to the KPZ and transitional regimes in GTASEP. Further limits can be taken of the transitional processes $\mathcal{X}_{\mathcal{A}_2\to\mathcal{N}}$ and $\mathcal{X}_{\mathcal{A}_1^\tau\to\mathcal{N}}$, which finally either return them to the KPZ regime or bring them to the DA regime, where the processes collapse into a single normal random variable $\mathcal{N}$.   We show these limits by arrows  indicating in which parameter the limit is performed.  The  limit of $\mathcal{X}^{(\tau)}_{\mathcal{A}_1\to\mathcal{N}}$ towards DA is performed in two steps via an intermediate  process $\mathcal{X}_1$. \label{fig: processes}}
\end{figure}

Now we are in position to make  rigorous statements about the limiting processes 
 that appear under the scalings (\ref{eq: x_n-x_0,flat},\ref{eq: x_n-x_0,step},\ref{eq: transitional ansatz}) and their relations with each other. The whole scheme is outlined in fig. \ref{fig: processes}. We obtain four main processes: $\mathcal{A}_1(u),\mathcal{A}_2(u)$  corresponding to Airy$_1$ and  Airy$_2$ processes that appear in the KPZ regime for alternating and step IC respectively and  $\mathcal{X}^{(\tau)}_{\mathcal{A}_1\to\mathcal{N}}$ and $\mathcal{X}_{\mathcal{A}_2\to\mathcal{N}}$ corresponding to the transitional regime for the same IC.   Then, we perform  further limits of the transitional processes towards the KPZ regime and the DA limit, yielding the  fully correlated normal random variables.

The limiting processes are also characterized by their finite dimensional distributions  obtained as  limits of  the exact distributions of  theorem \ref{K} and having the form of  Fredholm determinants of specific operators. In   the scaling limit the  particle coordinates  become continuous and the  operators involved into definitions of the $m$-point distributions  act  on functions on the disjoint union of $m$ copies of $\mathbb{R}$, i.e. on the set isomorphic to  $\{1,\dots,m\}\times\mathbb{R}$. The  action of  the operator, say $K$ again, on  functions
\begin{equation}
	(Kf)(n,x)=\sum_{i=1}^m \int_{\mathbb{R}} K(n,x;i,y)f(i,y)dy\label{eq: K action_2}
\end{equation}
is defined in terms of  kernel $K(n,x;i,y)$, which is also used to represent the Fredhom determinant in the form of  infinite series.
\begin{eqnarray}
	&&\det\left(\boldsymbol{1}-K\right)_{L^{2}((1,\dots,m)\times\mathbb{R})}\label{eq: Fredholm series 2}\\&&=\sum_{k=1}^\infty \frac{(-1)^k}{k!}\sum_{i_1,\dots,i_k=1}^m \int_{\mathbb{R}^k} \det \left\{ K(i_j,x_j;i_l,x_l) \right\}_{j,l=1}^m  d^kx.\notag
\end{eqnarray}
In the present paper we deal with two universal processes referred
to as Airy$_{1}$ and Airy$_{2}$ processes, corresponding to the
step and flat IC respectively. 
\begin{definition} 
\label{def:Airy}The  Airy$_{1}$ and Airy$_{2}$  processes are the random processes   $\mathcal{A}_{1}:\mathbb{R}\to\mathbb{R}$ and $\mathcal{A}_{2}:\mathbb{R}\to\mathbb{R}$, respectively,   defined by their finite dimensional distributions. For any  $m\in \mathbb{N}$, any $m$-point sets $r=$ ($r_{1},\dots ,r_{m})\subset\mathbb{R}$ and $(a_{1},\dots ,a_{m})\subset\mathbb{R}$ the joint distribution of  values of $(\mathcal{X}(r_1),\dots \mathcal{X}(r_m))$  with either $\mathcal{X}=\mathcal{A}_1$ or $\mathcal{X}=\mathcal{A}_2$  is given by the  Fredhom  determinant
\begin{equation}
\mathbb{P}\left(\bigcap_{k=1}^m\left(\mathcal{X}(r_{k})<a_{k}\right)\right)=\det\left(\boldsymbol{1}-P_{a}K_{\mathcal{X}}P_{a}\right)_{L^{2}(\{1,\dots,m\}\times\mathbb{R})}\label{eq: FDD fredholm}
\end{equation}
of the operator $P_{a}K_{\mathcal{X}}P_{a}$ with $m\times m$ matrix kernel 
$$K(i,x;j,y)=P_{a}(i,x)K_{\mathcal{X}}(r_i,x;r_j,y)P_a(j,y),\quad 1\leq i,j\leq m$$    composed of projectors  
\begin{equation}
	P_{a}(k,x)=\mathbbm{1}_{x\geq a_{k}}, \quad k=1,\dots,m\label{projector_2}
\end{equation}
and the kernel of an integral operator $K_{\mathcal{X}}$ being one of the kernels 
\begin{eqnarray}
K_{\mathcal{A}_{1}}(r_{i},x;r_{j},y) & = & -\frac{\mathbbold{1}_{r_{j}>r_{i}}}{\sqrt{4\pi(r_{j}-r_{i})}}\exp\left(-\frac{(y-x)^{2}}{4(r_{j}-r_{i})}\right)\label{eq:Airy1}\\
 &  & \!\!\!\!\!\!\!\!\!\!\!\!\!\!\!\!\!\!\!\!\!\!\!\!\!\!\!\!\!\!\!\!\!\!\!\!\!\!\!\!\!\!+\,\mathrm{Ai}\left((r_{j}-r_{i})^{2}+(y+x)\right)\exp\left(\frac{2}{3}(r_{j}-r_{i})^{3}+(r_{j}-r_{i})(y+x)\right),\nonumber 
\end{eqnarray}
or
\begin{eqnarray}
 & \!\!K_{\mathcal{A}_{2}}(r_{j},x;r_{i},y) & =\left\{ \begin{array}{ll}
\int_{0}^{\infty}d\xi e^{\xi(r_{j}-r_{i})}\mathrm{Ai}(\xi+x)\mathrm{Ai}(\xi+y), & r_{i}\geq r_{j}\\
-\int_{-\infty}^{0}d\xi e^{\xi(r_{j}-r_{i})}\mathrm{Ai}(\xi+x)\mathrm{Ai}(\xi+y), & r_{i}<r_{j}
\end{array}\right..\label{eq:Airy2}
\end{eqnarray}  
\end{definition}
respectively.
Here  $\mathrm{Ai}(z)$ is the Airy function that can be defined  via its integral representation 
\begin{equation}
\mathrm{Ai}(z)=\frac{1}{2\pi\mathrm{i}}\int_{\infty e^{-\mathrm{i}\phi}}^{\infty e^{\mathrm{i}\phi}}\exp\left({\frac{t^3}{3}-zt}\right)dt\label{eq: Airy}
\end{equation}
as an integral along the contour composed of two rays outgoing from the origin at angles $\pm \phi$  to the real axis, where 
$\phi$ can be chosen in the range $\phi\in(\pi/6,\pi/2)$ and most often is taken to be $\phi=\pi/3$, which ensures the most rapid  convergence.  This is the definition that will be used below. 

With these definitions in hand we are in position to make statements
about the KPZ scaling limit of particle positions. 
\begin{theorem}
\label{thm: KPZ convergence}Let $x_{n}^{\eta_{step}}(t)$ and $x_{n}^{\eta_{alt}}(t)$
be the position of a particle with number $n$ conditioned to step
and alternating IC respectively. We define the functions $j_{\infty}(c),\kappa_{f}(c),\kappa_{c}(c),c(\chi)$ and $\chi(\theta)$ as the unique solution of parametric equations (\ref{eq: c(z)}-\ref{eq: kappa_c})
obtained by eliminating the parameter $z_{c}$ varying in the range
$z_{c}\in(0,1)$.\footnote{Below the subscript in $z_{c}$ will refer to the word ``critical''\textsf{
}from the critical points, which will appear in two different contexts.} Then, as $t\to\infty$ the following limits hold in a sense of finite
dimensional distributions.

\textbf{Step IC: \cite{KnizelPetrovSaenz}}
\begin{align*}
-\lim_{t\to\infty}\frac{x_{n}^{\eta_{step}}(t)-t\chi(\theta_{u})}{t^{1/3}\kappa_{f}(c(\chi(\theta)))}= & \mathcal{A}_{2}(u),
\end{align*}
where 
\[
n=t\theta_{u}:=t\theta+2t{}^{2/3}\kappa_{c}(c(\chi(\theta)))u,
\]
the value of $\theta>0$ is such that $j'_{\infty}(0)>\chi(\theta)\geq j'(1)$.

\textbf{Alternating IC:
\[
-\lim_{t\to\infty}\frac{x_{n}^{\eta_{alt}}(t)+2n-2tj_{\infty}(1/2)}{(2t)^{1/3}\kappa_{f}(1/2)}=\mathcal{A}_{1}(u),
\]
}where
\[
n=t\theta+2(2t)^{2/3}\kappa_{c}(1/2)u.
\]
\end{theorem}
\begin{remark}
The powers of $2^{1/3}$ appearing in the coefficients is a normalization
necessary to make the definitions of $\kappa_{f}(c)$ and $\kappa_{c}(c)$
given below consistent with the above definitions of universal Airy
processes. This normalization is expected to be universal for interacting
particle systems, provided that the scaling constants are related
to the stationary state observables in the way to be specified in
the next section. 
\end{remark}

The transitional processes are defined as follows.
\begin{definition}\label{def: transition processes}
Transitional processes $\mathcal{X}_{\mathcal{A}_{2}\to\mathcal{N}}:\mathbb{R}_{\geq0}\to\mathbb{R}$
and $\mathcal{X}_{\mathcal{A}_{1}\to\mathcal{N}}^{(\tau)}:\mathbb{R}\to\mathbb{R}$ are defined by their  finite dimensional distributions. Let us fix any $m\in \mathbb{N}$ and an $m$-point subset $r=\{r_1,\dots,r_m\}$   of   $\mathbb{R}_{\geq 0}$ or  $\mathbb{R}$  for  the first or the second case respectively. Then the multi-point distributions 
of the transitional processes
 are given by the Fredholm determinant formula (\ref{eq: FDD fredholm})
with either $\mathcal{X}=\mathcal{X}_{\mathcal{A}_{2}\to\mathcal{N}}$ or $\mathcal{X}=\mathcal{X}_{\mathcal{A}_{1}\to\mathcal{N}}^{(\tau)}$  and  the kernel of $K_\mathcal{X}$ being  
\begin{align}\label{LimStep}
K_{\mathcal{A}_{2}\to\mathcal{N}}(r_{i},s_{1};r_{j},s_{2})\\
 & \!\!\!\!\!\!\!\!\!\!\!\!\!\!\!\!\!\!\!\!\!\!\!\!\!\!\!\!\!\!\!\!\!\!\!\!\!\!\!\!\!=\frac{1}{4\pi^{2}}\int_{\Gamma_{(-\mathrm{i}\infty,0^{-},\mathrm{i}\infty)}\!\!\!\!\!\!\!\!\!\!\!}dx_{1}\oint_{\Gamma_{0}}dx_{2}\frac{x_{1}\exp\left(\frac{x_{1}^{2}-x_{2}^{2}}{2}+\frac{r_{i}}{x_{1}}+s_{1}x_{1}-\frac{r_{j}}{x_{2}}-s_{2}x_{2}\right)}{x_{2}(x_{1}-x_{2})}\nonumber \\
 & \,\,\,\,\,\,\,\,\,\,\,\,\,\,\,\,\,\,-\mathbbold{1}_{r_{j}>r_{i}}\left[\mathbbold{1}_{s_{2}>s_{1}}\sqrt{\frac{r_{ji}}{s_{21}}}I_{1}(2\sqrt{s_{21}r_{ji}})+\delta(s_{21})\right]\nonumber 
\end{align}
or 
\begin{align}
K_{\mathcal{A}_{1}\to\mathcal{N}}^{(\tau)}(r_{i},s_{1};r_{j},s_{2})\\
 & \!\!\!\!\!\!\!\!\!\!\!\!\!\!\!\!\!\!\!\!\!\!\!\!\!\!\!\!\!\!\!\!\!\!\!\!\!\!\!\!\!\!\!\!\!\!=\tau\int_{\Gamma_{(-\mathrm{i}\infty,0^{-},\mathrm{i}\infty)}\!\!\!\!\!\!\!\!\!\!\!}\frac{dx}{2\pi \mathrm{i}}\exp\left[\tau^{2}\frac{x^{2}-x^{-2}}{2}+\tau\left(s_{1}x-s_{2}x^{-1}-r_{ji}(x+x^{-1})\right)\right]\nonumber \\
 & \!\!\!\!\!\!\!\!\!\!\!\!\!\!\!\!\!\!\!\!\!\!\!\!\!\!-\mathbbold{1}_{r_{j}>r_{i}}\left[\delta(s_{21}+r_{ji})+\mathbbold{1}_{s_{21}+r_{ji}>0}\frac{\tau I_{1}\left(2\tau\sqrt{r_{ji}\left(r_{ji}+s_{21}\right)}\right)}{\sqrt{1+s_{21}/r_{ji}}}\right]\!\!,\nonumber 
\end{align}
respectively, where $1\leq i,j\leq m $ and for brevity we used the notations $r_{ji}=r_{j}-r_{i}$
and $s_{ji}=s_{j}-s_{i}$. Here, $I_{1}(x)$ is the modified Bessel
function of the first kind defind by the integral
\begin{equation}
I_{n}(x)=\oint_{\Gamma_0} t^{-n-1}   e^{(x/2) (t+1/t)},
\end{equation}
and	$\delta(x)$ is the Dirac delta function.The integration contour $\Gamma_{0}$
is a counterclockwise loop closed around the origin and the contour
$\Gamma_{(-\mathrm{i}\infty,0^{-},\mathrm{i}\infty)}$ is parallel
to the imaginary axis going from $-\mathrm{i}\infty$ to $\mathrm{i}\infty$
leaving the origin and $\Gamma_{0}$ on the right, Fig.\ref{fig:Integration-contours}.  
We also explicitly  emphasize that in this case the Fredholm determinant is understood as the Fredholm sum (\ref{eq: Fredholm series 2}),  convergence of which will be proved below. Apparently, the kernels shown can not be used to define the trace class operators,  at least unless properly conjugated. The operator content of   Fredholm determinants with these kernels is yet to be understood.  

\begin{figure}
\center{\includegraphics[width=0.4\textwidth]{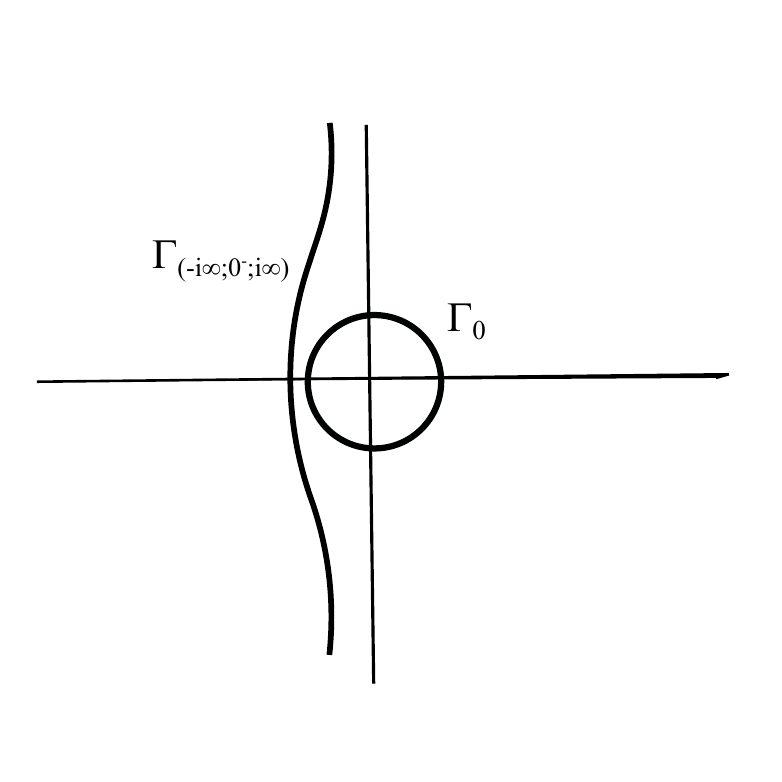}}
\caption{Integration contours\label{fig:Integration-contours}}
\end{figure}

Note that the second kernel depends on the time variables only via
difference parameter $r_{ji}$. Thus, $\mathcal{X}_{\mathcal{A}_{1}\to\mathcal{N}}^{(\tau)}(r)$
is stationary, while $\mathcal{X}_{\mathcal{A}_{2}\to\mathcal{N}}(r)$
is not. This is not unexpected as in the latter case the time parameter
$r$ describes the coordinate along the variable density profile,
while in the latter the density is constant. 

Then, we can formulate the statements about the convergence of particle
coordinates in the transitional regime. 
\end{definition}
\begin{theorem}
\label{thm: Transitional regim}Consider simultaneous limits $t\to\infty,\lambda\to\infty$,
such that $\tau_{\beta}=\lambda^{\beta-1}\sqrt{tp(1-p)}$, is constant.
Then, for any $0<\beta<1$ and $\theta>0$ the limit 
\[
\lim_{t\to\infty}\frac{pt-\left(x_{[r\lambda^{\beta}]}^{\eta_{step}}(t)+r\lambda^{\beta}\right)}{\sqrt{tp(1-p)}}=\mathcal{X}_{\mathcal{A}_{2}\to\mathcal{N}}(\tau_{\beta}r),\,\,\,r\in\mathbb{R}_{\geq0}
\]
holds in the sense of finite dimensional distributions.

Also, for $\beta=1/2$ we have

\[
\lim_{t\to\infty}\frac{pt-\left(x_{[r\sqrt{\lambda}]}^{\eta_{alt}}(t)+2r\sqrt{\lambda}\right)}{\sqrt{tp(1-p)}}=\mathcal{X}_{\mathcal{A}_{1}\to\mathcal{N}}^{(\tau_{1/2})}(r/\tau_{1/2}),\,\,\,r\in\mathbb{R}.
\]
\end{theorem}

As was announced above, the processes obtained are expected to interpolate
between the KPZ and DA regimes as $\tau_{\beta}$ varies in its range.
For $\mathcal{X}_{\mathcal{A}_{2}\to\mathcal{N}}(r)$ this would be
a statement about the behavior of the process at large and small
values of the ``time'' parameter $r$ respectively, while for $\mathcal{X}_{\mathcal{A}_{1}\to\mathcal{N}}^{(\tau)}(r)$
we need to formulate the limits in terms of two parameters $\tau$
and $r$. As usual the ingredients of the KPZ part include the large
scale deterministic part that should be extracted and the random part
characterized by the fluctuation and correlation scales. The scales
are supposed to be consistent with those of the KPZ regime described
above after returning from $r$ and $\tau$ back to the variables
of the original process. One can check that this is indeed the case
in the limits below. 
\begin{proposition}\label{Prop: KPZ tails}
KPZ tails:

Let $\mathcal{X}_{\mathcal{A}_{2}\to\mathcal{N}}(r)$ and $\mathcal{X}_{\mathcal{A}_{1}\to\mathcal{N}}^{(\tau)}(r)$ be the processes defined in Def.  \ref{def: transition processes}. The following limits hold in the  sense of finite dimensional distributions.
\begin{equation}
\left(2/3\right)^{1/3}\left(2r\right)^{1/9}\left(\mathcal{X}_{\mathcal{A}_{2}\to\mathcal{N}}\left(r_{u}\right)-\frac{3}{2}\left(2r_{u}\right)^{1/3}\right)\underset{r\to\infty}{\longrightarrow}\mathcal{A}_{2}(u)-u^{2},\label{eq:X->Airy_2}
\end{equation}
where $r_{u}:=r+u\left(2r\right)^{7/9}\left(\frac{3}{2}\right)^{2/3},$
and

\begin{equation}
\left(\frac{\tau}{3}\right)^{1/3}\left(\mathcal{X}_{\mathcal{A}_{1}\to\mathcal{N}}^{(\tau)}(u\,3^{2/3}\tau^{1/3})-\tau\right)\underset{\tau\to\infty}{\longrightarrow}\mathcal{A}_{1}(u).
\end{equation}
\end{proposition}
In the DA limit we expect that all particles move diffusively along
the same trajectory, that is to say that the random variables representing
the limiting process at different times are identical and normally
distributed. To attain the limit for $\mathcal{X}_{\mathcal{A}_{2}\to\mathcal{N}}(r)$,
we rescale the time $r$ by a factor $\epsilon,$ which is sent to
zero afterwards. As a result we indeed come to the process that is
represented by a single normal random variable independent of  ``time''
parameter. The situation is more delicate for $\mathcal{X}_{\mathcal{A}_{1}\to\mathcal{N}}^{(\tau)}(r)$
because of the extra parameter $\tau$. In this case we come to the
limit in two steps. In the first step we perform the limit $\tau\to0,$
that brings us to yet another random process $\mathcal{X}_{1}(r)$.
\begin{definition}\label{def: X_1}
The random process $\mathcal{X}_{1}:\mathbb{R}_{\geq 0}\to\mathbb{R}$ is defined by its  finite dimensional distributions
given by  the Fredhom determinant formula (\ref{eq: FDD fredholm})  with $\mathcal{X}(r)=\mathcal{X}_{1}(r)$ and the  kernel defined by  
\begin{equation}
K_{\mathcal{X}_{1}}(r_{i},s_{1};r_{j},s_{2})=\frac{e^{-\frac{1}{2}\left(s_{1}+r_{i}-r_{j}\right)^{2}}}{\sqrt{2\pi}}-\mathbbold{1}_{r_{j}>r_{i}}\delta(s_{21}+r_{ji}),  1\leq i,j\leq m\label{K_chi}
\end{equation}
\end{definition}
The process $\mathcal{X}_{1}(r)$ has normal one-point distribution,
while the multipoint distributions seem to be nontrivial, though have
not been studied yet. The subsequent limit is taken in the same way
as the one for $\mathcal{X}_{\mathcal{A}_{2}\to\mathcal{N}}(r)$,
yielding the similar result. Note that as $\mathcal{X}_{\mathcal{A}_{1}\to\mathcal{N}}^{(\tau)}(r)$
is stationary the limits taken refer to the vicinity of an arbitrary
point $r\in\mathbb{R}$, while the statement about $\mathcal{X}_{\mathcal{A}_{2}\to\mathcal{N}}(r)$
is the property of $r=0$ only. 
\begin{proposition}\label{Prop: DA tails}
DA tails: 

Let $\mathcal{X}_{\mathcal{A}_{2}\to\mathcal{N}}(r)$, $\mathcal{X}_{\mathcal{A}_{1}\to\mathcal{N}}^{(\tau)}(r)$ and $\mathcal{X}_{1}(r)$ be the processes defined in Defs.  \ref{def: transition processes},\ref{def: X_1}. The following limits hold in the sense of finite dimensional distributions

\begin{equation}
\mathcal{X}_{\mathcal{A}_{2}\to\mathcal{N}}(\epsilon r)\underset{\epsilon\to0}{\longrightarrow}\mathcal{N},\quad r \in\mathbb{R}_{\geq 0}, \label{eq:A2->N}
\end{equation}

\begin{equation}
\mathcal{X}_{\mathcal{A}_{1}\to\mathcal{N}}^{(\tau)}(r)\underset{\tau\to0}{\longrightarrow}\mathcal{X}_{1}(r),\quad r \in\mathbb{R} \label{eq:A2->Chi1}
\end{equation}

\begin{equation}
\mathcal{X}_{1}(\epsilon r)\underset{\epsilon\to0}{\longrightarrow}\mathcal{N},\quad r \in\mathbb{R} \label{eq:Chi1->N}
\end{equation}
where $\mathcal{N}$ is the standard normal random variable that does
not depend on $r$ anymore. 
\end{proposition}
At the first glance the limits (\ref{eq:A2->N}) and (\ref{eq:Chi1->N})
look like the statement about the one-point distribution at one point
$r=0$. We want to emphasize, however, that all the above limits are
stated in terms of finite-dimensional distributions of all orders.
The statement like this could be the consequence of the continuity
of sample paths, if we had one a priori. Technically, the limits (\ref{eq:A2->N}) and (\ref{eq:Chi1->N})
follow from  the convergence of the finite-dimensional distributions to those of the  processes $\mathcal{X}_{\mathcal{N}}$ and  $\mathcal{X}_{\mathcal{N},+}$ defined as follows.
\begin{definition}\label{def: normal}
The random processes $\mathcal{X}_{\mathcal{N},+}:\mathbb{R}_{\geq 0}\to\mathbb{R}$ and $\mathcal{X}_{\mathcal{N}}:\mathbb{R}\to\mathbb{R}$ are defined by their  finite dimensional distributions given by  the Fredhom determinant formula (\ref{eq: FDD fredholm})  with $\mathcal{X}=\mathcal{X}_{\mathcal{N},+}$ or $\mathcal{X}=\mathcal{X}_{\mathcal{N}}$
and the  same kernel   
\begin{eqnarray}
	K_{\mathcal{N}}(r_{i},s_{1};r_{j},s_{2})  =\frac{\exp\left(-\frac{1}{2}s_{1}^{2}\right)}{\sqrt{2\pi}}-\mathbbm{1}_{r_j> r_i}\delta(s_2-s_1),   1\leq i,j\leq m, \label{Gauss full}
\end{eqnarray}
for both cases.
\end{definition}
Then, statements (\ref{eq:A2->N}) and (\ref{eq:Chi1->N})  follow from the fact that the values of both processes at different points are trivially correlated each having the normal one-point distribution.
\begin{proposition}\label{prop: identical}
Let $\mathcal{X}_{\mathcal{N},+}$ and $\mathcal{X}_{\mathcal{N}}$ be the random processes defined in Def. \ref{def: normal}. For any $m\in \mathbb{N}$ and any $m$-point subset $r=(r_1,\dots,r_m)$ of either $\mathbb{R}_{\geq 0}$ or  $\mathbb{R}$ for the former or the latter process respectively the identity
\begin{equation}
\mathcal{X}(r_1)=\dots=\mathcal{X}(r_m)
\end{equation}
holds almost surely and 
\begin{equation}
\mathcal{X}(r_1)\sim \mathcal{N},
\end{equation}
for both $\mathcal{X}=\mathcal{X}_{\mathcal{N},+}$ and $\mathcal{X}=\mathcal{X}_{\mathcal{N}}$.
\end{proposition}
The statement that the Fredholm determinant (\ref{eq: FDD fredholm})  with the kernel similar to (\ref{Gauss full}) defines the joint distribution 
of  trivially correlated random variables is a consequence of Prop. \ref{KN}   proved in the end of this paper. The kernel (\ref{Gauss full}) can be obtained as a limit from the kernel describing a  particle obeying the Ornstein-Uhlenbeck process, which in turn was found in \cite{imamura2007dynamics} as a limiting kernel for the tagged particle dominated by a defect particle in TASEP with slow particles, see formula   (3.46) in \cite{imamura2007dynamics}.

The   kernel (\ref{LimStep}) that yields the distribution interpolating between the Tracy-Widom and normal distributions is to be compared with the kernel from  
\cite{baik2005phase}   appearing in the  BBP distribution. Although they look different, there are similarities that make us wonder about possible connections between two transitions. Let us briefly discuss the resemblance with the example of  the  TASEP with step initial conditions and  finitely many slower defect particles \cite{imamura2007dynamics}. An effect of the defect particles is a formation of  slow particle clusters, aka shock waves, each moving as a single particle, i.e. diffusively. As a result, similarly our findings fluctuations of  the tagged particle can be either of KPZ type or Gaussian  depending  on whether the tagged particle is involved into the shock or not.     The crossover between the two distributions is given by the BBP distribution. Its most general form corresponds to the case of the hopping rates of the defect particles being only slightly different from the non-defect ones with the difference measured on the scale  $O(t^{-1/3})$. In this case the kernel of the Fredholm determinant defining the BBP distribution looks like the Airy kernel   plus an additional  finite rank part, with the rank equal to the number of defect particles.
Equivalently it can be represented in  an integral form  similar to the one of the Airy kernel, but with as many  additional poles  
in the integrand encoding the shifts. The poles disappear, when the hopping rates of the slow particles equalize with those of the rest of the system.  
In our case we also have different  poles in the integrand within the initial  exact kernel (\ref{eq:K_step}) owing to  different hopping probabilities in our model. However,  the order of these poles becomes  infinite in the scaling limit and the merging of different poles   leads to   
essential singularities in the integrand. It is likely that to obtain a similar behavior of the kernel in the system with slow particles one could possibly try to consider
 the number of slow particles growing to infinity, which scales together with vanishing difference between hopping rates of normal and 
 defect particles. This scenario  at least seems capable to produce  singularities of similar type. 
 
 Also, when in the TASEP with slow defect particles all the hopping rates of the defect particles  are distinct on the scale $O(1)$, only the slowest particle is relevant, and the BBP kernel degenerates to a simpler kernel appeared before in the studies of the polynuclear growth with sources \cite{imamura2004fluctuations,imamura2005polynuclear}. The one-point distribution defined as the Fredholm determinant with the latter kernel in turn interpolates between the so called GOE$^2$ \cite{baik2000limiting,forrester2000painlev}  and Tracy-Widom distributions. It has a form of the Airy kernel plus a rank one part given by the product of Airy function and the integrated Airy function. Remarkably, by observing that $x_1/(x_2(x_1-x_2))= 1/x_2+1/(x_1-x_2)$ the double integral part of the kernel (\ref{LimStep}) can be split into  infinite rank and rank one parts.\footnote{We thank  the anonymous referee for  pointing at this fact.} These are the parts, which converge to the main parts of the Airy kernel (\ref{eq:Airy2}) and of the  Gaussian kernel (\ref{Gauss full})   under the limits (\ref{eq:X->Airy_2},\ref{eq:A2->N}) respectively, while the other ones vanish. On the other hand, in the KPZ limit (\ref{eq:X->Airy_2})  the vanishing rank one part acquires the form of  the product of Airy function and the integrated Airy function 
 identical to that from the mentioned reduced BBP distribution  \cite{imamura2007dynamics}. The only difference with \cite{imamura2007dynamics} is that this part is accompanied by a  constant coefficient that makes the rank one part  vanishingly  small  comparing to the Airy kernel.  This may be an indication that our results have more in common with the BBP transition than we could find here. For example, it is an  interesting question whether it is possible to find another scaling, under which the exact kernel 
  (\ref{eq:K_step}) would again similarly converge to the sum of the Airy kernel and the rank one part described with the two being of the same order. We leave this question for further work.

 It is also worth mentioning that the  kernel of the form similar to (\ref{LimStep})   was obtained  
 in \cite{kuijlaars2011non} in studies of the double  scaling limit of the non-intersecting Bessel paths. That kernel  has an identical functional form with  (\ref{LimStep}) up to the fact 
 that the parameters responsible for the random variables and for the parameter of the process are interchanged.  Whether this  similarity is a pure coincidence or it has a deeper roots
 is also the matter for further investigation.

\section{Determinantal point process and exact distributions}
\label{sec: Determinantal point process}
In this section we prove Theorem \ref{K}. The starting point is the
determinant formula for Green function proved in \cite{gtasep}. The
Green function $G(X|Y;t)$ is the probability for particles of a finite configuration
$\eta_{t}$ at time $t$ to have coordinates $X={x_{1}>x_{2}>\dots>x_{N}}$,
given the coordinates $Y={y_{1}>y_{2}>\dots>y_{N}}$ of particles
of initial configuration $\eta_{0}$.
\begin{theorem}
The Green function $G(X|Y;t)$ has a determinantal form 
\begin{eqnarray}
G(X|Y;t) & = & \lambda^{N(X)}\det(F_{i-j}(x_{N+1-i}-y_{N+1-j},t))_{1\leq i,j\leq N},\label{g}
\end{eqnarray}
where $N(X)$ is the number of pairs of neighboring occupied sites
in the final configuration $X$, and the functions $F_{n}(x,t)$ have
integral representation 
\begin{eqnarray}
F_{n}(x,t) & = & (\nu-1)\oint_{\Gamma_{0,1}}\frac{du}{2\pi \mathrm{i}}\frac{(1-u)^{n-x-1}(1-\mu u)^{t}}{u^{n}(1-\nu u)^{n-x+t+1}}\label{F}
\end{eqnarray}
where the integration contour $\Gamma_{0,1}$ is a simple loop counterclockwise
oriented, which has the poles $u=0,1$ of the integrand inside and
the others outside. 
\end{theorem}
The key observation that allows a calculation of joint distributions
of particle positions is that the Green function being a probability
measure on the set of $N-$particle configurations is a marginal
of a measure on a bigger space 
\[
\mathcal{D}=\{x_{i}^{n}\in\mathbb{Z},0\le i\le n\le N|x_{i}^{n}>x_{i-1}^{n}\}
\]
of configurations characterized by $\left(N+1\right)(N+2)/2$ coordinates,
which turns out to be the determinantal process. To every $\boldsymbol{x}\in\mathcal{D}$
let us assign a measure 
\begin{equation}
\mathcal{M}(\boldsymbol{x})=\mathrm{const}\bigg(\prod_{n=1}^{N-1}\det(\phi(x_{i}^{n},x_{j+1}^{n+1}))_{0\leq i,j\leq n}\bigg)\det\left(\Psi_{N-j}^{N}(x_{i}^{N})\right)_{1\leq i,j\leq N},\label{eq: M}
\end{equation}
where we define the functions 
\begin{equation}
\Psi_{j}^{N}(x):=F_{-j}(x-y_{N-j},t),\quad j=0,\dots,N-1\label{eq:Psi vs F}
\end{equation}
and
\begin{equation}
\phi(x,y):=\begin{cases}
\nu-1, & y\geq x,\\
\nu, & y=x-1,\\
0, & y\leq x-2,
\end{cases}\label{eq: phi(x,y)}
\end{equation}
 and the constant ensures a unit normalization. Then we have the the following
lemma. Its proof comes back to \cite{Nagao_Sasamoto_2004} and \cite{BFPS_2007}
and follows the line of \cite{BFS}, where the details can be found.
\begin{lemma}
\label{The-Green-function-vs-measure}The Green function is the marginal
of $\mathcal{M}$
\end{lemma}
\[
G(X|Y;t)=\mathcal{M}(x_{1}^{n}=x_{n},n=1,\dots,M|x_{0}^{i}=-\infty,i=1,\dots,N),
\]
conditioned to $x_{0}^{i}=-\infty,$ that is to say that $\phi(x_{0}^{i},\cdot)=(\nu-1).$ 
\begin{svmultproof}
The proof is based on direct evaluation of the sum in r.h.s. of 
\[
G(X|Y;t)=\lambda^{1-N}\sum_{\left\{ \boldsymbol{x\in}\mathcal{D}:x_{1}^{n}=x_{n},x_{0}^{n}=-\infty,n\in1\dots,N\right\} }W(\boldsymbol{x})
\]
with 
\[
W(\boldsymbol{x})=\bigg(\prod_{n=1}^{N-1}\det(\phi(x_{i}^{n},x_{j+1}^{n+1}))_{0\leq i,j\leq n}\bigg)\det(F_{1-j}(x_{i}^{N}-y_{N+1-j},t))
\]
that uses simple matrix operations within determinants and the recurrent
relation 
\[
F_{n+1}(x,t)=(\phi*F_{n})(x,t).
\]
\end{svmultproof}

The next step is the generalization of Eynard-Mehta theorem proved
in \cite{Borodin_Rains} and applied to the TASEP in \cite{BFPS_2007}.
It states that the conditioned $\mathcal{M}$ is the determinantal
process and provides an explicit recipe for writing its correlation
kernel. We state it here already reduced for our particular case, in
which the Lindstr\"{o}m-Guessel-Viennot matrix corresponding to the
process on $\mathcal{D}$ is upper-triangular non-degenerate  matrix. 
\begin{theorem}
\label{thm: Eynard-Mehta}The conditional measure $\mathcal{M}(\boldsymbol{\cdot}|x_{0}^{i}=-\infty)$
of the form (\ref{eq: M}) is the determinatal process. To define
its correlation kernel we define functions 
\begin{eqnarray*}
\phi^{(n_{1},n_{2})}(x_{1},x_{2}) & := & \mathbbold{1}_{n_{2}>n_{1}}\phi_{n_{1}}*\cdots*\phi_{n_{2}-1}(x_{1},x_{2}),
\end{eqnarray*}
where $\phi_{n}(x,y):=\phi(x,y)$ from (\ref{eq: phi(x,y)}) and the
subindex is used to keep memory about the spaces this function connect,

\begin{equation}
	\Psi_{n-j}^{n}(x):=\phi^{(n,N)}*\Psi_{N-j}^{N}(x),\quad 1 \leq n<N,0\leq j\leq N-1\label{eq: Psi^n =00003D phi*Psi^N}
\end{equation}
and functions 
\[
\Phi_{n-k}^{n}(x)\in{\rm span}\{1,x,\ldots,x^{n-1}\}
\]
 with $k=1,\dots,n$ are polynomials of degree $\left(n-1\right)$
fixed by the orthogonality condition
\begin{equation}
\sum_{x\in Z}\Psi_{n-l}^{n}(x)\Phi_{n-k}^{n}(x)=\delta_{k,l},\quad1\leq k,l\leq n,\label{orthogonalization}
\end{equation}
and $\Phi_{0}^{n}(x)=const$. Then, under assumption that the matrix
$M$ with matrix elements 
\[
M_{ij}=(\phi_{N-i}*\phi^{(N+1-i;N)}*\Psi_{j-1}^{N})(x_{0}^{N-i})
\]
 is upper-triangular and non-degenerate with correlation kernel 
\begin{equation}
K(n_{1},x_{1};n_{2},x_{2})=-\phi^{(n_{1},n_{2})}(x_{1},x_{2})+\sum_{k=1}^{n_{2}}\Psi_{n_{1}-k}^{n_{1}}(x_{1})\Phi_{n_{2}-k}^{n_{2}}(x_{2}).\label{eqKernelFinal}
\end{equation}
\end{theorem}
\begin{remark}
The proof of upper-triangular form and non-degeneracy of the matrix
$M$ requires defining it in terms of a deformed functions $\phi_{n}(x),$
which ensures the convolution series to converge and the contours
of their integral representation to be nested. Then the formula (\ref{eqKernelFinal})
for the kernel is restored using arguments based on analytic continuation.
The whole procedure is developed in \cite{BFS} for
the TASEP with parallel update. We refer the Reader to that paper
for details of the proof.

An important corollary of $\mathcal{M}$ being the determinantal process
is the Fredholm determinant form of finite-dimensional distributions
of particle positions. Taking into account lemma \ref{The-Green-function-vs-measure}
we obtain.
\end{remark}
\begin{corollary}
\label{Kernel} Consider $m$ out of $N$ particles with indices $\sigma(1)<\dots<\sigma(m)$.
The joint distribution of their positions $x_{\sigma(1)}(t)>\dots>x_{\sigma(m)}(t)$,
conditioned to initial configuration $Y$ is 
\begin{align}
\mathbbm{P}\Big(\bigcap_{k=1}^{m}\big\{ x_{\sigma(k)}(t) & \geq a_{k}\big\}|\{x_{i}(0)\}_{i=1,\dots,N}=Y\Big)\\
 & =\det(\mathbbm{1}-\bar{P}_{a}K\bar{P}_{a})_{\ell^{2}(\{\sigma(1),\ldots,\sigma(m)\}\times\mathbbm{Z})}.\nonumber 
\end{align}
\end{corollary}
We want to apply Theorem \ref{thm: Eynard-Mehta} to two particular
cases of IC: 
\[
y_{i}=-i
\]
 and 
\[
y_{i}=-2i,
\]
with $i=1,\dots,N.$ Note that like in usual TASEPs the motion of
a particle in GTASEP is independent of the particles to the left of
it. Therefore, the finite $N$ formulas of any multipoint distributions
for particles with numbers less that $N$ coincide with the formulas
for infinite $N.$ For the second case we also finally concentrate
on particles with such large numbers $i\gg1$ that they forget that
the starting configuration is bounded from the right, thus, reproducing
the situation of infinite alternating IC. 

We start from finding the explicit form of $\phi^{(n_{1},n_{2})}(x_{1},x_{2})$
and $\Psi_{n-j}^{n}(x)$. According to (\ref{eq:Psi vs F},\ref{eq: phi(x,y)})
and (\ref{eq: Psi^n =00003D phi*Psi^N}) they are obtained by a repeated
convolution of $\phi(x,y)$ and $\Psi_{N-j}^{N}(x)$ respectively
with several copies of $\phi(x,y)$, starting with the integral representations
of the formers. This is reduced to summing geometric series under
the integrals, which yields 

\begin{equation}
\phi^{(n_{1},n_{2})}(x_{1},x_{2})=\mathbbold{1}_{n_{2}>n_{1}}\oint_{\Gamma_{0,1}}\frac{du}{2\pi \mathrm{i}}\frac{(\nu-1)(1-u)^{x_{2}-x_{1}+n_{2}-n_{1}-1}}{u^{n_{2}-n_{1}}(1-\nu u)^{x_{2}-x_{1}+n_{2}-n_{1}+1}},\label{eq: eq:  phi^(n1,n2)}
\end{equation}

\begin{equation}
\Psi_{j}^{n}(x)=(\nu-1)\oint_{\Gamma_{0,1}}\frac{du}{2\pi \mathrm{i}}\frac{u^{j}(1-u)^{y_{n-j}-x-j-1}(1-\mu v)^{t}}{(1-\nu u)^{t+y_{n-j}-x-j+1}}.\label{eq: Psi^n}
\end{equation}
The series consist of terms $\left(\left(1-\nu u\right)/\left(1-u\right)\right)^{x}$,
with $x$ running up to plus infinity. For the series to converge
the inequality $|\left(1-\nu u\right)/\left(1-u\right)|<1$ must hold.
The relation $|\left(1-\nu u\right)/\left(1-u\right)|=1$ defines
a contour, which is a circle of radius $1/(1+\nu)$ with center at
$u=1/(1+\nu)$. The convergence then takes place at any contour having
this circle inside. At the same time we still have to keep the pole
$u=1/\nu$ outside of the contour. This conditions define $\Gamma_{0,1}$.
Note that having zero inside is superfluous for the definition of
$\Psi_{j}^{n}(x)$ with $j\geq0$, as the pole at zero is absent.
However, the final formula for the kernel includes also those with
negative $j$ still defined by (\ref{eq: Psi^n =00003D phi*Psi^N}),
where the pole at zero appears inside the contour.

Next, one has to find corresponding set of $\Phi_{j}^{n}(x)$. As
usual, we make an educated guess, which is checked against the consistency
with (\ref{orthogonalization}) afterwards. Let us consider step and
alternating IC separately.

\subsection{Step IC}
\begin{lemma}
\label{PsiPhi2} For 
\begin{equation}
y_{k}=-k,\quad k=1,\ldots,n
\end{equation}
functions $\Psi_{j}^{n}(x)$
and $\Phi_{j}^{n}(x)$ have the following integral representations
\begin{equation}
\Psi_{j}^{n}(x)=\frac{(\nu-1)}{2\pi \mathrm{i}}\oint_{\Gamma_{0,1}}du\frac{u^{j}(1-\mu u)^{t}(1-\nu u)^{x+n-1-t}}{(1-u)^{x+n+1}},\label{eq2}
\end{equation}
and 
\begin{equation}
\Phi_{j}^{n}(x)=\frac{1}{2\pi \mathrm{i}}\oint_{\Gamma_{0}}dv\frac{(1-v)^{x+n}(1-\nu v)^{t-x-n}}{v^{j+1}(1-\mu v)^{t}}.\label{eq3}
\end{equation}
for $j=0,\dots,n-1.$
\end{lemma}
\begin{svmultproof}
The formula for $\Psi_{j}^{n}$ is just obtained by substituting the
IC. The functions $\Phi_{j}^{n}(x)$ defined in (\ref{eq3}) are
obviously polynomials in $x$ of the degree not greater than $j$.
Let us check that the orthonormality relation (\ref{orthogonalization})
holds. For $j\geq0$ and $x<-n,$ there is no poles inside the contour
of integration. Therefore, the summation can be restricted to the
terms with $x\geq-n.$ Thus,
\begin{eqnarray*}
\sum_{x\in\mathbb{Z}}\Phi_{j}^{n}(x)\Psi_{k}^{n}(x) & = & \frac{\nu-1}{(2\pi \mathrm{i})^{2}}\oint_{\Gamma_{0}}dv\oint_{\Gamma_{0,1}}du\frac{(1-v)^{n}(1-\nu v)^{t-n}}{v^{j+1}(1-\mu v)^{t}}\\
 & \times & \frac{u^{k}(1-\mu u)^{t}(1-\nu u)^{n-1-t}}{(1-u)^{n+1}}\sum_{x=-n}^{\infty}\left(\frac{(1-v)(1-\nu u)}{(1-u)(1-\nu v)}\right)^{x},
\end{eqnarray*}
where the convergence of the series in r.h.s. implies the constraint
$|\frac{1-v}{1-\nu v}|<|\frac{1-u}{1-\nu u}|$ on the integration
contours, which is fulfilled if $\Gamma_{0}$ is inside $\Gamma_{0,1}$.
The sum in the r.h.s is evaluated to 
\[
\left(\frac{(1-v)(1-\nu u)}{(1-u)(1-\nu v)}\right)^{-n}\frac{(1-u)(1-\nu v)}{(1-\nu)(v-u)}.
\]
Now the pole at $u=1$ has disappeared. Instead, there is a simple
pole at $u=v$ that yields 
\begin{eqnarray}
\sum_{x\in\mathbb{Z}}\Phi_{j}^{n}(x)\Psi_{k}^{n}(x) & = & \frac{1}{2\pi \mathrm{i}}\int_{\Gamma_{0}}dvv^{k-j-1}=\delta_{j,k}.
\end{eqnarray}
\end{svmultproof}

\hphantom{}
\begin{svmultproof}
\textit{of the first part of Theorem \ref{K}.} We substitute (\ref{eq2},
\ref{eq3}) to (\ref{eqKernelFinal}).
\begin{align}
\sum_{k=1}^{\infty}\Psi_{n_{1}-k}^{n_{1}}(x_{1})\Phi_{n_{2}-k}^{n_{2}}(x_{2})=\frac{(1-\nu)}{(2\pi \mathrm{i})^{2}}\oint_{\Gamma_{1,0}}du\oint_{\Gamma_{0}}dv\\
\times\frac{u^{n_{1}}(1-\mu u)^{t}(1-\nu u)^{n_{1}+x_{1}-t-1}(1-v)^{x_{2}+n_{2}}}{v^{n_{2}}(1-\mu v)^{t}(1-\nu v)^{n_{2}+x_{2}-t}(1-u)^{x_{1}+n_{1}+1}}\frac{1}{(v-u)}.\nonumber 
\end{align}
 Since $\Phi_{j}^{n}(x)=0$ for $j<0$, we extend the summation
over $k$ up to $\infty$. We can interchange the order of summation
and integration provided that the contours satisfy $\left|v/u\right|<1$.
Then we compute the geometric series and get rid of the pole at $u=0$
for the price of getting a new simple pole at $u=v$. The residue
at this pole gives an integral over $\Gamma_{0}$, which is nonzero
when $n_{2}>n_{1}$, with the same integrand as in the definition
(\ref{eq: eq:  phi^(n1,n2)}) of $\phi^{(n_{1},n_{2})}(x_{1},x_{2})$.
Within the sum (\ref{eqKernelFinal}) it exactly cancels the part
of $\phi^{(n_{1},n_{2})}(x_{1},x_{2})$ coming from the pole at $u=0$.
Thus we obtain the $\phi^{*(n_{1},n_{2})}(x_{1},x_{2})$ , defined
by the integral over $\Gamma_{1}$ and the double integral part, where
the integration in $u$ is over $\Gamma_{1}$ as well. This concludes
the proof. 
\end{svmultproof}

\subsection{Alternating IC}
\begin{lemma}
\label{PsiPhi} For 
\begin{equation}
y_{k}=-2k, \quad k=1,\ldots,n \label{eq: IC, alt}
\end{equation}
 function $\Psi_{j}^{n}(x)$
and $\Phi_{j}^{n}(x)$ have following integral representation :
\begin{equation}
\Psi_{j}^{n}(x)=\frac{(\nu-1)}{2\pi \mathrm{i}}\oint_{\Gamma_{0,1}}du\frac{u^{j}(1-\mu u)^{t}(1-\nu u)^{x+2n-j-1-t}}{(1-u)^{x+2n-j+1}},\label{eq4}
\end{equation}
and 
\begin{equation}
\Phi_{j}^{n}(x)=\frac{1}{2\pi \mathrm{i}}\oint_{\Gamma_{0}}dv\frac{(1-2v+\nu v^{2})(1-v)^{x+2n-j-1}}{(1-\nu v)^{x+2n-j-t+1}v^{j+1}(1-\mu v)^{t}},\label{eq5}
\end{equation}
for $j=0,\dots,n-1.$ In particular, $\Phi_{0}^{n}(x)=1$.
\end{lemma}
\begin{svmultproof}
\ref{PsiPhi}. The formula for $\Psi_{j}^{n}(x)$ is obtained by
substituting the IC. Now we prove that the orthonormality
relation (\ref{orthogonalization}) holds. For $j\geq0$ and $x<-2n+k$
there is no poles at $u=0,1$ in $\Psi_{j}^{n}(x)$ and we can restrict
the sum to $x\geq-2n+k$. Thus 
\begin{eqnarray*}
\sum_{x\in \mathbb{Z}}\Phi_{j}^{n}(x)\Psi_{k}^{n}(x) & = & \frac{(\nu-1)}{(2\pi \mathrm{i})^{2}}\oint_{\Gamma_{0}}dv\oint_{\Gamma_{0,1}}du\frac{(1-2v+\nu v^{2})(1-v)^{2n-j-1}}{v^{j+1}(1-\mu v)^{t}(1-\nu v)^{2n-j-t+1}}\\
 & \times & \frac{u^{k}(1-\nu u)^{2n-k-1-t}}{(1-\mu u)^{-t}(1-u)^{2n-k+1}}\sum_{x=-2n+k}^{\infty}\left(\frac{(1-v)(1-\nu u)}{(1-u)(1-\nu v)}\right)^{x}.
\end{eqnarray*}
The series convergence requires the constraint on the integration
paths $|\frac{1-v}{1-\nu v}|<|\frac{1-u}{1-\nu u}|$, which suggests
$\Gamma_{0}$ to be inside $\Gamma_{0,1}$. The sum equals 
\begin{equation}
\left(\frac{(1-v)(1-\nu u)}{(1-u)(1-\nu v)}\right)^{-2n+k}\frac{(1-u)(1-\nu v)}{(1-\nu)(v-u)}.
\end{equation}
Now, the pole at $u=1$ has disappeared and instead of it there is
a simple pole at $u=v$. Thus, the integral in $u$ is just the residue
at $u=v$, leading to 
\begin{eqnarray}
\sum_{x\in \mathbb{Z}}\Phi_{j}^{n}(x)\Psi_{k}^{n}(x) & = & \frac{1}{2\pi \mathrm{i}}\oint_{\Gamma_{0}}dv\frac{(1-2v+\nu v^{2})}{(1-\nu v)^{2}}\left(\frac{v(1-v)}{1-\nu v}\right)^{k-j-1}\nonumber \\
 & = & \frac{1}{2\pi \mathrm{i}}\int_{\Gamma_{0}}dzz^{k-j-1}=\delta_{j,k}
\end{eqnarray}
where we used the variable change $z=\frac{v(1-v)}{1-\nu v}$.
\end{svmultproof}

\phantom{}
\begin{svmultproof}
\textit{of the second part of Theorem \ref{K}.} Let us substitute
(\ref{eq4}, \ref{eq5}) to (\ref{eqKernelFinal}). Since $\Phi_{j}^{n}(x)=0$
for $j<0$, we can extend the sum in $k$ up to $\infty$. The sum
can be taken inside the integrals if the integration contours satisfy
$\left|\frac{1-\nu u}{u(1-u)}\frac{v(1-v)}{1-\nu v}\right|<1$. Summation
of geometric series yields 
\begin{eqnarray*}
 &  & \sum_{k=1}^{\infty}\Psi_{n_{1}-k}^{n_{1}}(x_{1})\Phi_{n_{2}-k}^{n_{2}}(x_{2})=\frac{1}{(2\pi \mathrm{i})^{2}}\oint_{\Gamma_{0,1}}du\oint_{\Gamma_{0}}dv(\nu-1)\\
 &  & \,\,\,\,\,\,\,\,\,\,\,\,\,\times\frac{u^{n_{1}}(1-\mu u)^{t}(1-\nu u)^{n_{1}+x_{1}-t}(1-v)^{x_{2}+n_{2}}(1-2v+\nu v^{2})(1-\nu v)}{v^{n_{2}}(1-\mu v)^{t}(1-\nu v)^{x_{2}+n_{2}+1-t}(1-u)^{x_{1}+n_{1}+1}(v-u)(\frac{1-v}{1-\nu v}-u)}.
\end{eqnarray*}
Both simple poles $u=v$ and $u=\frac{1-v}{1-\nu v}$ are inside the
integration contour $\Gamma_{0,1}$, and there is no pole at $u=0$.
Separating the contribution from the pole at $u=v$ we obtain 
\begin{eqnarray}
\sum_{k=1}^{n_{2}}\Psi_{n_{1}-k}^{n_{1}}(x_{1})\Phi_{n_{2}-k}^{n_{2}}(x_{2}) & = & \widetilde{K}(n_{1},x_{1};n_{2},x_{2})\label{eq2.55}\\
 & + & \frac{(\nu-1)}{2\pi \mathrm{i}}\oint_{\Gamma_{0}}\frac{(1-v)^{n_{2}+x_{2}-n_{1}-x_{1}-1}}{v^{n_{2}-n_{1}}(1-\nu v)^{n_{2}+x_{2}-n_{1}-x_{1}+1}}dv.\nonumber 
\end{eqnarray}
Moreover, we also have 
\begin{eqnarray}\label{phi_alt}
\phi^{(n_{1},n_{2})}(x_{1},x_{2}) & = & \phi^{*(n_{1},n_{2})}(x_{1},x_{2})\\
 & + & \frac{(\nu-1)}{2\pi \mathrm{i}}\oint_{\Gamma_{0}}\frac{(1-v)^{n_{2}+x_{2}-n_{1}-x_{1}-1}}{v^{n_{2}-n_{1}}(1-\nu v)^{n_{2}+x_{2}-n_{1}-x_{1}+1}}dv.\nonumber 
\end{eqnarray}
The two last terms cancel each other. 

Thus, we have obtained the kernel, which being substituted  into  formulas (\ref{eq: Fredholm series},\ref{Prob-1}) yields the joint distributions of 
distances travelled by a subset of  $n$ particles starting from positions (\ref{eq: IC, alt}). Note that since the  evolution of a particle is independent of  particles to  the left of it, this kernel in fact  provides the finite dimensional distributions of particle positions in GTASEP conditioned to a semi-infinite alternating initial configuration 
\begin{equation}
	\eta_{\mathrm{alt}}^{< 0}:=\{\eta_{-i}=\mathbbm{1}_{i\in2\mathbb{N}}\}_{i\in\mathbb{Z}}.\notag
\end{equation} 
Note that the formula for $\phi^{*(n_{1},n_{2})}(x_{1},x_{2})$ is manifestly translationally invariant, i.e. it is invariant with respect to simultaneous shift of  $n_1,n_2$ by $1$ and $x_1,x_2$ by $(-2)$, while  $ \widetilde{K}(n_{1},x_{1};n_{2},x_{2})$ is not, having a memory of the position of the right end of the initial configuration. The remaining argument shows that if we shift the reference point deep into the bulk of the occupied part of the lattice this memory is lost.

Let us consider  the finite dimensional distributions of GTASEP    conditioned to $\eta(0)=	\eta_{\mathrm{alt}}^{< 0}$  shifting the reference point  by $2N$ steps  to the left.
\begin{equation}
	n_k \to n_k+N, \quad a_i\to a_i-2N
\end{equation} 
They  can be identified  with distributions of positions of particles in GTASEP conditioned to the shifted semi-infinite alternating initial configuration  $$\eta_{\mathrm{alt}}^{< 2N}:=\{\eta_{2N-i}=\mathbbm{1}_{i\in2\mathbb{N}}\}_{i\in\mathbb{Z}},$$ where particles occupy every second site of the lattice to the left of the site $2N$,  
\begin{eqnarray}
	\mathbbm{P}\Big(\bigcap_{k=1}^{m}\big\{ x_{n_{k}+N}(t)\geq a_{k}-2N\big\}|\eta(0)=\eta_{\mathrm{alt}}^{< 0}\Big)\label{eq: shift}\\
	=\mathbbm{P}\Big(\bigcap_{k=1}^{m}\big\{ x_{n_{k}}(t)\geq a_{k}\big\}|\eta(0)=\eta_{\mathrm{alt}}^{< 2N}\Big).\notag
\end{eqnarray}
Here we  imply that the  numbering of particles   within the system conditioned to $\eta_{\mathrm{alt}}^{< 2N}$ is also shifted accordingly: $\{x_{k}(0)=-2k\}_{k\in \mathbb{N}-N}.$  If the r.h.s. of (\ref{eq: shift}) has a limit as $N\to\infty$ while keeping $a_k$ and $n_k$ finite, this will be exactly  the  distribution associated with infinite alternating initial configuration $\eta_{\mathrm{alt}}$, which we are looking for. On the other hand, the l.h.s. is  convergent since it does not depend on $N$, when $N$ is large enough. Indeed,  the values of the arguments   $x_1,x_2$  of the kernel  $\widetilde{K}(n_{1},x_{1};n_{2},x_{2})$ of (\ref{eq2.55}), which contribute to the    Fredholm  series representation of the l.h.s. of (\ref{eq: shift}),  are bounded from above by  $x_1,x_2<\max_{k=1}^m a_k-2N$  due to the projectors (\ref{projector}). Substituting this bound in the place of  $x_1$ together with $n_1 \to n_1+N$ into $\widetilde{K}(n_{1},x_{1};n_{2},x_{2})$ we observe that the pole at $u=1$ is guaranteed to disappear, when $N>\max_{k=1}^m (a_k+n_k)+1.$
 Computing the remaining residue at $u=(1-v)/(1-\nu v)$ we arrive at the $N$-independent translationally invariant formula  (\ref{K_alt}) for the kernel involved into the Fredhom determinant representation of the limiting distributions.
\end{svmultproof}

\section{Asymptotic analysis: KPZ regime}
\label{sec: Asymptotic analysis: KPZ regime}

We would like to analyze the asymptotics of the Fredholm determinants
understood as a sum (\ref{eq: Fredholm series}) as $t\to\infty$. To this end, we study the $t\to\infty$ limit of
the kernel. For this limit to be exchangeable with the summation one
should use arguments based on the uniform convergence and integrability
of the kernel in terms of new rescaled variables. 

\subsection{Step initial configuration }

\subsubsection*{Expansion near the double saddle point}

Let us write the kernel in the form 
\begin{eqnarray}
K_{t}^{step}(n_{1},x_{1};n_{2},x_{2})=(1-\nu)\frac{\mathbbold{1}_{r_{2}>r_{1}}}{2\pi \mathrm{i}}\oint_{\Gamma_{1}}\frac{e^{t(f(\chi_{2},\theta_{2},v)-f(\chi_{1},\theta_{1},v))}}{(1-\nu v)(1-v)}dv\label{f kernel}\\
+\frac{(1-\nu)}{(2\pi \mathrm{i})^{2}}\oint_{\Gamma_{1}}du\oint_{\Gamma_{0}}dv\frac{e^{t(f(\chi_{2},\theta_{2},v)-f(\chi_{1},\theta_{1},u))}}{(1-\nu u)(1-u)}\frac{1}{(v-u)}\nonumber 
\end{eqnarray}
where 
\begin{equation}\chi_{i}=x_{i}/t,\,\,\,\theta_{i}=n_{i}/t \label{chi_i,theta_i}\end{equation} 
for $i=1,2$, and
we introduce the function 
\begin{equation}
f(\chi,\theta,u)=(\theta+\chi)\ln(1-u)-\theta\ln u-(\theta+\chi-1)\ln(1-\nu u)-\ln(1-\mu u).\label{f}
\end{equation}

An essential part of $t\to\infty$ asymptotical analysis of the sum
(\ref{eq: Fredholm series}) is an evaluation of integrals in (\ref{f kernel})
in the saddle point approximation. Of course the location of the saddle
points depends on the running summation indices. In particular, in
the double integral part these are two saddle points of the same function
$f(\chi,\theta,v)$, one for each integration variable. 

In the KPZ scaling regime the asymptotic behavior of the whole sum
is dominated by the values of the indices, where these two saddle
points coalesce into a double saddle point. For  the single integral
part it is also the case. 

The position $z_{c}$ of the double saddle point is defined by the
conditions 

\begin{equation}
f^{(0,0,1)}(\chi,\theta,z_{c})=0,f^{(0,0,2)}(\chi,\theta,z_{c})=0,\label{eq:dsp}
\end{equation}
where the superscripts denote the numbers of derivations with respect
to corresponding variables. Note that we again use the notation $z_{c}$
for the quantity, which is seemingly different from what it has been
reserved for. However, let us look at (\ref{eq:dsp}) more carefully.
These are two polynomial equations for $z_{c}$
of degrees three and six. Their consistency impose a constraint on
values of $\theta$ and $\chi$. Solving the pair of equations as
a linear system for $\theta$ and $\chi$ we can express them in terms
of the location of the double saddle point $z_{c}$. It is not a surprising
coincidence that we arrive at the formulas (\ref{eq: chi},\ref{eq: theta})
obtained from the analysis of the stationary state.
In view of this and to avoid multiplication of notations we use $z_{c}$
for the location of the double saddle point, implying that it is defined
by its functional dependence (\ref{eq: chi},\ref{eq: theta}) on $\theta$
and $\chi$. 

Let us make an expansion of the function $f(\chi,\theta,z)$ near
$z_{c}$. The vicinity of the double saddle point that brings dominant
contribution into the integrals is of order of $(z-z_{c})\sim t^{-1/3}.$
In addition, we suggest that the values of $\theta$ and $\chi$ vary
near their large scale positions as 
\begin{equation}
\theta_{r}:=\theta+2r\kappa_{c}t^{-1/3},\text{\ensuremath{\chi}}_{r,s}:=\chi(\theta_{r})-s\kappa_{f}t^{-2/3},\label{eq:theta_r, chi_r,s}
\end{equation}
where $\chi(\theta)$ is the function defined parametrically by (\ref{eq: chi},\ref{eq: theta})
for $z_{c}\in(0,1)$ (see the proof of Lemma \ref{lem: uniform estimate phi psi} below), the variables $r$ and $s$ characterize the
displacements of order of $t^{2/3}$ and $t^{1/3}$ of the corresponding
quantities from their macroscopic positions on the scale of order
of $t$, and the constants $\kappa_{f}$ and $\kappa_{c}$ are yet
to be defined. (Unlike the previous formulas, see e.g. (\ref{f kernel}),
the subscripts $r$ and $s$ in the notations $\theta_{r},\text{ \ensuremath{\chi}}_{r,s}$
just introduced refer to positions in the correlation and fluctuation
scales respectively. These notations will be used from now on unless
a different meaning is stated explicitly.)

The expansion of the function $f(\chi,\theta,z)$ looks as follows{
\begin{align}
f(\chi_{r,s},\theta_{r},z_{c}+u) & =f(\chi_{r,s},\theta_{r},z_{c}(\theta_{r})+\delta z_{r}+u)\label{eq:f-expansion}\\
 & \simeq f(\chi_{r,s},\theta_{r},z_{c}(\theta_{r}))+(\delta z_{r}+u)f^{(0,0,1)}(\chi_{r,s},\theta_{r},z_{c}(\theta_{r}))\nonumber \\
 & \,\,\,\,\,\,\,\,\,\,\,\,\,\,\,\,\,\,\,\,\,\,\,\,\,\,\,\,\,+\frac{(\delta z_{r}+u)^{2}}{2!}f^{(0,0,2)}(\chi_{r,s},\theta_{r},z_{c}(\theta_{r}))\nonumber \\
 & \,\,\,\,\,\,\,\,\,\,\,\,\,\,\,\,\,\,\,\,\,\,\,\,\,\,\,\,\,\,\,\,\,\,\,\,\,\,\,\,+\frac{(\delta z_{r}+u)^{3}}{3!}f^{(0,0,3)}(\chi_{r,s},\theta_{r},z_{c}(\theta_{r}))\nonumber \\
 & \!\!\!\!\!\!\!\!\!\!\!\!\!\!\!\!\!\!\!\!\!\!\!\!\!\!\!\!\!\!\!\!\!\!\!\!\!\!\!\!\!\!\!\!\!\simeq f(\chi_{r,s},\theta_{r},z_{c}(\theta_{r}))-(\delta z_{r}+u)s\kappa_{f}t^{-2/3}f^{(1,0,1)}(\chi(\theta),\theta,z_{c})\nonumber \\
 & \!\!\!\!\!\!\!\!\!\!\!\!\!\!\!\!\!\!\!\!\!\!\!\!\!\!\!\!\!\!\!\!\!\!\!\!\!\!\!\!\!\!\!\!\!\!-f^{(0,0,3)}\left(\chi(\theta),\theta,z_{c}\right)\left(\frac{4\left(z_{c}'(\theta)r\kappa_{c}\right)^{3}}{3t}-\frac{2u\left(z_{c}'(\theta)r\kappa_{c}\right)^{2}}{t^{2/3}}+\frac{u^{2}z_{c}'(\theta)r\kappa_{c}}{t^{1/3}}-\frac{u^{3}}{3!}\right)\nonumber 
\end{align}
}where $\delta z_{r}=z_{c}(\theta)-z_{c}(\theta_{r})\simeq-2r\kappa_{c}z_{c}'(\theta)/t^{1/3}$
and we keep the terms up to the order $O(1)$ assuming that $u\sim t^{-1/3}$.
The function $z_{c}'(\theta)=1/\theta'(z_{c})$ is obtained from differentiating
(\ref{eq: theta}). If we now make the variable change 
\begin{equation}
u\to ut^{1/3}\left|f^{(0,0,3)}(\chi(\theta),\theta,z_{c})/2\right|^{1/3}\label{eq: u->ut^1/3}
\end{equation}
 and set 
\begin{align}
\kappa_{f} & =\frac{\left|f^{(0,0,3)}(\chi(\theta),\theta,z_{c})\right|^{1/3}}{2^{1/3}\left|f^{(1,0,1)}(\chi(\theta),\theta,z_{c})\right|},\label{eq: kappa_f saddle point}\\
\kappa_{c} & =\frac{\theta'(z_{c})}{2^{2/3}\left|f^{(0,0,3)}(\chi(\theta),\theta,z_{c})\right|^{1/3}},\label{eq: kappa_c saddle point}
\end{align}
we obtain 
\[
\mathrm{(\ref{eq:f-expansion})}\simeq f(\chi_{r,s},\theta_{r},z_{c}(\theta_{r}))-t^{-1}\left((r^{2}-s)u-ru^{2}+\frac{u^{3}}{3}+rs-\frac{r^{3}}{3}\right),
\]
where we took into account that $f^{(0,0,3)}(\chi(\theta),\theta,z_{c})<0$,
$f^{(1,0,1)}(\chi(\theta),\theta,z_{c})<0$ and $\theta'(z_{c})>0$.
An explicit substitution of $f(\chi,\theta,z),\chi(\theta)$ and $\theta(z)$
to (\ref{eq: kappa_f saddle point},\ref{eq: kappa_c saddle point})
reproduce the formulas (\ref{eq: kappa_f},\ref{eq: kappa_c}) obtained
from the scaling arguments. Substituting this expansion into the formula
(\ref{f kernel}) we obtain

{
\begin{align}
\frac{\kappa_{f}}{t^{1/3}}K_{t}^{step}(n_{1},x_{1};n_{2},x_{2}) & \simeq\exp\left(t\left(\tilde{f}(r_{2},s_{2})-\tilde{f}(r_{1},s_{1})\right)\right)\label{eq:Airy_2 Kernel asymptotics}\\
\times\left(-\mathbbold{1}_{r_{2}>r_{1}}\int_{-\mathrm{i}\infty}^{+\mathrm{i}\infty}\right. & \frac{du}{2\pi \mathrm{i}}e^{\left(r_{2}-r_{1}\right)u^{2}-(r_{2}^{2}-r_{1}^{2}-s_{2}+s_{1})u}\nonumber \\
+\int_{\infty e^{\frac{\pi\mathrm{i}}{3}}}^{\infty e^{-\frac{\pi\mathrm{i}}{3}}}\frac{du}{2\pi\mathrm{i}} & \int_{\infty e^{-\frac{2\pi\mathrm{i}}{3}}}^{\infty e^{+\frac{2\pi\mathrm{i}}{3}}}\left.\frac{dv}{2\pi\mathrm{i}}\frac{e^{\left(u^{3}-v^{3}\right)/3-r_{1}u^{2}+r_{2}v^{2}+\left(r_{1}^{2}-s_{1}\right)u-\left(r_{2}^{2}-s_{2}\right)v}}{v-u}\right).\nonumber 
\end{align}
}Here we deformed the integration contours to the steepest descent
ones and limited the integration to small segments in a vicinity of
the double saddle point. After the variable change and sending $t$
to infinity these segments become the rays that approach the origin
at angles $\pm\pi/3,\pm2\pi/3$ with the real axis in the double integral
part and parallel to the imaginary axes in the single integral one.
Up to the factor $\exp\left[t\left(\tilde{f}(r_{2},s_{2})-\tilde{f}(r_{1},s_{1})\right)\right]$,
where $\tilde{f}(r,s)=f(\chi_{r,s},\theta_{r},z_{c}(\theta_{r}))+t^{-1}\left(\frac{r^{3}}{3}-rs\right),$
this formula is an alternative form of the extended Airy kernel, see
e.g. \cite{WeissFerrariSpohn2017}. Note also that a multiplication of the kernel
by the factor $e^{t\left(\tilde{f}(r_{2},s_{2})-\tilde{f}(r_{1},s_{1})\right)}$
results in conjugation of the corresponding operator, $K\to DKD^{-1}$,
with a diagonal operator $D$ and, hence, does not affect the value
of the Fredholm determinant.

\subsubsection*{Convergence}

To prove the convergence of the Fredholm determinant we first obtain
the $t\to\infty$ estimate of both double integral and single integral
parts of the kernel. As a result we obtain the Airy$_{2}$ kernel
plus the corrections of two sorts. First these are $O(t^{-1/3})$
corrections. They are integrable in the rescaled variables $s_{1},s_{2}$
and thus give contribution into Fredholm sum, which vanishes in the
limit $t\to\infty.$ The other corrections are exponentially small
in $t$, though their dependence on $s_{1},s_{2}$ is not controlled.
To control the kernel, where it is exponentially small, we prove the
large deviation bounds.

To analyse the kernel we use the representation 
\begin{equation}
K(n_{1},x_{1};n_{2},x_{2})=-\phi^{*(n_{1},n_{2})}(x_{1},x_{2})+\widetilde{K}^{step}(n_{1},x_{1};n_{2},x_{2})\label{K=phi+ K}
\end{equation}
where instead of working with the final expression (\ref{eq:K_step})
for the double integral part $\widetilde{K}^{step}(n_{1},x_{1};n_{2},x_{2})$
we return to the sum 
\begin{equation}
\widetilde{K}^{step}(n_{1},x_{1};n_{2},x_{2})=\sum_{k=1}^{n_{2}}\tilde{\Psi}_{n_{1}-k}^{n_{1}}(x_{1})\Phi_{n_{2}-k}^{n_{2}}(x_{2})\label{eq:tilda K=00003Dsum}
\end{equation}
 of products of two functions 
\begin{align}
\tilde{\Psi}_{\theta t-j}^{\theta t}(\chi t) & =\frac{\nu-1}{2\pi\mathrm{i}}\oint_{\Gamma_{1}}\frac{\exp\left(-tf(\chi,\theta,u)-j\ln u\right)}{\left(1-u\right)(1-\nu u)}du,\label{eq: Psi int}\\
\Phi_{\theta t-j}^{\theta t}(\chi t) & =\frac{1}{2\pi\mathrm{i}}\oint_{\Gamma_{0}}\frac{\exp\left(tf(\chi,\theta,u)+j\ln u\right)}{u}du.\label{eq: Phi int}
\end{align}
Here we use the function $\tilde{\Psi}$ different from $\Psi$ in
integration contour, which now encloses the pole at $z=1$ only. We
thus exclude the pole at $z=0$, whose contribution is transferred
to the single integral part of the kernel, so that we work with $\phi^{*}$
rather than $\phi$ below. The uniform estimate for $\widetilde{K}^{step}(n_{1},x_{1};n_{2},x_{2})$
follows from similar estimates for $\tilde{\Psi}$ and $\Phi$. 
\begin{lemma}
\label{lem: uniform estimate phi psi}Given $r>0$ and $\underline{s}\in\mathbb{R}$
fixed, let us take
\[
\theta_{r}:=\theta+2r\kappa_{c}t^{-1/3},\text{\ensuremath{\chi}}_{r,s}:=\chi(\theta_{r})-s\kappa_{f}t^{-2/3},\,\,\,\mathrm{and}\,\,\,j=aqt^{1/3}
\]
with
\begin{equation}
a=z_{c}\left|f^{(0,0,3)}(\chi(\theta),\theta,z_{c})/2\right|^{1/3}.\label{eq: a}
\end{equation}
Then, the there exist $\delta>0$, such that estimates

\begin{align*}
\Phi_{t\theta_{r}-j}^{t\theta_{r}}(t\chi_{r,s})= & \,\,t^{-1/3}a^{-1}e^{tf(\chi_{r,s},\theta_{r},z_{c}(\theta_{r}))+aqt^{1/3}\ln z_{c}}e^{-rq}\\
\times & \left(\mathrm{Ai}\left(s+q\right)+O(e^{-\delta t})+O\left(t^{-1/3}e^{-\delta_{1}\left(s+q\right)}\right)\right)\\
\tilde{\Psi}_{t\theta_{r}-j}^{t\theta_{r}}(t\chi_{r,s})= & \,\,t^{-1/3}\kappa_{f}^{-1}e^{-tf(\chi_{r,s},\theta_{r},z_{c}(\theta_{r}))-aqt^{1/3}\ln z_{c}}e^{rq}\\
\times & \left(\mathrm{Ai}(s+q)+O(e^{-\delta t})+O\left(t^{-1/3}e^{-\delta_{1}\left(s+q\right)}\right)\right),
\end{align*}
hold uniformly for $s>\underline{s}$ and $j>0$ with any $\delta_{1}>0.$
\end{lemma}
\begin{svmultproof}
\textsl{(Method of steepest descent) }The proof uses nowadays standard
estimates of the saddle point method following mainly the line of
\cite{GTW}. 

As the integrands of the kernel integral representation are the exponentials
of the function $f(\chi,\theta,z)$, we first look at the analytic
structure of this function. It has logarithmic singularities at the
points $z=0,1,1/\nu,1/\mu$. Our goal is to deform the contours $\Gamma_{0}$
and $\Gamma_{1}$ closed around $z=0$ and $z=1$ respectively into
the steepest descent contours. 

First, we need to locate the saddle points defined by 
\[
f^{(0,0,1)}(\theta,\chi,z)=0.
\]
This yields a cubic polynomial equation with real coefficients, which
has either all three roots real or one real and two complex conjugate.
To locate them let us first look at the case when two roots coincide.
As was discussed above, it follows from  (\ref{eq:dsp}) that, when the parameters $\chi$ and $\theta$
are related by (\ref{eq: chi},\ref{eq: theta}), i.e. $\chi=\chi(\theta)$,
and $0<\theta<p/(1-\mu)$, the two roots meet in the double saddle
point, $z_{-}=z_{+}=z_{c}\in(0,1)$. That the parametric dependence (\ref{eq: chi},\ref{eq: theta}) indeed defines  a single valued monotonously decreasing  function $\chi(\theta)$ is justified by inequality
\begin{equation}
	\frac{d\chi(\theta)}{d\theta}=-\frac{1}{c(\theta)}<-1,\label{eq:d chi/d theta <-1}
\end{equation}
where  $c(\theta)$ is defined by (\ref{eq: c(z)}). Also, it is easy to find the
third root
\[
z_{3}(\chi=\chi(\theta))=\frac{1}{\nu\mu z_{c}^{2}}
\]
in this case. 

The coefficients of the cubic polynomial depend on $\chi$ linearly.
Therefore, as $\chi$ varies away from $\chi(\theta)$, the two roots
move along the real axis, merging at $z_{c}$ when $\chi=\chi(\theta)$,
and then go away from the real axis as complex conjugate pair. Investigating
the behavior of $f(\theta,\chi,z)$ near the singularities we conclude
that when the two extrema of $\Re f(\theta,\chi,z)$
are on the real axis between zero and one, $z_{\pm}\in(0,1)$, the
minimum $z_{-}$ is on the left of the maximum $z_{+}$, i.e. $z_{-}<z_{+}$,
see fig \ref{fig: contours (double integral)} .
\begin{figure}
\centerline{\includegraphics[clip,width=0.8\textwidth]{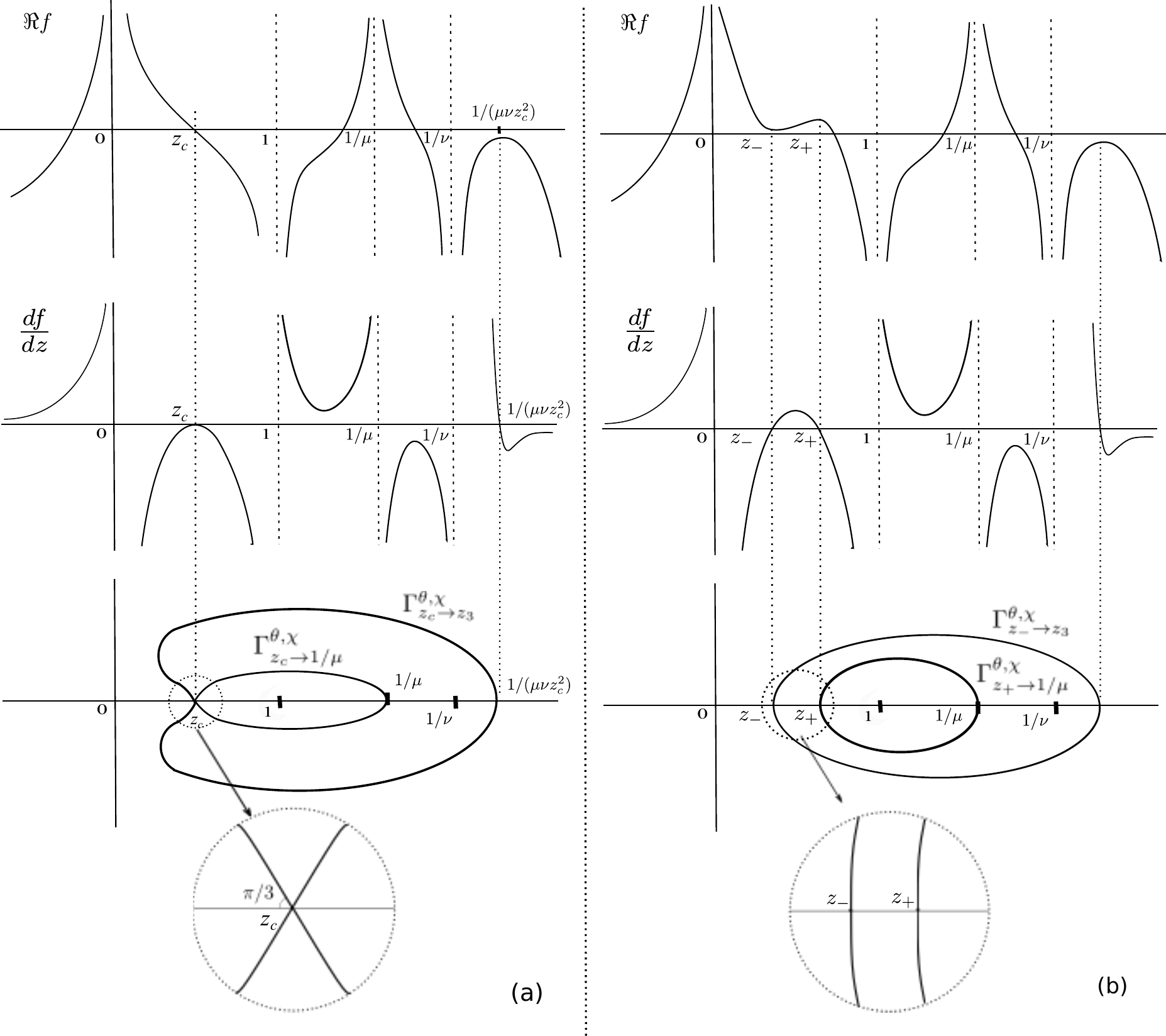}}
\caption{\label{fig: contours (double integral)}Schematic illustration of the behavior of $\Re f(\theta,\chi,z)$, of the derivative $f^{(0,0,1)}(\theta,\chi,z)$ and of the 
the steepest descent contours  (a)  $\Gamma_{z_{c}\to z_3}^{\theta,\chi}$ and
$\Gamma_{z_{c}\to1\mu}^{\theta,\chi}$ in the case $\chi=\chi(\theta)$ and (b) $\Gamma_{z_{-}\to z_3}^{\theta,\chi}$ and
$\Gamma_{z_{+}\to 1/\mu}^{\theta,\chi}$ in the case $\chi<\chi(\theta).$ }
\end{figure}
 Correspondingly from the sign of 
\begin{equation}
\frac{dz_{\pm}}{d\chi}=\frac{1-\nu}{\left(1-\nu z_{\pm}\right)\left(1-z_{\pm}\right)f^{(0,0,2)}(\theta,\chi,z_{\pm})}\label{eq:dzpm/d chi}
\end{equation}
coinciding with the sign of $f^{(0,0,2)}(\theta,\chi,z_{\pm})$ we
see that as $\chi$ decreases down from $\chi(\theta)$, $z_{-}$
and $z_{+}$ move along the real axis away from $z_{c}$ towards the
singularities at $z=0$ and $z=1,$ respectively. The left saddle
point $z_{-}$ asymptotically approaches the origin as $\chi\to-\infty.$
The right saddle point $z_{+}$ crosses $z=1$, when $\chi=-\theta$
and moves further to the right as $\chi$ continues decreasing. However
at $\chi=-\theta$ the extremum of $\Re f(\theta,\chi,z)$ at $z=z_{+}$
changes from maximum to minimum because the singularity of $f(\theta,\chi,z)$
changes the sign. That is an indication of the fact that when $\chi<-\theta$
the point $z=1$ becomes zero of the term $e^{-tf(\theta,\chi,z)}$
rather than a pole, and the corresponding integral vanishes.

When $\chi<\chi(\theta)$, the contours $\Gamma_{0}$ and $\Gamma_{1}$
in the double integral part of the kernel can be deformed into the
steepest descent and ascent contours $\Gamma_{z_{-}}^{\theta,\chi}$
and $\Gamma_{z_{+}}^{\theta,\chi}$ respectively , which are the stationary
phase contours being simple closed curves defined by equation
\[
\Gamma_{z_{\pm}}^{\theta,\chi}=\left\{ z:\Im f(\theta,\chi,z)=0\right\} .
\]
with $z_{-}$ or $z_{+}$ being the points in $(0,1)$ where the contours
cross the real axis. The steepest descent and ascent contours starting
from the saddle points must be closed either via another saddle point
or via a singularity, since $\Re f(\theta,\chi,z)$ is monotonous
everywhere on stationary phase contours except at these points. Only
one such a possibility exists in the range of interest of $\chi$.
Specifically, for general $\chi<\chi(\theta)$ the contour $\Gamma_{0}$ is deformed
to the steepest descent contour $\Gamma_{z_{-}\to z_{3}}^{\mathrm{\theta,\chi}}$
starting at $z_{-}$ and being closed via the third saddle point $z_{3}\in(1/\nu,\infty)\cup(-\infty,0)$
located on the positive or negative parts of the real axis when $\nu>0$
and $\nu<0$, respectively. In the latter case, the contour $\Gamma_{0}$
should be deformed via infinity, where the integrand is regular. 

The steepest ascent version of $\Gamma_{1}$ is the contour $\Gamma_{z_{+}\to1/\mu}^{\theta,\chi}$
outgoing from $z_{+}$, looping around $z=1$ and terminating at $z=1/\mu.$
For generic values of $\theta$ and $-\theta<\chi<\chi(\theta)$ the steepest
descent and ascent contours cross the real axis at $z_{-}$ and $z_{+}$ respectively at the
angle $\pi/2$, while when $\chi=\chi(\theta)$ the contours approach
the double saddle point $z_{c}$ at angles divisible by $\pm\pi/3$
and $\pm2\pi/3$. 

As was noted above, when $\chi<-\theta$, the integral along $\Gamma_{1}$
vanishes, since zero of the integrand is at $z=1$ in this case. Also,
it is easy to argue that the contours $\Gamma_{z_{-}\to z_{3}}^{\mathrm{\theta,\chi}}$
and $\Gamma_{z_{+}\to1/\mu}^{\theta,\chi}$ corresponding to different
values of $\chi<\chi(\theta)$ are always nested in the same way as
for $\chi=\chi(\theta)$. Indeed, the contours separate the domains
with opposite signs of $\Im$$f(\theta,\chi,z).$ Observing that $\mathrm{sgn}(\Im\log\left[(1-u)/(1-\nu u)\right]=\mathrm{-sgn}(\Im u)$,
we see that as $\chi$ decreases, the contours should move outward
with respect to the domain between them to compensate the change of
$\Im f(\theta,\chi,z)$.

When $\chi>\chi(\theta)$ the two roots $z_{\pm}$ turn into a pair of complex conjugate roots, and the picture  
of the steepest descent contours becomes  substantially different. Thus,  the arguments below  based on the picture  described fail in that case. Fortunately, by conditions of the lemma the positive values of   $(\chi-\chi(\theta))$  we are interested in  are as small as $O(t^{-1/3})$. For this case of the variable $s$ taking values in  bounded sets we will need only the first part of the analysis, 
which is  insensitive to the sign of  $(\chi-\chi(\theta))$,   being an extension of  $\chi=\chi(\theta)$ case.

\textsl{(Bounded sets) }Suppose first $s\in[\underline{s},\bar{s}]$
and $q\in[0,\underline{q}]$ for some $\bar{s}>\underline{s}$ and
$\underline{q}\geq0$. We outline the proof for $\Phi_{t\theta_{r}-j}^{t\theta_{r}}.$
The proof for $\tilde{\Psi}_{t\theta_{r}-j}^{t\theta_{r}}$ is completely
analogous. The integration contour we use is the steepest descent
contour $\Gamma_{z_{c}\to z_{3}}^{\mathrm{\theta,\chi(\theta)}}$
of the function $f(\chi(\theta),\theta,z)$, when the two saddle points
have merged into the double saddle point, $z_{-}=z_{+}=z_{c}.$ Then,
for some small $\epsilon>0$ we drop the part of the integral over
$\Gamma_{z_{c}\to z_{3}}^{\mathrm{\theta,\chi(\theta)}}$ beyond the
$\epsilon-$neighborhood $U_{\epsilon}(z_{c})=\{z:|z-z_{c}|<\epsilon\}$
of the double saddle point. For the contour being steepest descent
this yields the error of order of $O\left(\exp(t\left(f(\chi(\theta),\theta,z_{c})-\delta\right))\right).$ 

Limiting the integration to the part of the contour inside $U_{\epsilon}(z_{c})$
we use the approximation for the integrand, which yields 
\[
\frac{1}{2\pi\mathrm{i}}\int_{\Gamma_{z_{c}\to z_{3}}^{\mathrm{\theta,\chi(\theta)}}\bigcap U_{\epsilon}(z_{c})}\frac{\exp\left(tf_{app}+aqt^{1/3}\left(\ln z_{c}+u/z_{c}\right)\right)}{z_{c}}du
\]
where we use the notation $f_{app}$ for the Taylor approximation
of $f(\chi_{r,s},\theta_{r},z_{c}+u)$ from the r.h.s. of (\ref{eq:f-expansion}).
After making the variable change (\ref{eq: u->ut^1/3}) this integral
becomes 
\[
t^{-1/3}a^{-1}e^{tf(\chi_{r,s},\theta_{r},z_{c}(\theta_{r}))+aqt^{1/3}\ln z_{c}}e^{rq}\int_{t^{1/3}\epsilon e^{\mathrm{i}\left(\frac{2\pi}{3}+\epsilon_{1}\right)}}^{t^{1/3}\epsilon e^{-\mathrm{i}\left(\frac{2\pi}{3}+\epsilon_{1}\right)}}e^{-\frac{(u-r)^{3}}{3}+(s+q)(u-r)}du.
\]
Here we replaced the the upper and lower half of the contour by two
segments of rays approaching the origin at the angles $\pm\left(\frac{\pi}{3}+\epsilon_{1}\right),$
where $\epsilon_{1}$ is an $\epsilon-$dependent constant, which
can be made small by choosing the $\epsilon$ small enough. Finally,
shifting the integration contour by $r$ horizontally for the price
of the error of order of $O(e^{-t\epsilon^{3}})$ coming from the
boundary of $U_{\epsilon}$ and sending $t$ to infinity we arrive
at the integral representation of the Airy function (\ref{eq: Airy})

To estimate the error coming from the approximation we first note that
the Taylor expansion (\ref{eq:f-expansion}) is obtained with error 

\begin{align}
f(\chi_{r,s},\theta_{r},z_{c}+u)-f_{app}(\chi_{r,s},\theta_{r},z_{c}+u)\,\,\,\,\,\,\,\,\,\,\,\,\,\,\,\,\,\,\,\,\,\,\,\,\,\,\,\,\,\,\,\,\,\,\,\,\,\,\,\,\,\,\,\,\,\,\,\,\,\,\,\,\,\,\,\,\,\,\label{eq: f error}\\
=c_{0}t^{-4/3}+c_{1}t^{-1}u+c_{2}t^{-2/3}u^{2}+c_{3}t^{-1/3}u^{3}+c_{4}u^{4},\nonumber 
\end{align}
 where $c_{0},\dots,c_{4}$ are some constants. Then, the corrections
to the integrand satisfy 

\begin{align}
 & \left|e^{tf+(aqt^{1/3}-1)\ln(z_{c}+u)}-e^{tf_{app}+(aqt^{1/3}-1)(\ln z_{c}+u/z_{c})}\right|\nonumber \\
\leq & \left|e^{tf_{app}+(aqt^{1/3}-1)\ln z_{c}}\right|\left|e^{c_{0}t^{-1/3}+c_{1}|u|+c_{2}t^{1/3}|u|^{2}+c_{3}t^{2/3}|u|^{3}+tc_{4}|u|^{4}}-1\right|\label{eq: error double}\\
\leq & \left|e^{tf_{app}+(aqt^{1/3}-1)\ln z_{c}}\right|e^{c_{0}t^{-1/3}+\epsilon\left(c_{1}+c_{2}t^{1/3}|u|+c_{3}t^{2/3}|u|^{2}+tc_{4}|u|^{3}\right)}\nonumber \\
\times & \left|c_{0}t^{-1/3}+c_{1}|u|+c_{2}t^{1/3}|u|^{2}+c_{3}t^{2/3}|u|^{3}+tc_{4}|u|^{4}\right|,\nonumber 
\end{align}
where the first inequality uses the estimate (\ref{eq: f error})
and the second uses the inequality $|e^{x}-1|\leq xe^{x}$ for $x>0$
and the fact that the integrand is limited to $|u|<\epsilon$. The
modulus of the difference of  $\Phi_{t\theta_{r}-j}^{t\theta_{r}}$
and its approximation is majorized by the integral of the first line
of this expression over the contour $\Gamma_{z_{c}\to z_{3}}^{\mathrm{\theta,\chi(\theta)}}\bigcap U_{\epsilon}$.
Making the variable change (\ref{eq: u->ut^1/3}) and sending $t\to\infty$
we observe that the integral of the last two lines r.h.s. of (\ref{eq: error double})
is convergent for $\epsilon$ small enough and is $$O(t^{-2/3}e^{tf(\chi_{r,s},\theta_{r},z_{c}(\theta_{r}))+aqt^{1/3}\ln z_{c}})=O\left(\left|\Phi_{t\theta_{r}-aqt^{1/3}}^{t\theta_{r}}(t\chi_{r,s})\right|t^{-1/3}\right).$$

\textsl{(Arbitrary sets) }The next step is to extend this estimate
to arbitrary values of $s$ and $j$. We first prove the statements
for particular case $j=0$ and then extend to arbitrary $j>0$. To perform
the analysis of the integrals in the case $\chi<\chi(\theta)$ we
use the steepest descent and ascent integration contours $\Gamma_{z_{-}\to z_{3}}^{\mathrm{\theta,\chi}}$
and $\Gamma_{z_{+}\to1/\mu}^{\theta,\chi}$ for $\Phi$ and $\tilde{\Psi}$.
The corresponding integrals hence are bounded by the maxima of the
integrands at these contours. To show that the points $z_{-},z_{+}\in(0,1)$
defined by $f^{(0,0,1)}(\chi,\theta,z_{\pm})=0$ are the minimum and
maximum of $f(\chi,\theta,z)$ respectively we note that 

\begin{equation}
f^{(0,0,2)}(\chi,\theta,z_{\pm})\lessgtr0\label{eq: f''<>0}
\end{equation}
 when $\chi<\chi(\theta)$ for $z_{-}$ and for $-\theta<\chi<\chi(\theta)$
for $z_{+}.$ To this end we note that though the derivatives $dz_{\pm}/d\chi$
diverge as $\chi\to\chi(\theta)$, $z_{-}$ and $z_{+}$ are smooth
functions of 
\[
\zeta=\sqrt{\chi(\theta)-\chi}.
\]
 In terms of $\zeta$ we have, 
\begin{equation}
\left.\frac{df^{(0,0,2)}(\chi(\theta)-\zeta^{2},\theta,z_{\pm})}{d\zeta}\right|_{\zeta=0}=\left.\frac{dz_{\pm}}{d\zeta}\right|_{\zeta=0}f^{(0,0,3)}(\chi(\theta),\theta,z_{c}),\label{eq: df''/d zeta}
\end{equation}
where 
\[
f^{(0,0,3)}(\chi(\theta),\theta,z_{c})=-\frac{2(1-\mu)(\mu-\nu)\left(1-\mu\nu z_{c}^{3}\right)}{(1-z_{c})z_{c}(1-\mu z_{c})^{3}(1-\nu z_{c})\left(1-\nu z_{c}^{2}\right)}<0
\]
and 
\begin{align}
\left.\frac{dz_{\pm}}{d\zeta}\right|_{\zeta=0} & =\pm\alpha,\label{eq: dz_pm/dzeta}
\end{align}

$$
\mathrm{with}\qquad\alpha=\sqrt{\frac{2f^{(1,0,1)}(\chi(\theta),\theta,z_{c})}{f^{(0,0,3)}(\chi(\theta),\theta,z_{c})}}=\sqrt{\frac{(1-\nu)z_{c}(1-\mu z_{c})^{3}\left(1-\nu z_{c}^{2}\right)}{(1-\mu)(\mu-\nu)\left(1-\mu\nu z_{c}^{3}\right)}}>0.
$$
As $f^{(0,0,2)}(\chi(\theta),\theta,z_{c})$ is zero, while its derivative
in $\zeta$ is not, the inequalities (\ref{eq: f''<>0}) hold for
small values of $\zeta>0$. Furthermore, they can be extended to the
whole domain of interest, because the opposite would imply that $f^{(0,0,2)}(\chi(\theta),\theta,z)$
vanishes at more then one point in $(0,1)$, i.e. $z_{c}$ that solves
eqs.(\ref{eq:dsp}) for given $\chi$ and $\theta$ is not unique.
However, both $\chi(z_{c})$ and $\theta(z_{c})$ defined by (\ref{eq: chi},\ref{eq: theta})
are monotonous functions of $z_{c}\in(0,1)$, which can be seen by
direct differentiation, and, hence, are one-to-one. 

From here we conclude that $f(\chi,\theta,z)$ is increasing when
$z_{-}<z<z_{+},$ and there exist $\delta$ such that 
\[
\left|f(\chi,\theta,z_{\pm})-f(\chi,\theta,z_{c})\right|>\delta.
\]

Thus, we first state that given $\theta>0$, $\epsilon>0$ and $\chi<\chi(\theta)-\epsilon$
there exists $\delta$, such that 
\begin{align}
\tilde{\Psi}_{t\theta}^{t\theta}(t\chi) & =O\left(e^{-tf(\chi,\theta,z_{c}(\theta))-t\delta}\right),\label{eq: Psi e^-delta t}\\
\Phi_{t\theta}^{t\theta}(t\chi) & =O\left(e^{tf(\chi,\theta,z_{c}(\theta))-t\delta}\right).\label{eq:eq: Phi e^-delta t}
\end{align}

Second, we note that we can limit the integration by small $\epsilon$-vicinities
$U_{\epsilon}(z_{\pm})$ of the the critical points, introducing another
error of order of $O(e^{-\delta_{1}t})$. 

Third, within $U_{\epsilon_1}(z_{\pm})$ and for $\chi(\theta)-\chi<\epsilon_2$ with some small $\epsilon_1,\epsilon_2>0$
we can approximate $f(\chi,\theta,z)$ by its Taylor expansion near
the critical points, with coefficients given by expansions in $\zeta$
. Using (\ref{eq: df''/d zeta}) and (\ref{eq: dz_pm/dzeta}) we obtain
\begin{equation}
z_{\pm}=z_{c}\pm\alpha\zeta+O(\zeta^{2}).\label{eq: z_pm exp}
\end{equation}
After substituting this into $f(\chi,\theta,z_{\pm})$ and its derivatives
with respect to the last argument we have. 
\begin{align}
f(\chi,\theta,z_{\pm})= & f(\chi(\theta),\theta,z_{c})+\zeta^{2}f^{(1,0,0)}(\chi,\theta,z_{c})\label{eq:fpm}\\
 & \mp\frac{2}{3}\alpha f^{(1,0,1)}(\chi,\theta,z_{c})\zeta^{3}+O\left(\zeta^{4}\right)\nonumber \\
= & f(\chi,\theta,z_{c})\pm\frac{2}{3}\frac{\zeta^{3}}{\kappa_{f}^{3/2}}+O\left(\zeta^{4}\right),\nonumber \\
f^{(0,0,1)}(\chi,\theta,z_{\pm}) & =0,\label{eq:fpm'}\\
f^{(0,0,2)}(\chi,\theta,z_{\pm}) & =\pm\alpha f^{(0,0,3)}(\chi,\theta,z_{c})\zeta+O(\zeta^{2})\nonumber \\
 & =\mp2\left(\frac{\left|f^{(0,0,3)}(\chi,\theta,z_{c})\right|}{2}\right)^{2/3}\frac{\zeta}{\sqrt{\kappa_{f}}}+O(\zeta^{2}),\nonumber \\
f^{(0,0,3)}(\chi,\theta,z_{\pm}) & =f^{(0,0,3)}(\chi,\theta,z_{c})+O\left(\zeta\right)\label{eq:fpm''}
\end{align}
The first equation here is obtained by integrating relation $df(\chi,\theta,z_{\pm})/d\chi=f^{(1,0,0)}(\chi,\theta,z_{\pm})$,
where all the dependence on $\chi$ in r.h.s. enters only through
eq. (\ref{eq: z_pm exp}). 

Using the above estimates, let us substitute the Taylor expansion
for $f(\chi,\theta,z)$ into the integral formula of $\Phi_{t\theta}^{t\theta}(t\chi)$
with $\zeta=\sqrt{s\kappa_{f}}t^{-1/3}<\sqrt{\epsilon_2}$, assuming that $s>0$ and $\epsilon_2$ is arbitrarily small.
\begin{align*} 
\Phi_{t\theta}^{t\theta}(t\chi) & \simeq O\left(e^{t\left(f(\chi,\theta,z_{c})-\delta\right)}\right)+e^{tf(\chi,\theta,z_{c})-\frac{2}{3}s^{3/2}+O(t^{-1/3}s^{2})}\\
\times & \oint_{\Gamma_{z_{-}\to z_{3}}^{\mathrm{\theta,\chi}}\bigcap U_{\epsilon_1}(z_{-})}e^{t\left(z-z_{-}\right)^{3}\frac{f^{(0,0,3)}(\chi,\theta,z_{c})}{6}+\left(z-z_{-}\right)^{2}\left(\frac{f^{(0,0,3)}(\chi,\theta,z_{c})}{2}\right)^{2/3}t^{2/3}\sqrt{s}}\\
\times & e^{O\left(t^{1/3}s(z-z_{-})^{2}\right)+O\left(t^{2/3}\sqrt{s}(z-z_{-})^{3}\right)+O\left(t\left(z-z_{-}\right)^{4}\right)}\frac{dz}{2\pi\mathrm{i}z}.
\end{align*}
Then, we make the variable change
\begin{eqnarray}
x=(z-z_-)\left(tf^{(0,0,3)}(\chi,\theta,z_c))/2\right)^{1/3}+\sqrt{s}\notag\\\simeq (z-z_c)\left(tf^{(0,0,3)}(\chi,\theta,z_c))/2\right)^{1/3}\label{eq: x-varchange}
\end{eqnarray}
in the integral and deform the contour for it to pass via the point $x=0$ ($z=z_c$ in the original variables). 
If we   neglect the corrections,  we find the integrand coinciding with the  one appearing from integral representation of Airy function (\ref{eq: Airy}). We would like to make sure that the deformed integration contour also matches that definition. To this end, 
we note that  the  contour $\Gamma_{z_{-}\to z_{3}}^{\theta,\chi}$   leaves the domain  $U_{\epsilon_1}(z_{-})$  through the points 
$z_-+\epsilon_1\exp({\pm \mathrm{i}(2\pi/3-\phi_1)}),$ where the angle $\phi_1$ can be made small by choosing $\zeta$ small enough, in which case $\phi_1=\sqrt{3}\alpha\zeta/(2\epsilon_1)+O\left( \zeta^2/\epsilon_1^2\right)$. The deformation changes the 
angles to $\pm(2\pi/3+\phi_2)$ where $\phi_2=O(\zeta^2/\epsilon_1^2)$,  which can be made small by choosing $\epsilon_1$ and $\epsilon_2$ such that the ratio $\epsilon_2/\epsilon_1^2$   is small. This ensures that the integration contour in $x$-variable
consists of two segments of rays approaching the real axis at angles close to $\pm\pi/3$, which matches with (\ref{eq: Airy}). The segments   become the infinite rays in the limit $t\to\infty$.
\begin{figure}
	\centerline{\includegraphics[width=0.45\textwidth]{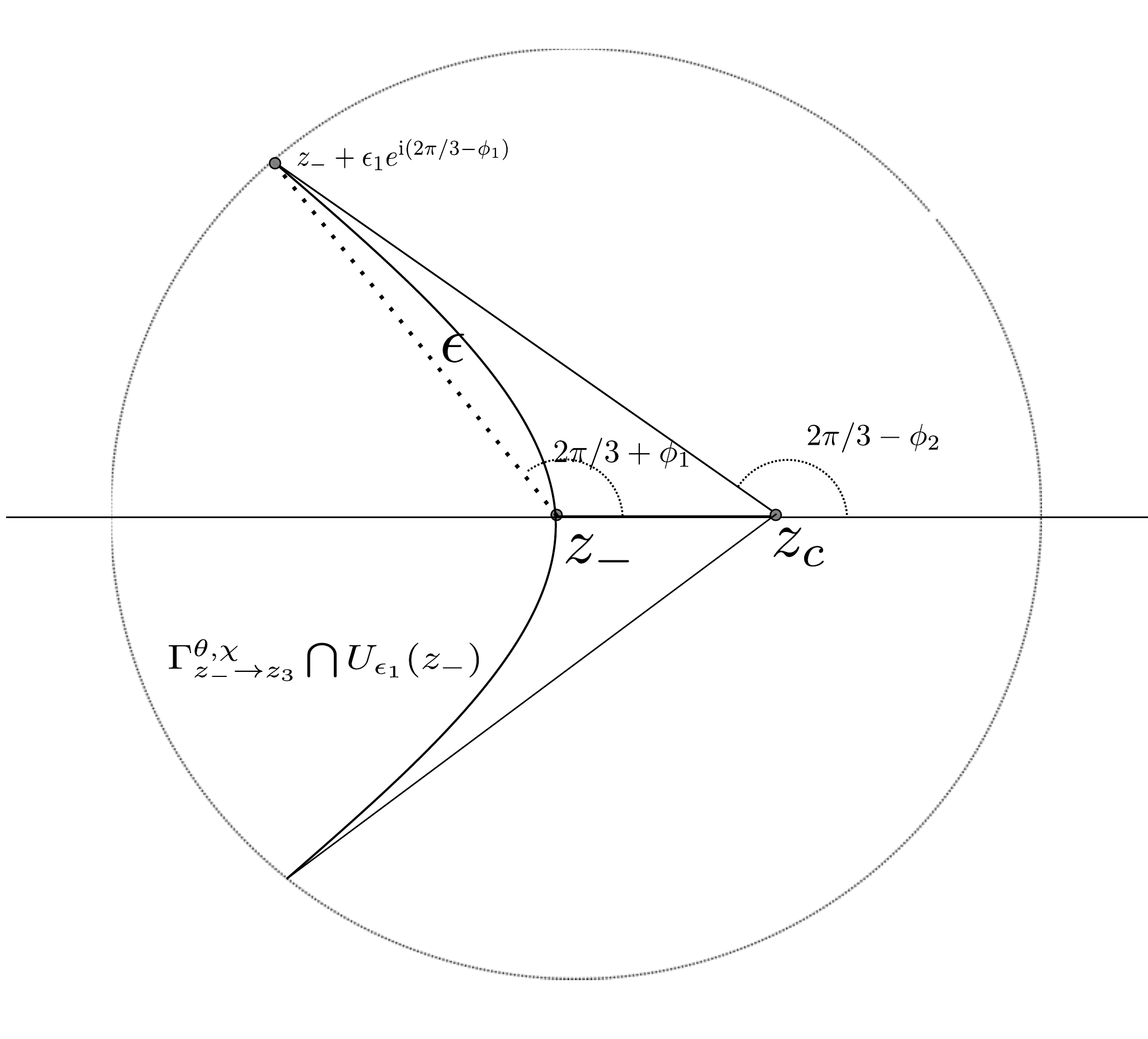}}
	\caption{The steepest descend contour in  $\epsilon_1$-vicinity of $z_-$ and  its deformation  passing through $z_c$. \label{fig: triangle}}
\end{figure}
Taking into account that with $st^{-2/3}=O(\epsilon_2)$ the estimate
\[
\left|e^{-\frac{2}{3}s^{3/2}+O(t^{-1/3}s^{2})}-e^{-\frac{2}{3}s^{3/2}}\right|=O(t^{-1/3}s^{2}e^{-\left(\frac{2}{3}-\sqrt{\epsilon_2}\right)s^{3/2}})=O(t^{-1/3}e^{-\delta_{1}s})
\]
holds with any $\delta_{1}>0$, we obtain 
\begin{equation}
\Phi_{t\theta}^{t\theta}(t\chi)=t^{-1/3}a^{-1}e^{tf(\chi,\theta,z_{c})}\left(\mathrm{Ai}(s)+O(e^{-\delta t})+O(t^{-1/3}e^{-\delta_{1}s})\right).\label{eq:Phi uniform}
\end{equation}
Similarly for $\tilde{\Psi}$ we have
\begin{equation}
\tilde{\Psi}_{t\theta}^{t\theta}(t\chi)=t^{-1/3}\kappa_{f}^{-1}e^{-tf(\chi,\theta,z_{c})}\left(\mathrm{Ai}(s)+O(e^{-\delta t})+O(t^{-1/3}e^{-\delta_{1}s})\right).\label{eq:Psi uniform}
\end{equation}
To extend these results to $\tilde{\Psi}_{t\theta-j}^{t\theta}(t\chi)$
and $\Phi_{t\theta-j}^{t\theta}(t\chi)$ with $j>0$ we note that
from (\ref{eq: Psi int},\ref{eq: Phi int}) and (\ref{f}) that

\begin{align*}
\tilde{\Psi}_{t\theta-j}^{t\theta}(t\chi)=\tilde{\Psi}_{t(\theta-\xi)}^{t(\theta-\xi)}(t(\chi+\xi)),\,\,\,\Phi_{t\theta-j}^{t\theta}(t\chi) & =\Phi_{t(\theta-\xi)}^{t(\theta-\xi)}(t(\chi+\xi)),
\end{align*}
where we introduce notation $\xi=j/t$. Thus, the estimate of $\tilde{\Psi}_{t\theta-j}^{t\theta}(t\chi)$
and $\Phi_{t\theta-j}^{t\theta}(t\chi)$ is reduced to the previously
studied case of $j=0$ with $\theta$ and $\chi$ replaced by $(\theta-\xi)$
and $(\chi+\xi)$ respectively. 

Suppose first that $\chi(\theta)>\chi$ and $\xi>\epsilon$.
Due to (\ref{eq:d chi/d theta <-1}),  there exists $\epsilon_{1}>0$ such that 
\[
\chi(\theta-\xi)-(\chi+\xi)>\epsilon_{1},
\]
and hence, from (\ref{eq:eq: Phi e^-delta t},\ref{eq: Psi e^-delta t})
\begin{align*}
\Phi_{t\theta-j}^{t\theta}(t\chi) & =O(e^{tf(\chi+\xi,\theta-\xi,z_{c}(\theta-\xi))-t\delta}),\\
\tilde{\Psi}_{t\theta-j}^{t\theta}(t\chi) & =O(e^{-tf(\chi+\xi,\theta-\xi,z_{c}(\theta-\xi))-t\delta})
\end{align*}
in this case. Otherwise, when $\xi<\epsilon$, we apply  the formulas
(\ref{eq:Phi uniform},\ref{eq:Psi uniform}) with $\theta$ and $\chi$
replaced by $(\theta_{r}-\xi)$ and $(\chi_{r,s,}+\xi)$ respectively.
Using the approximations 
\begin{align*}
\chi_{r,s}+\xi & =\chi(\theta_{r}-\xi)+(\chi(\theta)-\chi(\theta_{r}-\xi)-s\kappa_{f}t^{-2/3})\\
 & =\chi(\theta_{r}-\xi)-t^{-2/3}\kappa_{f}\left(s+q\left(1+O(t^{-1/3})+O(\xi)\right)\right),
\end{align*}
where we set set $\xi=t^{-2/3}q\frac{\kappa_{f}c(\theta)}{c(\theta)+1}=t^{-2/3}aq,$
and 
\begin{align*}
f(\chi_{r,s}+\xi,\theta_{r}-\xi,z_{c}(\theta_{r}-\xi)) & =f(\chi_{r,s},\theta_{r},z_{c}(\theta_{r}-\xi))+\xi\ln\left(z_{c}(\theta_{r}-\xi)\right)\\
=f(\chi_{r,s},\theta_{r},z_{c}(\theta_{r}))+ & t^{-2/3}q\left(a\ln(z_{c}(\theta))+t^{-1/3}r+O(t^{-2/3})+O(\xi)\right)
\end{align*}
we arrive at the statement of the lemma . 
\end{svmultproof}

With this estimates in hand we can estimate the double integral part
of the kernel.
\begin{corollary}
\label{Cor: Ksteptilda estimate} Under conditions of lemma \ref{lem: uniform estimate phi psi} there exists $\delta>0,$ such that
for $t\to\infty$ the identity
\begin{align*}
t^{1/3}\kappa_{f}e^{t\left(f(\chi_{1},\theta_{1},z_{c}(\theta_{1}))-f(\chi_{2},\theta_{2},z_{c}(\theta_{2}))\right)}\widetilde{K}^{step}(n_{1},x_{1};n_{2},x_{2})\\
=\int_{0}^{\infty}e^{-q(r_{2}-r_{1})}\mathrm{Ai}(s_{1}+q)\mathrm{Ai}(s_{2}+q)dq+O\left(e^{-\delta t}\right) & +O\left(e^{-\delta_{1}\left(s_{1}+s_{2}\right)}t^{-1/3}\right)
\end{align*}
holds for any $\delta_{1}>0.$ 
\end{corollary}
\begin{svmultproof}
The proof is based on the uniformity of the above estimates, where
one should choose $\delta_{1}>(r_{1}-r_{2})/2$, and the fact that
there is at most $O(t)$ summands in the sum (\ref{eq:tilda K=00003Dsum}).
Finally, the sum converges to an integral as $t\to\infty$. 
\end{svmultproof}

\begin{lemma}
\label{lem: Diffusive-part estimate}(Uniform estimate of the diffusive part of the kernel)

Given $r_2>r_1$, let $\theta_i=\theta_{r_i}$ and $\chi_i=\chi_{r_i,s_i}$ for $i=1,2$  and $\theta_r$ and $\chi_{r,s}$ be as in 
(\ref{eq:theta_r, chi_r,s}). Then, the estimate
\begin{align*}
e^{t(f(\chi_{1},\theta_{1},z(\theta_{1}))-f(\chi_{2},\theta_{2},z(\theta_{2})))}\phi^{*(n_{1},n_{2},)}(x_{1},x_{2})\\
=\frac{t^{-1/3}}{2\kappa_{f}\sqrt{\pi(r_{2}-r_{1})}}e^{-\frac{\left(s_{2}+s_{1}\right)\left(r_{2}-r_{1}\right)}{2}+\frac{2}{3}\left(\frac{r_{2}-r_{1}}{2}\right)^{3}-\frac{\left(s_{1}-s_{2}\right)^{2}}{4(r_{2}-r_{1})}}\\
\times\left(1+O\left(t^{-1/3}e^{\epsilon\left(s_{1}-s_{2}\right)^{2}}\right)\right) & +O\left(e^{-\epsilon t^{1/3}\frac{1}{2}(r_{2}-r_{1})}\right).
\end{align*}
holds for any $s_1,s_2$ and any small $\epsilon>0.$ 
\end{lemma}
\begin{svmultproof}
The integral representation of the single integral part including the
conjugation is given by 
\begin{equation}
e^{tf(\chi_{1},\theta_{1},z(\theta_{1}))-f(\chi_{2},\theta_{2},z(\theta_{2}))}\phi^{*(n_{1},n_{2},)}(x_{1},x_{2})=\mathbbold{1}_{n_{2}>n_{1}}\oint_{\Gamma_{1}}\frac{dv}{2\pi\mathrm{i}}\frac{(\nu-1)e^{t^{2/3}h(v)}}{(1-\nu v)(1-v)},\label{eq: phi conj}
\end{equation}
where 
\begin{equation}
h(v)=t^{1/3}\left(f(\chi_{1},\theta_{1},z_{c}(\theta_{1}))-f(\chi_{2},\theta_{2},z_{c}(\theta_{2}))-f(\chi_{1},\theta_{1},v)+f(\chi_{2},\theta_{2},v)\right).\label{eq:h(v)}
\end{equation}
We separated $t^{1/3}$ to ensure that $h(v)$ in its effective range
survives in the limit $t\to\infty$ likewise the other equations below.
In addition to the dependence of $h(v)$ on its argument, there is
also dependence on $\chi_{1},\chi_{2}$ and $\theta_{1},\theta_{2}$,
which we omit in the notation for brevity In fact the $v-$dependent
part of the integrand depends only on relative coordinates 
\[
\delta\theta_{21}=\theta_{2}-\theta_{1},\,\,\,\delta\chi_{21}=\chi_{2}-\chi_{1}.
\]
By definition the whole expression is nonzero when 
\begin{equation}
\delta\theta_{21}>0.\label{eq:delta theta>0}
\end{equation}
Also, the integral is nonzero, when the integrand has a pole at $v=1$,
i.e. 
\begin{equation}
\delta\theta_{21}+\delta\chi_{21}\leq0.\label{eq:delta theta+delta chi<0}
\end{equation}
 The stationary points of the integrand are defined by the equation
\begin{align}
h'(v) & =t^{1/3}\left(\frac{(\delta\chi_{21}+\delta\theta_{21})\left(\nu-1\right)}{(1-v)(1-\nu v)}-\frac{\delta\theta_{12}}{v}\right)=0.\label{eq:phi stat}
\end{align}
This yields a quadratic equation for $v$ that always has two real
roots within the range of $\delta\theta_{21}$ and $\delta\chi_{21}$
specified. One can see that one of the roots is always in $(0,1)$
and the other is either in $(1,\infty)$ or in $(-\infty,0).$ 

This can be seen by using a parametrization
\[
\delta\theta_{21}+\delta\chi_{21}=-e^{2y}\delta\theta_{21},\quad \nu=1-\left(\frac{\sinh x}{\sinh y}\right)^{2}
\]
in terms of two parameters $x>0$ and $y\in(-\infty,\infty)$ consistent
with (\ref{eq:delta theta>0},\ref{eq:delta theta+delta chi<0}).
Then, the roots of (\ref{eq:phi stat}) are 
\[
u_{\pm}=\frac{e^{\pm x}\sinh y}{\sinh(y\mp x)},
\]
whose behavior  as functions of $y$ is shown in fig. \ref{fig: u_pm}.
Looking at the second derivative 
\[
h''(u_{\pm})=\pm t^{1/3}\delta\theta_{21}2e^{\pm2x-y}\frac{\coth x\sinh^{2}(x\pm y)}{\sinh y},
\]
we can see that it is positive at $u_{+}$, when $y<0$, and at $u_{-}$,
when $y>0$. 
\begin{figure}
\centerline{\includegraphics[width=0.5\textwidth]{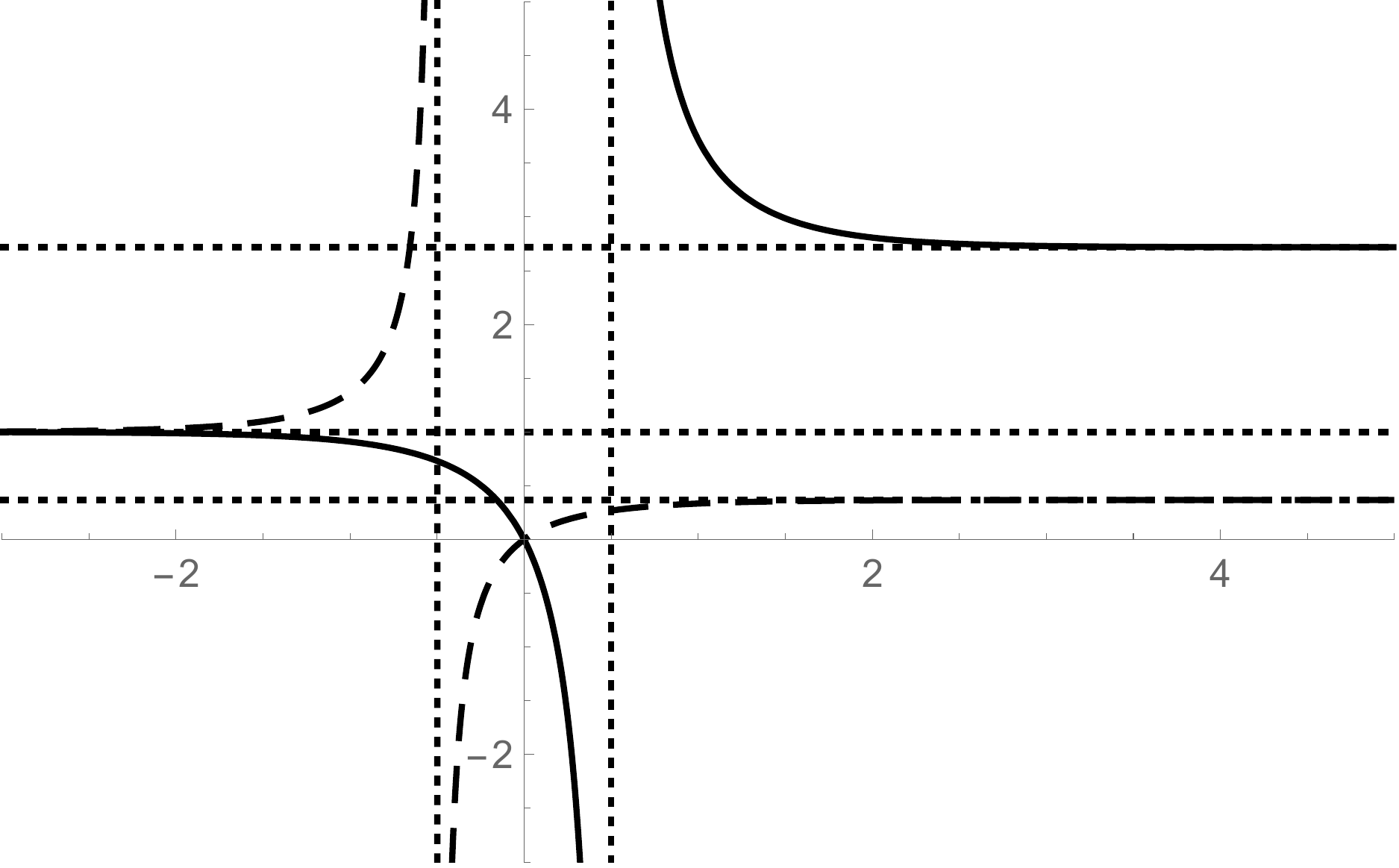}}\caption{\label{fig: u_pm}The stationary points $u_{+}$(solid line) and $u_{-}$(dashed
line) as functions of $y$ with $x$ fixed. The roots $u_{\pm}$ diverge
at $y=\pm x$ respectively (vertical dotted lines). The roots approach
the limiting values $u_{\pm}\to1$ as $y\to-\infty$ and $u_{\pm}\to e^{\pm2x}$
as $y\to+\infty$ (horizontal dotted lines). }
\end{figure}
The stationary phase contour $\Gamma_{u_{+}\to u_{-}}$ defined by
$\Im h(v)=0$ passes through both roots crossing the real axis traversally.
In the two cases mentioned it is a simple loop around one of the poles
of $h(u),$ either $u=1$ or $u=0$ respectively. In the former case
it suits for $\Gamma_{1}$ as is and in the latter it can be obtained
from $\Gamma_{1}$ by transforming via infinity, where $h(u)$ is
regular. 

It is clear from the above consideration that the critical point bringing
the dominant contribution into the integral, i.e. the maximum of $h(v)$
at the steepest descent contour is always the root in $(0,1)$. We
now denote this root $z^{*}$. It is a function of $\delta\chi_{21}$
for $\delta\theta_{21}$ fixed.

Once we have fixed the steepest descent contour, we expect that the
order of the integral (\ref{eq: phi conj}) is $O(e^{t^{2/3}h(z^{*})}).$

Remember that within the Fredholm determinant $\chi_{1}$ and $\chi_{2}$ are
the rescaled running summation indices. Our first goal is to limit
the range of their values by the domain, where the value of $h(z^{*})$
is close to the maximal one. An attempt to locate the maxima by looking
for the values of $\chi_{1}$ and $\chi_{2}$ where the corresponding
derivatives simultaneously vanish fails unless $\theta_{1}=\theta_{2}.$
Note, however, that the range of $\chi_{1}$ and $\chi_{2}$ within
the sum representing the Fredholm determinant is bounded from above.
Therefore, it is enough to argue that $h(z^{*})$ decays in the direction
of summation. To this end, let us introduce auxiliary variables 
\begin{equation}
\mathcal{X}=\frac{\chi_{1}+\chi_{2}+\theta_{1}+\theta_{2}}{2},\,\,\,\Delta=t^{1/3}\frac{\delta\chi_{21}+\delta\theta_{21}}{2}. \label{eq: X,Delta}
\end{equation}
Observe that the $h(z)$ can be split  into parts depending either
on $\mathcal{X}$ or on $\Delta$ and $z$ or on neither of them. 
\[
h(z)=h_{1}(\Delta,z)+\mathcal{X}h_{2}+h_{3}
\]
Here 
\begin{align*}
h_{1}(\Delta,z) & =\Delta\ln\left(\left(\frac{1-z}{1-\nu z}\right)^{2}\frac{1-\nu z_{c}(\theta_{1})}{1-z_{c}(\theta_{1})}\frac{1-\nu z_{c}(\theta_{2})}{1-z_{c}(\theta_{2})}\right)\\
 & \,\,\,\,\,\,\,\,\,\,\,\,\,\,\,\,\,\,\,\,\,\,\,\,\,\,\,\,\,\,\,\,\,\,\,\,\,\,\,\,-t^{1/3}\delta\theta_{21}\ln\left(\frac{z}{\sqrt{z_{c}(\theta_{1})z_{c}(\theta_{2})}}\right)
\end{align*}
 and 
\begin{align*}
h_{2} & =t^{1/3}\ln\left(\frac{1-z_{c}(\theta_{1})}{1-\nu z_{c}(\theta_{1})}\frac{1-\nu z_{c}(\theta_{2})}{1-z_{c}(\theta_{2})}\right)=\\
 & =\frac{2\text{\ensuremath{\kappa_{c}}}(r_{2}-r_{1})(1-\nu)z_{c}'(\theta)}{(1-z_{c}(\theta))(1-\nu z_{c}(\theta))}+O(t^{-1/3}) =\frac{(r_{2}-r_{1})}{\kappa_{f}}+O(t^{-1/3})>0\\
\end{align*}
\begin{align*}
h_{3} & =t^{1/3}\left(\frac{\theta_{1}+\theta_{2}}{2}\ln\frac{z_{c}(\theta_{2})}{z_{c}(\theta_{1})}+\ln\left(\frac{1-\nu z_{c}(\theta_{1})}{1-\mu z_{c}(\theta_{1})}\frac{1-\mu z_{c}(\theta_{2})}{1-\nu z_{c}(\theta_{2})}\right)\right)\\
 & =\frac{2\left(r_{1}-r_{2}\right)\kappa_{c}}{\theta'(z_{c})}\left(\frac{\mu}{1-\mu z_{c}(\theta)}-\frac{\nu}{1-\nu z_{c}(\theta)}+\frac{\theta}{z_{c}(\theta)}\right)+O(t^{-1/3})
\end{align*}

When $z=z^{*}$, the function $h_{1}(\Delta,z^{*})$ achieves a maximum
at a single point $\Delta_{c}$. Note that in addition to the explicit
$\Delta-$dependence in $h_{1}(\Delta,z)$, it depends on $\Delta$
through $z^{*}$. Recalling that $z=z^{*}$ is a critical point of $h_{1}(\Delta,z)$,
we obtain the equation for $\Delta_{c}$ in the form 
\[
\left.\frac{dh_{1}(\Delta,z^{*})}{d\Delta}\right|_{\Delta=\Delta_{c}}=h_{1}^{(1,0)}(\Delta_{c},z^{*})=0.
\]
This yields an equation
\[
g(z^{*}(\Delta_{c}))=\frac{1}{2}\left(g(z_{c}(\theta_{1}))+g(z_{c}(\theta_{2}))\right),
\]
where $z_{c}^{*}$ is $z^{*}$ evaluated at $\Delta=\Delta_{c}$ and
\[
g(z)=f^{(1,0,0)}(\chi,\theta,z)=\ln\frac{1-z}{1-\nu z}.
\]
Solving the equation perturbatively in powers of $t^{-1/3}$ up to
the third order terms we obtain 
\begin{align}
z_{c}^{*} & =z_{c}(\theta)+\ensuremath{\kappa_{c}}(r_{1}+r_{2})z_{c}'(\theta)t^{-1/3}\nonumber \\                                                                                                                                                                                                                                                                                                                         
 & +\frac{1}{2}\kappa_{c}^{2}\left(\frac{(r_{2}-r_{1})^{2}\left(z_{c}'(\theta\right)^{2}g''(z_{c}(\theta))}{g'(z_{c}(\theta))}+2\left(r_{1}^{2}+r_{2}^{2}\right)z_{c}''(\theta)\right)t^{-2/3}\label{eq: z*=00005BDelta_c=00005D}\\
 & +O(t^{-1})\nonumber 
\end{align}
To find $\Delta_{c}$ explicitly we substitute $z_{c}^{*}$ back to
the stationary point equation (\ref{eq:phi stat}), which can be recast
in the following form.

\begin{align*}
\Delta_{c} & =\frac{t^{1/3}\delta\theta_{21}}{2z^{*}g'(z^{*})},
\end{align*}
 which together with (\ref{eq: z*=00005BDelta_c=00005D}) yields 
\begin{equation}
\Delta_{c}=\kappa_{c}(r_{2}-r_{1})(1+\chi'(\theta))+\kappa_{c}^{2}(r_{2}^{2}-r_{1}^{2})\chi''(\theta)t^{-1/3}+O\left(t^{-2/3}\right),\label{eq:Delta_c}
\end{equation}
 where we exploit the relation 
\[
\chi'(\theta)=-\frac{1}{c(\theta)}=\frac{1}{z_{c}(\theta)g'(z_{c}(\theta))}-1.
\]
following from the formula (\ref{eq: c(z)}) of the density $c$
and the relation (\ref{eq: thetalegandre}). 

It is also not difficult to check that at $h_{1}(\Delta,z^{*})$ is
concave in $\Delta$. That the second derivative
\begin{align}
\frac{d^{2}h_{1}(\Delta,z^{*})}{d\Delta^{2}} & =-\frac{\left(h_{1}^{(1,1)}(\Delta,z^{*})\right)^{2}}{h_{1}^{(0,2)}(\Delta,z^{*})}\label{eq: d2h/dDelta2}\\
 & =-\frac{(1-\nu)^{2}}{\delta\theta_{21}t^{1/3}\left(1-1/z^{*}\right)^{2}\left(1-\nu z^{*}\right)^{2}-2\Delta(1-\nu)\left(1+\nu-2\nu z^{*}\right)},\nonumber 
\end{align}
 is negative is easily seen from the $x-y$ parametrization 
\[
(\ref{eq: d2h/dDelta2})=\frac{\left|\coth y-1\right|\tanh x}{4\Delta}<0
\]
and from the fact that $\Delta<0.$ Thus, $\Delta_{c}$ is indeed
the maximum. 

Now, we are in position to push forward the necessary estimates. First
we recall that within the Fredholm sum (\ref{eq: Fredholm series}) the
rescaled summation indices decrease down from $\chi_{i}=\chi(\theta_{i})+O(t^{-1/3}),\,\,i=1,2$.
Obviously the integral (\ref{eq: phi conj}) of interest is close
to maximal, when $\mathcal{X}=\mathcal{X}_{0}:=\left(\chi(\theta_{1})+\chi(\theta_{2})+\theta_{1}+\theta_{2}\right)/2$
and $\Delta=\Delta_{c}$, being $O(e^{t^{2/3}(h_{1}(\Delta_{c},z^{*}(\Delta_{c}))+\mathcal{X}_{0}h_{2}+h_{3})})$. 
Given $\epsilon>0$, when $\mathcal{X}_{0}-\mathcal{X>\epsilon}t^{-1/3}$,
it is less at least by the factor $e^{-\frac{1}{2}(r_{2}-r_{1})(s_{2}+s_{1})}=O(e^{-\epsilon t^{1/3}\frac{1}{2}(r_{2}-r_{1})})$.

Let us consider the range of $\chi_{1},\chi_{2}$ where this inequality
does not hold, i.e. when $\mathcal{X}_{0}-\mathcal{X<}\epsilon t^{-1/3},$
i.e.
\begin{equation}
-a_{i}t^{-2/3}<\chi(\theta_{i})-\chi_{i}<\epsilon t^{-1/3},\,\,\,i=1,2.\notag
\end{equation}
Obviously in this range
 \begin{equation}
|\Delta-\Delta_{c}|<\epsilon \label{eq: |Delta-Delta_c|<epsilon}
\end{equation}
holds. Here,
one can approximate $h(z)$ by the Taylor expansion. The three leading
coefficients of the Taylor expansion are 
\begin{eqnarray}
h(z^{*}) & =&h(z_{c}^{*})+\frac{1}{2}\left.\frac{d^{2}h_{1}(\Delta,z^{*})}{d\Delta^{2}}\right|_{\Delta=\Delta_{c}}\left(\Delta-\Delta_{c}\right)^{2}+O\left(\left(\Delta-\Delta_{c}\right)^{3}\right) \label{eq: h(z^*)}\\
&= & t^{-2/3}\left(\frac{2}{3}\left(\frac{r_{2}-r_{1}}{2}\right)^{3}-\frac{\left(s_{2}+s_{1}\right)\left(r_{2}-r_{1}\right)}{2}-\frac{\left(s_{1}-s_{2}\right)^{2}}{4(r_{2}-r_{1})}\right)\notag\\&&+O\left(\left(\Delta-\Delta_{c}\right)^{3}\right)\notag\\
h'(z^{*}) & =&0\\
h''(z^{*}) & =&2^{1/3}|f^{(0,0,3)}(\chi(\theta),\theta,z_{c}(\theta))|^{2/3}(r_{2}-r_{1})+O\left(\Delta-\Delta_{c}\right).
\end{eqnarray}
The most technical part here is the evaluation  $h(z^{*}_c)$ in the
first line. To evaluate $h(z_{c}^{*})$ we use the expansion (\ref{eq:f-expansion})
of (\ref{eq:h(v)}) and the formula (\ref{eq: z*=00005BDelta_c=00005D})
for $z_{c}^{*}.$ The second derivative of $h_{1}(\Delta,z^{*})$
in $\Delta$ is given in (\ref{eq: d2h/dDelta2}), where we should
set $z^{*}=z_{c}^{*}$, and for $\Delta_{c}$ we use (\ref{eq:Delta_c}).

Let us denote the approximate value of $h(z^*)$ given by (\ref{eq: h(z^*)}) without  the correction  as $h_{app}(z^*)$ . Then, 
\begin{eqnarray}
	\left|e^{t^{2/3}h(z^{*})}-e^{t^{2/3}h_{app}(z^{*})}\right|&=&\left|e^{t^{2/3}h_{app}(z^{*})}\right|\left|1-e^{O\left(t^{2/3}(\Delta-\Delta_c)^3\right)}\right|\label{eq: h_app-h 1}\\
	&\leq&\left|e^{t^{2/3}h_{app}(z^{*})+O\left(t^{2/3}(\Delta-\Delta_c)^3\right)}O\left(t^{2/3}(\Delta-\Delta_c)^3\right) \right|,\notag 
\end{eqnarray}
from where we obtain
\begin{eqnarray}
	&&\left|e^{t^{2/3}h(z^{*})}-e^{-\frac{\left(s_{2}+s_{1}\right)\left(r_{2}-r_{1}\right)}{2}+\frac{2}{3}\left(\frac{r_{2}-r_{1}}{2}\right)^{3}-\frac{\left(s_{1}-s_{2}\right)^{2}}{4(r_{2}-r_{1})}}\right|\,\,\,\,\,\,\,\,\,\,\,\,\,\,\,\,\,\,\,\,\,\,\,\,\label{eq: h_app-h 2}\\
	&\leq&O\left(e^{-\frac{\left(s_{2}+s_{1}\right)\left(r_{2}-r_{1}\right)}{2}+\frac{2}{3}\left(\frac{r_{2}-r_{1}}{2}\right)^{3}-\frac{\left(s_{1}-s_{2}\right)^{2}}{4(r_{2}-r_{1})}(1-\epsilon c_{1})}t^{-1/3}\left(s_{1}-s_{2}\right)^{3}\right)\notag
\end{eqnarray}
with some $\epsilon-$independent constant $c_{1}>0$. 
Here we used  the identity $$\left(\Delta-\Delta_{c}\right)^{3}=t^{-1}\left(s_{1}-s_{2}\right)^{3}\kappa_{f}^{3}+O((s_{1}-s_{2})^2t^{-4/3})$$ following 
 from (\ref{eq: X,Delta},\ref{eq:Delta_c}) to estimate the pre-exponential factor and  the inequality 
   $$\left|\Delta-\Delta_{c}\right|^{3}\leq\epsilon t^{-2/3}\left(s_{1}-s_{2}\right)^{2}\kappa_{f}^{2}$$
following from (\ref{eq: |Delta-Delta_c|<epsilon}) to estimate the exponent. As $\epsilon$
can be chosen arbitrarily small, this is obviously an integrable
function in the range $(s_{1},s_{2})\in[\underline{s},\infty)^{2}$.

For the integral part we have
\begin{align*}
(\nu-1)\int_{\Gamma_{1}} & \frac{\exp\left(t^{2/3}\left(h(v)-h(z^{*})\right)\right)}{(1-\nu v)(1-v)}\frac{dv}{2\pi\mathrm{i}}\\
 & =(\nu-1)\int_{\Gamma_{u_{+}\to u_{-}}}\frac{e^{t^{2/3\left(2^{-2/3}|f^{(0,0,3)}(\chi(\theta),\theta,z_{c}(\theta))|^{2/3}(r_{2}-r_{1})\right)(v-z^{*})^{2}}}}{(1-\nu v)(1-v)}\\
 & \,\,\,\,\,\,\,\,\,\,\,\,\,\,\,\,\,\,\,\,\,\,\,\,\,\,\,\,\,\,\,\,\,\,\,\,\,\,\,\,\times e^{O\left(t^{1/3}\left(s_{1}-s_{2}\right)\right)(v-z^{*})^{2}+O\left(t^{2/3}\right)\left(v-z^{*}\right){}^{3}}\frac{dv}{2\pi\mathrm{i}}\\
 & =\frac{t^{-1/3}}{2\kappa_{f}\sqrt{\pi(r_{2}-r_{1})}}\left(1+O\left(e^{-\delta t^{2/3}}\right)+O\left(t^{-1/3}(s_{1}-s_{2})\right)\right).
\end{align*}
Multiplying this integral by the estimate of $e^{t^{2/3}h(z^{*})}$
we come to the statement of the lemma. 
\end{svmultproof}

To prove the convergence of the Fredholm determinant we need the Kernel
to be an integrable in variables $s_{1},s_{2}.$ The above uniform
estimates are effective, when $s_{i}\ll t^{2/3}$ in the former case
and when $s_{i}\ll t^{1/3}$ in the latter. Otherwise the limiting
expression becomes comparable to the exponentially small correction,
which have not been controlled yet. The large deviation estimates
are necessary to fill this gap. In fact the large deviation estimate
for the diffusive term was already obtained in course of proof of
the lemma \ref{lem: Diffusive-part estimate}.
\begin{corollary}
\label{cor: ldb diffusive }For $\left(s_{1}+s_{2}\right)>\epsilon t^{1/3}$
and $t$ large enough
\[
e^{tf(\chi_{1},\theta_{1},z(\theta_{1}))-f(\chi_{2},\theta_{2},z(\theta_{2}))}\phi^{*(n_{1},n_{2},)}(x_{1},x_{2})\leq e^{-(r_{2}-r_{1})\frac{s_{1}+s_{2}}{2}}.
\]
\end{corollary}
The following lemma gives the one for the double integral part of
the kernel.
\begin{lemma}
\label{lem: large deviation bound Ksteptilda}(Large deviation bound
for the main part of the kernel.)

There exists small $\epsilon>0$, such that for $(s_{1},s_{2})\in[\underline{s},\infty)^{2}\backslash[\underline{s},\epsilon t^{2/3})^{2}$
and $t$ large enough 
\[
e^{t\left(f(\chi_{1},\theta_{1},z_{c}(\theta_{1}))-f(\chi_{2},\theta_{2},z_{c}(\theta_{2}))\right)}\widetilde{K}^{step}(n_{1},x_{1};n_{2},x_{2})\leq e^{-\delta(s_{1}+s_{2})},
\]
for some $\delta>0$. 
\end{lemma}
\begin{svmultproof}
We first note that $\widetilde{K}^{step}(n_{1},x_{1};n_{2},x_{2})$
written as a sum is majorized by the maximal summand times the total
number of summands, which is $O(t).$ Thus keeping the terms exponentially
decaying with $t$ in the variable range under consideration and ignoring
the power law factors we have 
\begin{eqnarray}
e^{t\left(f(\chi_{1},\theta_{1},z_{c}(\theta_{1}))-f(\chi_{2},\theta_{2},z_{c}(\theta_{2}))\right)}\left|\sum_{k=1}^{n_{2}}\tilde{\Psi}_{n_{1}-k}^{n_{1}}(x_{1})\Phi_{n_{2}-k}^{n_{2}}(x_{2})_{i}\right| & \label{eq: maxestimate},\\
\leq t \theta_2\exp\left(t\max_{\xi\in(0,\theta_{2})}\left(f_{2}^{-}(\xi)-f_{1}^{+}(\xi)\right)\right)\notag
\end{eqnarray}
where 
\[
f_{i}^{\pm}(\xi)=f(\chi_{i}+\xi,\theta_{i}-\xi,z_{\pm}(\theta_{i}-\xi))-f(\chi_{i},\theta_{i},z_{c}(\theta_{i}))
\]
 for $i=1,2$. Suppose that 
\begin{equation}
\chi(\theta)-\chi_{i}>\epsilon,\,\,\,i=1,2.\label{eq: chi_i-chi(theta)>epsilon}
\end{equation}

 To maximize the difference in the exponent, we note that
\begin{align}
\frac{d}{d\xi}\left(f_{2}^{-}(\xi)-f_{1}^{+}(\xi)\right)= & \ln\frac{z_{-}(\chi_{2}+\xi,\theta_{2}-\xi)}{z_{+}(\chi_{1}+\xi,\theta_{1}-\xi)}\label{eq: d/dxi(f2-f1)}\\
= & \ln\frac{z_{-}(\chi_{2}+\xi,\theta-\xi)}{z_{+}(\chi_{1}+\xi,\theta-\xi)}+O\left(t^{-1/3}\right)<0.\nonumber 
\end{align}
 Indeed, due to (\ref{eq:d chi/d theta <-1})
\[
\chi(\theta-\xi)-(\chi_{i}+\xi)=(\chi(\theta-\xi)-\chi(\theta)-\xi)+(\chi(\theta)-\chi_{i})>\epsilon,
\]
and for $z_{+}(\theta)$ $(z_{-}(\theta)$) being monotonously decreasing
(increasing) function the inequality (\ref{eq: d/dxi(f2-f1)}) holds.
This suggests that the maximum is achieved at $\xi=0.$ At $\xi=0$
the the maximized expression is monotonously increasing in $\chi$
as it is a difference of increasing and decreasing parts, whose derivatives
are 
\begin{equation}
\frac{d}{d\chi_{i}}\left(f_{i}^{\pm}(0)\right)=\ln\left(\frac{\left(1-z_{\pm}\right)\left(1-\nu z_{c}(\theta)\right)}{\left(1-\nu z_{\pm}\right)\left(1-z_{c}(\theta)\right)}\right)\lessgtr0.\label{eq:dfpm/dchi}
\end{equation}
Furthermore, as follows from 
\[
\frac{d^{2}}{d\chi_{i}^{2}}\left(f_{i}^{\pm}(0)\right)=\frac{dz_{\pm}}{d\chi}\frac{\nu-1}{\left(1-z_{\pm}\right)\left(1-\nu z_{\pm}\right)}\gtrless0
\]
 $f_{i}^{+}(0)$ and $f_{i}^{-}(0)$ are convex and concave functions
of $\chi_{i}$ respectively. Thus, the estimate for arbitrary $\chi_{i}<\chi(\theta)-\epsilon$
can be bounded by that with $\chi_{i}=\chi(\theta)-\epsilon$, i.e.
\begin{align*}
f_{i}^{\pm}(0) & \gtreqless\left.f_{i}^{\pm}(0)\right|_{\chi_{i}=\chi(\theta)-\epsilon}+\left.\frac{df_{i}^{\pm}(0)}{d\chi_{i}}\right|_{\chi_{i}=\chi(\theta)-\epsilon}(\chi_{i}-\chi(\theta)+\epsilon)\\
 & =\pm\frac{2}{3}\frac{\epsilon^{3/2}}{\kappa_{f}^{3/2}}\mp\left(\frac{\sqrt{\epsilon}}{\kappa_{f}^{3/2}}+O(\epsilon)\right)(\epsilon-s_{i}\kappa_{f}t^{-2/3})+O(\epsilon^{2})\\
 & =\mp\frac{1}{3}\frac{\epsilon^{3/2}}{\kappa_{f}^{3/2}}\pm s_{i}\sqrt{\frac{\epsilon}{\kappa_{f}}}t^{-2/3}+O(\epsilon^{2})+O(\epsilon s_{i}t^{-2/3}),
\end{align*}
where the first summand in the first line is approximated using (\ref{eq:fpm})
and the derivative in the second line is obtained from (\ref{eq:dfpm/dchi})
and (\ref{eq: z_pm exp}). Hence 
\begin{align*}
f_{2}^{-}(0)-f_{1}^{+}(0) & \leq\left(\frac{2}{3}\frac{\epsilon^{3/2}}{\kappa_{f}^{3/2}}-t^{-2/3}\sqrt{\frac{\epsilon}{\kappa_{f}}}\left(s_{2}+s_{1}\right)\right)\left(1+O(\sqrt{\epsilon})\right)\\
 & \leq-\frac{2t^{-2/3}}{3}\sqrt{\frac{\epsilon}{\kappa_{f}}}\left(s_{2}+s_{1}\right)\left(1+O(\sqrt{\epsilon})\right),
\end{align*}
where we used inequalities (\ref{eq: chi_i-chi(theta)>epsilon}), equivalent to  $\kappa_f s_i t^{-2/3}>\epsilon$ for large $t$.
It follows then, that for any $\delta>0$ and large enough $t$ 
\[
\theta_2t\exp\left[t\left(f_{2}^{-}(0)-f_{1}^{+}(0)\right)\right]\leq e^{-\delta(s_{1}+s_{2})}.
\]
 If one of $\chi_{i}$ is inside the $\epsilon-$vicinity of $\chi(\theta)$,
the same argument is used for the other variable, while for $f_{i}^{\pm}(0)$
we use (\ref{eq:fpm}), that yields 
\[
f_{i}^{\pm}(0)=\pm\frac{2}{3}s_{i}^{3/2}t{}^{-1}+O(\epsilon^{2}).
\]
Then for example for $s<\chi_{1}-\chi(\theta)<\epsilon$ and we have 

\begin{align*}
f_{2}^{-}(0)-f_{1}^{+}(0) & \leq\left(\frac{1}{3}\frac{\epsilon^{3/2}}{\kappa_{f}^{3/2}}-\frac{2}{3}s_{1}^{3/2}t{}^{-1}-t^{-2/3}\sqrt{\frac{\epsilon}{\kappa_{f}}}s_{2}\right)\left(1+O(\sqrt{\epsilon})\right)\\
 & \leq-\left(\frac{2t^{-2/3}}{3}\sqrt{\frac{\epsilon}{\kappa_{f}}}s_{2}+\frac{2}{3}s_{1}^{3/2}t{}^{-1}\right)\left(1+O(\sqrt{\epsilon})\right),
\end{align*}
which leads to the same estimate.
\end{svmultproof}

Now we are in position to complete the proof of the first part of
Theorem \ref{thm: KPZ convergence}. 
\begin{svmultproof}
First we note that there exists $\delta>0,$ such that 
\[
K(n_{1},x_{1};n_{2},x_{2})\leq\mathrm{const}\,e^{-\delta\left(s_{1}+s_{2}\right)}.
\]
Indeed, for the corrections and for the diffusive part this directly
follows from the lemmas \ref{lem: uniform estimate phi psi}-\ref{lem: large deviation bound Ksteptilda}.The
main part for is bounded by
\begin{align*}
\left|\int_{0}^{\infty}e^{-q(r_{2}-r_{1})}\mathrm{Ai}(s_{1}+q)\mathrm{Ai}(s_{2}+q)dq\right|\\
\leq & \mathrm{\delta_{1}}\int_{0}^{\infty}e^{-q(r_{2}-r_{1})}e^{-\delta\left((s_{1}+q)+(s_{2}+q)\right)}dq\\
\leq & \mathrm{const\,}e^{-\delta(s_{1}+s_{2})},
\end{align*}
where the first inequality follows from the asymptotics of Airy function
$\mathrm{Ai}(x)=O\left(e^{-\frac{2}{3}x^{3/2}}\right)$ for $x\to\infty$ and
its boundedness, which in particular suggests that for any $\delta_{1}>0$
there exists $\delta$, such that $\mathrm{Ai}(x)<\delta e^{-\delta_{1}x}$
for $x>0$, while the integral converges if $\delta>r_{1}-r_{2}.$
Thus, by Hadamard inequality 
\begin{align} \label{Hadamard1}
\det_{1\leq i,j\leq n}K^{step}(n_{1},x_{1};n_{2},x_{2})\leq n^{n/2}\delta^{n}\prod_{i=1}^{n}e^{-\delta_{1}s_{i}},
\end{align}
and hence 
\begin{align}\label{Hadamard2}
\sum_{x_{1}\leq a_{1}}\cdots\sum_{x_{n}\leq a_{n}}\det_{1\leq i,j\leq n}K^{step}(n_{1},x_{1};n_{2},x_{2})\leq n^{n/2}\mathrm{const}^{n}.
\end{align}
This ensures the absolute convergence of the Fredholm sum (\ref{eq: Fredholm series})
Then the summation in $x_{1},\dots,x_{n}$ and where estimates from
lemmas \ref{lem: uniform estimate phi psi}-\ref{lem: large deviation bound Ksteptilda}
inserted into the sum and taking the limit $t\to\infty$ yield the
Fredholm determinant with the Airy$_{2}$ kernel in the form 
\begin{align*}
K_{\mathcal{A}_{2}}(r_{1},s_{1};r_{2},s_{2}) & =-\mathbb{I}_{r_{2}>r_{1}}\left(4\pi\right)^{-1/2}e^{-\frac{\left(s_{2}+s_{1}\right)\left(r_{2}-r_{1}\right)}{2}+\frac{2}{3}\left(\frac{r_{2}-r_{1}}{2}\right)^{3}-\frac{\left(s_{1}-s_{2}\right)^{2}}{4(r_{2}-r_{1})}}\\
 & +\int_{0}^{\infty}e^{q(r_{1}-r_{2})}\mathrm{Ai}(s_{1}+q)\mathrm{Ai}(s_{2}+q)dq
\end{align*}
 plus the corrections vanishing in the limit $t\to\infty$. The latter
formula can be reduced to the the definition (\ref{eq:Airy2}) using
the relation \cite{Johansson1}
\[
\int_{-\infty}^{\infty}e^{q(r_{1}-r_{2})}\mathrm{Ai}(s_{1}+q)\mathrm{Ai}(s_{2}+q)dq=\left(4\pi\right)^{-1/2}e^{-\frac{\left(s_{2}+s_{1}\right)\left(r_{2}-r_{1}\right)}{2}+\frac{2}{3}\left(\frac{r_{2}-r_{1}}{2}\right)^{3}-\frac{\left(s_{1}-s_{2}\right)^{2}}{4(r_{2}-r_{1})}}.
\]
\end{svmultproof}

\subsection{Alternating initial configuration}

The ideas used  to prove  the alternating IC case are very similar to that of the step IC. The asymptotic analysis of the kernels is performed with the steepest descent method. 
To prove the convergence of the Fredholm sum (\ref{eq: Fredholm series}) we use the uniform convergence of the kernel on bounded sets of the summation variables 
and the large deviation bounds.  The difference is in technical details, which we sketch below.

\subsubsection*{Saddle points and contour of steepest descent}

We start the asymptotic analysis from the expression for the kernel  (\ref{K_alt}) written in the form
\begin{eqnarray}
	K_{t}^{alt}(n_{1},x_{1};n_{2},x_{2})=(1-\nu)\frac{\mathbbold{1}_{r_{2}>r_{1}}}{2\pi i}\oint_{\Gamma_{1}}\frac{e^{t(f(\chi_{2},\theta_{2},v)-f(\chi_{1},\theta_{1},v))}}{(1-\nu v)(1-v)}dv\label{eq: KaltF}\\+\frac{1}{2\pi i} \oint_{\Gamma_0} \frac{dv}{v}e^{ t f(\chi_{2},\theta_{2}, v )-t f\left(\chi_1,\theta_1,\frac{1-v}{1-\nu v}\right)},
	\nonumber
\end{eqnarray}
where  $\chi_{i},\theta_{i}$ for $i=1,2$    have the same meaning as in (\ref{chi_i,theta_i})  and the function  $f(\chi,\theta,v)$ was defined in (\ref{f}).
As before, we will refer to the first integral as the diffusive part  and to the second integral as the main part of the kernel.

The diffusive part integral is the same  as in the case of step IC. Thus,  we concentrate on the main part of the kernel.
Let us make the variable change  \begin{eqnarray}
\omega&=&2\theta_1+2\theta_2+\chi_1+\chi_2 -2v(1/2),\label{eq: omega}\\ 
\psi&=&\chi_1-\chi_2, 
\end{eqnarray}  
where $$v(1/2)=2 j_\infty(1/2)=\frac{\mu-\nu}{(1-\mu +\sqrt{1-\nu})\sqrt{1-\nu}}$$ is the particle velocity at $c=1/2$, obtained by 
substituting the  solution $z_c\in[0,1]$ of eq. (\ref{eq: c(z)}),
\begin{eqnarray}
z_c=\frac{1}{1+\sqrt{1-\nu}},\label{eq: z_c(1/2)}
\end{eqnarray}
into the formula (\ref{eq: j_infty}).

Then, the  exponent of the integrand   has the following form
\begin{eqnarray}
	g(\omega,\psi, u)&=&\left( v(1/2)+\frac{\omega-\psi}{2}\right)\ln\frac{1-u}{1-\nu u} \label{eq: g} \\
&-&\left(v(1/2)+\frac{\omega+\psi}{2}\right)\ln u	+\ln \left (\frac{(1-\nu u)(1-p+p u)}{(1-\mu u)}\right).\nonumber
\end{eqnarray}
Referring to  the conjecture (\ref{eq: x_n-x_0,flat}) for the distance traveled by particle  we  assume the following scaling  for  $\theta_i=\theta_{r_i}$ and $\chi_i=\chi_{r_i,s_i}$,   
\begin{eqnarray}
	\theta_{r}&=&\frac{v(1/2)}{2}+2^{5/3}r\kappa_{c }t^{-1/3},\label{eq:theta_r alt}\\
	\chi_{r,s}&=&-2^{8/3} r\kappa_{c }t^{-1/3}-2^{1/3}s \kappa_{f }t^{-2/3}.\label{eq:chi_r,s alt}
\end{eqnarray}
The variables $r$ and $s$ characterize the displacements of order of $t^{2/3}$ and $t^{1/3}$
of the corresponding quantities from their macroscopic positions, and, as before, we will use notations $\theta_{r_i},\chi_{r_i,s_i}$ instead of   $\theta_i,\chi_i$, when we want to emphasize the dependence of the latter  on these variables. 
Here and further within this subsection we imply that $\kappa_{c}=\kappa_{c}(1/2)$ and  $\kappa_{f}=\kappa_{f}(1/2)$ are constants obtained by substituting $z_c$ from (\ref{eq: z_c(1/2)}) to (\ref{eq: kappa_f},\ref{eq: kappa_c}). A useful relation to be used in further calculations is 
\begin{equation}
\kappa_{c}=\frac{\sqrt{1-\nu}}{2}\kappa^2_{f}.\label{kcf}
\end{equation}
In terms of $\omega, \psi$ this scaling has the form
\begin{eqnarray}
&\omega&=-2^{1/3}(s_1+s_2) \kappa_{f } t^{-2/3},\label{omega scaling}\\
&\psi&=2^{8/3}r_{21}\kappa_{c } t^{-1/3}-2^{1/3}(s_1-s_2) \kappa_{f} t^{-2/3},\label{psi scaling}
\end{eqnarray}
where $r_{21}=r_2-r_1$. Note that in (\ref{eq:theta_r alt},\ref{eq:chi_r,s alt})  we have chosen the macroscopic reference value of particle numbers  $\theta=v(1/2)/2$  
 such that the  corresponding  value of the coordinates is $\chi=0$.
Then, within  the Fredholm sum (\ref{eq: Fredholm series})      we will have $\chi_i\leq0$ in most of the summation range,
 may be with exception of small $O(t^{-2/3})$ vicinity of $\chi_i=0$. This suggests that   $\omega\leq 0$ and $|\psi|\leq -\omega$, when $|\omega|$ and $|\psi|$ are larger than $O(t^{-2/3})$ and $O(t^{-1/3})$ respectively.

To estimate the main part of the kernel in the limit   $t\to\infty$ we have to locate the saddle points of the integrand in the range of  values of $\omega$ and $\psi$.
To this end, let us consider  the function  $g(\omega, \psi , z)$. It has logarithmic singularities at the
points $z= 0,1,1/\nu,1/\mu$  and $z=\eta$, where $\eta=(p-1)/p$. It also has four critical points defined by  equation 
\begin{equation}
g^{(0,0,1)}(\omega, \psi, z)=0 ,\label{g001}\\
\end{equation}
which can be recast in the form of a degree 4 polynomial equation  with real coefficients that  has    either 4 real or a pair of real and a pair of  complex conjugate or two pairs of complex conjugate roots. Of course, the roots depend on $\omega$ and $\psi$.   

Two roots may coincide, when $z$ simultaneously satisfies another equation 
\begin{eqnarray}
g^{(0,0,2)}(\omega, \psi, z)=0\label{g002}, 
\end{eqnarray}
in which case we have a double saddle point of the inetgrand.  It should obviously be real.
The consistency of equations (\ref{g001},\ref{g002}) implies a constraint on values of $\psi$ and $\omega$ having the form of a system of two linear equations, which can be solved 
to obtain   $\omega=\omega(z)$ and $\psi=\psi(z)$ as functions of  position of the double saddle point $z$.
In particular, it can be readily checked that  the position of the double saddle point $z=z_c$, where $z_c$ is defined by  (\ref{eq: z_c(1/2)}), yields 
$\omega(z_c)=\psi(z_c)=0$. It is more difficult to verify that this is the only solution $z\in [0,1]$ satisfying $\omega(z)\leq0$. To this end, 
we represent the rational function $\omega(z)$ as a single fraction. Its denominator  factorizes into a product of positive factors and the numerator is divisible by $(z-z_c)^2$. Finally, the quotient of the numerator divided by  $(z-z_c)^2$ is a fifth degree polynomial in the variable $\sqrt{1-\nu}$ with manifestly positive coefficients. For the explicit expressions being cumbersome, we omit them.

Given two roots coinciding  at  $z_c$, we can find the other two roots   from  equation 
\begin{eqnarray}\label{two_roots}
v^2-v\left(\frac{1+\sqrt{1-\nu}}{\mu}+1\right)+\frac{1+\sqrt{1-\nu}}{\mu \nu}=0.
\end{eqnarray}
This equation has either two complex conjugate or two real roots, when  $\nu> 0$ or $\nu<0$ respectively. In the latter case, one of the roots  is located in  $(1/\nu,\eta)$ and the other in  $(1/\mu,+\infty)$.  

\begin{figure}
	\centerline{\includegraphics[clip,width=0.8\textwidth]{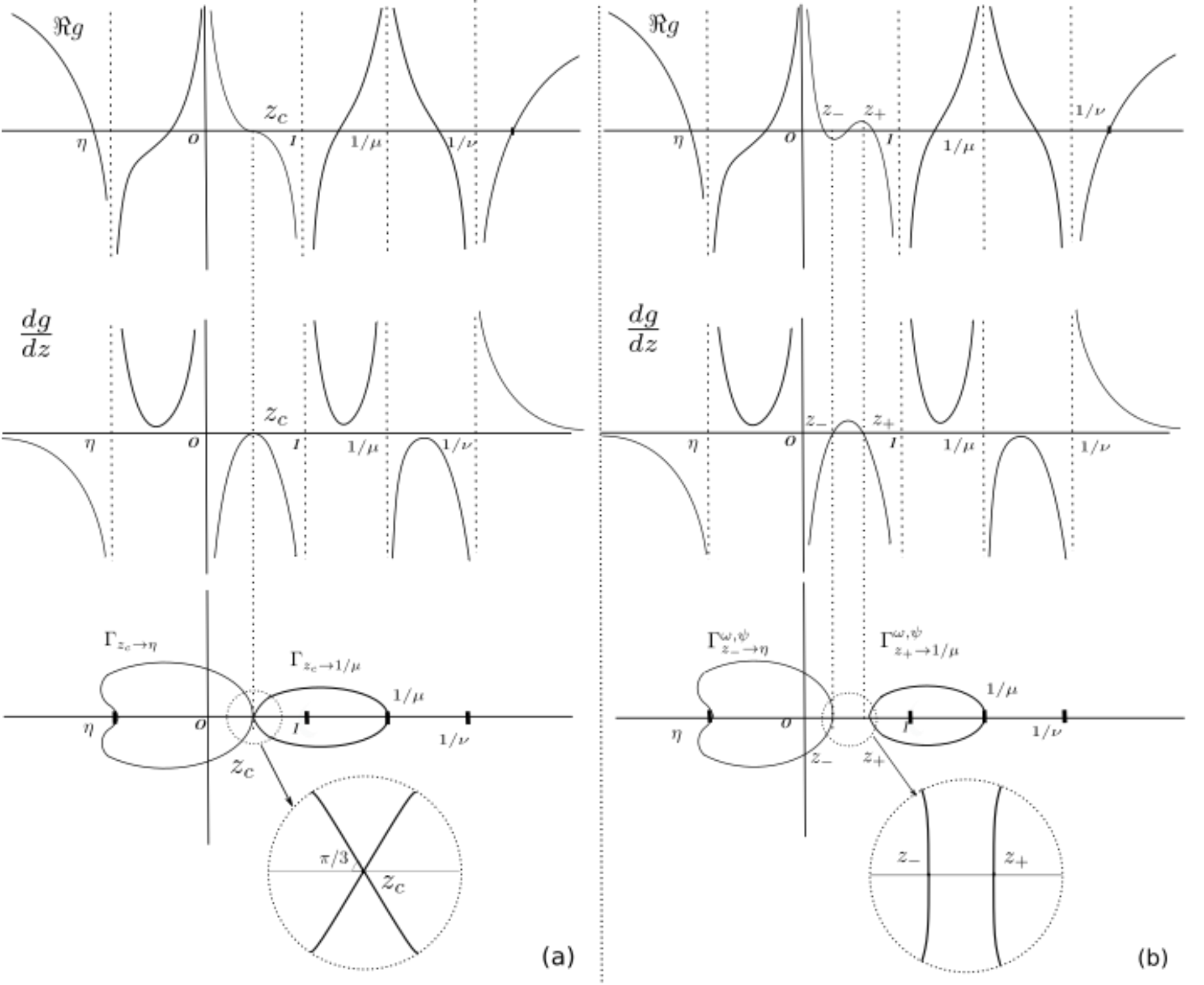}}
	\caption{Schematic illustration of behavior of $\Re g(\omega, \psi , u)$, $d g(\omega, \psi , u)/d u$ and the steepest descent and ascent contours  (a)  $\Gamma_{z_{c}\to\eta},\Gamma_{z_{c}\to1/\mu}$ for $(\omega,\psi)=(0,0)$, when the two critical points are merged together, and  (b)  $\Gamma_{z_{-}\to\eta}^{\omega,\psi},\Gamma_{z_{+}\to1/\mu}^{\omega,\psi}$ for   $\omega<0$, when there are two real critical points  and $z_-<z_c<z_{+}$. 	The behavior of the steepest descent contours in the vicinity to critical points is shown in circles. \label{fig: contours_1}}
\end{figure}

Thus, when   $\omega=0, \psi=0$ the function $g(0, 0, z)$ has the double saddle point at $z_c$.
As $\omega$ decreases or $\psi$ deviates from zero,  the double saddle point splits into two critical points $z_+>z_-$ located on the real line. One can  see from fig. \ref{fig: contours_1} that the point $z_{-}$ ($z_+$) is a minimum (maximum) along the real axis of the function $g(\omega, \psi , z)$.

To characterize the location of the roots $z_\pm$ consider first the case $\omega=0$. The equation (\ref{g001}) always has a solution $z_1=z_c$ independent of $\psi$. Indeed, the simpler equation $g^{(0,0,1)}(0,\psi,z)-g^{(0,0,1)}(0,0,z)=0$ leads to a quadratic equation for $z$ with one of the roots being $z=z_c$, while we have already shown that the second term in the l.h.s. vanishes at $z=z_c$.    Since the  double saddle point in $[0,1]$ exists only when  $\psi=0$, the second root $z_2$ of (\ref{g001}) should deviate from $z_c$ as $\psi$ goes away from zero. The derivative 
\begin{equation}
\frac{\partial z_2}{\partial \psi}=-\frac{g^{(0,1,1)}(0,\psi,z_2)}{g^{(0,0,2)}(0,\psi,z_2)}=\frac{ 1-2 z_2+\nu  z_2^2}{2 (1-z_2) z_2 (1-\nu  z_2)g^{(0,0,2)}(0,\psi,z_2)}>0
\end{equation}
is positive, since the numerator is positive (negative), when $z_2<z_c (z_2>z_c)$, as well as $g^{(0,0,2)}(0,\psi,z_2)>0 (<0)$ for $z_2$ being minimum (maximum) in this case. 
Thus, when $\omega=0$, $z_-=\min\{z_2,z_c\}$ and $z_+=\max\{z_2,z_c\}$.   Note that the ratio in the r.h.s becomes $0/0$ indeterminate  as $z_2\to z_c$, which however can be resolved using the L'Hospital's rule showing that the derivative is actually continuous at this point. 

To probe the regime $\omega<0$ we show that  the sign of the derivative 
\begin{equation}
\frac{\partial z_\pm}{\partial \omega}=-\frac{g^{(1,0,1)}(0,\psi,z_\pm)}{g^{(0,0,2)}(\omega,\psi,z_\pm)}=\frac{1-2\nu z_\pm +\nu z_\pm^2}{2 (1-z_\pm) z_\pm (1-\nu  z_\pm)g^{(0,0,2)}(\omega,\psi,z_\pm)}\lessgtr 0,
\end{equation}
coincides with that of $g^{(0,0,2)}(0,\psi,z_2)$. In other words as $\omega$ decreases from zero, both roots move away from $z_c$ towards $z_-=0$ and $z_+=1$ respectively. Note that when $\psi,\omega \to 0$ and hence $z_\pm\to z_c$ the denominator vanishes, while the numerator stays finite. This is a signature of  a non-analytic dependence of the roots on $\omega$ at $\omega=\psi=0$. 

We also note that  $z_-$ and $z_+$  cross the points $z_-=0$ and $z_+=1$ respectively, when the expressions in parenthesis being the coefficient of the first and second logarithms in  
(\ref{eq: g}) change their sign. As follows from fig.~\ref{fig: contours_1}, then the extrema  at these points change  from minima to maxima and vice versa. For $z_-$, this leads to disappearance of the pole inside the integration contour, which causes the integral vanishing. Also, the change of the extremum at $z_+$  is not relevant for our  discussion, since we will use only the contour passing through $z_-$.    

Finally, looking at the structure of the lines in complex  plain, where the imaginary part of $g(0, 0 ,z)=0$ vanishes, we conclude that    two stationary phase contours  $\Gamma_{z_c \to\eta}$ and  $\Gamma_{z_c\to1/\mu}$ 
 can be drawn through $z_c$  being the simple loops closed around $z=0$ and $z=1$  via the singularities at $z=\eta$ and $z=1/\mu$ respectively, see fig.~\ref{fig: contours_1}. Being the steepest descent and ascent contours for $\Re g(0, 0 ,z)$  they approach the double saddle point $z=z_c$ at angles $\pm 2\pi/3$ and $\pm \pi/3$ to the real axis.  The general structure is preserved by  continuity as $z_-$ and $z_+$ go away from $z_c$. Then, we have two steepest descent contours, which we refer to as $\Gamma_{z_-\to\eta}^{\omega,\psi}$ and $\Gamma_{z_+\to1/\mu}^{\omega,\psi}$, that cross the real axis perpendicularly. The contours  $\Gamma_{z_c \to\eta}$ and $\Gamma_{z_-\to\eta}^{\omega,\psi}$ are those we  use below in the  steepest descent analysis of the main part of the kernel.

\subsubsection*{Convergence}

\begin{lemma}
	\label{lem: uniform estimate g2}Given $r_1, r_2>0,$ $\underline{s} \in\mathbb{R}$, let  $n_i=t\theta_{r_i}$ and $x_i=t\chi_{r_i,s_i}$ for $i=1,2$ where 
	$\theta_r$ and $\chi_{r,s}$ are defined by (\ref{eq:theta_r alt}) and (\ref{eq:chi_r,s alt}). Then, there exist $\delta>0$, such that estimates 
	\begin{align*}
		\tilde{K}_{t}^{alt}(n_{1},x_{1};n_{2},x_{2})=&
		\kappa_{f}(2t)^{-1/3}z_c^{x_2-x_1} e^{\frac{2}{3}r_{21}^{3}+r_{21}(s_{2}+s_{1})}\\ \times& \Big( \mathrm{Ai}\left(r_{21}^{2}+s_{2}+s_{1}\right)+O(e^{-\delta t})+O(t^{-1/3}e^{-\delta_1(s_1+s_2)}) \Big)
	\end{align*}
	hold uniformly for $s_i>\underline{s}$ with any $\delta_{1}>0$.
\end{lemma}
\begin{remark}
		The factor 
	\begin{equation}
	e^{tg(\omega, \psi , z_c)}=z_c^{x_2-x_1} \label{conjugation}
	\end{equation}
coming with the kernel here and below is the conjugation  that  does not affect the value of the Fredholm determinant.  	
\end{remark}
\begin{svmultproof}
	
	\textsl{(Bounded sets) }Suppose first $s_1, s_2 \in[\underline{s},\bar{s}]$
	for some $\bar{s}>\underline{s}$ .
	The integration contour we use is the steepest descent
	contour $\Gamma_{z_{c}\to \eta}$
	of the function $g(0, 0 ,z)$. Then,
	for some small $\epsilon>0$ we can drop the part of the integral along
 $\Gamma_{z_{c}\to \eta}$ beyond the $\epsilon-$neighborhood $U_{\epsilon}(z_{c})=\{z:|z-z_{c}|<\epsilon\}$
	of the saddle point. For the contour being the steepest descent
	this yields an error of order of $O\left(\exp(-\delta_{\epsilon}t)\right).$
	
	Limiting the integration to the part of the contour inside $U_{\epsilon}(z_{c})$ we use the Taylor expansion  
	\begin{eqnarray}
	g(\omega, \psi , u)=-\psi \ln z_c+t^{-1}\left(-\frac{1}{3}y^3+y^2(r_2-r_1)+y(s_1+s_2)\right)\nonumber\\
	+O\left(y^4t^{-4/3}, y^3t^{-4/3}, y^2t^{-4/3}\right),\label{eq: g-Taylor}
	\end{eqnarray}
   where we make a variable change 
$ y=t^{1/3}\kappa_{f} 2^{1/3}(u-z_c)/z_c$.
   
   Using the notation  $g_{app} $ for the part of  (\ref{eq: g-Taylor}) without correction we have	
\begin{eqnarray}
	\int_{\Gamma_{z_{c}\to\eta}\bigcap U_{\epsilon}(z_{c})}&&\frac{\exp\left(tg_{app}\right)}{z_{c}}\frac{du}{2\pi\mathrm{i}}\\
	&&=\frac{ z_c^{x_2-x_1}}{(2t)^{1/3} \kappa_{f}}\int_{t^{1/3}\epsilon e^{-\mathrm{i}\left(\frac{2\pi}{3}+\epsilon_{1}\right)}}^{t^{1/3}\epsilon e^{\mathrm{i}\left(\frac{2\pi}{3}+\epsilon_{1}\right)}} e^{\frac{1}{3}(y^3+y^2r_{21}+y(s_1+s_2)}\frac{dy}{2	\pi i },\nonumber
\end{eqnarray}
where we replaced the upper and the lower halfs of the  contour by two
	segments of rays approaching the origin at the angles $\pm\left(\frac{\pi}{3}+\epsilon_{1}\right),$
	where $\epsilon_{1}$ is an $\epsilon-$ dependent constant, which
	can be made small by choosing the $\epsilon$ small enough. Making another variable change $y\to (y-r_{21})$, shifting the integration contour by $r_{21}$ in horizontal direction for  the price of another exponentially 
	small correction and using   the integral representation of the Airy function  (\ref{eq:Airy1}) we arrive at  the expression reproducing the statement  of the lemma up to  correction terms. 
	
	Keeping in mind  the exponentially small in $t$ corrections acquired in course of the above derivation we should also account for the corrections coming form Taylor approximation (\ref{eq: g-Taylor}). This argument  literally reproduces  
	that used in the case of step IC (see e.g. similar estimate in lemma \ref{lem: uniform estimate phi psi}, eqs. (\ref{eq: f error},\ref{eq: error double})) and results in  corrections of the form  $O(t^{-2/3}e^{t g(\omega,\psi, z_c)})$.
	
	\textsl{(Arbitrary sets) }The next step is to extend this estimate
	to large values of $s_1,s_2$ . Suppose first that $(s_1+s_2)>\epsilon t^{2/3} $ for some small $\epsilon>0$. To analyze the main part of the kernel we
	use the steepest descent integration contour $\Gamma_{z_-\to\eta}^{\omega,\psi}$.
	The integral hence is bounded by the maximum of the
	integrand at this contour, attained at the point $z_-=z_{-}(\omega,\psi) \in(0,z_c)$, i.e. 
\begin{equation}
e^{-t g(\omega,\psi, z_c)}\tilde{K}_{t}^{alt}(n_{1},x_{1};n_{2},x_{2})=O\left(e^{t (g(\omega,\psi, z_-)-g(\omega,\psi, z_c))}\right).\label{eq: tilde_K bound exp}
\end{equation}
In  the case under consideration   $\omega< - \epsilon_0 $ with $\epsilon_0\simeq\kappa_f 2^{1/3} \epsilon$, and  we assume that $|\psi|\leq -\omega$.
We want to show  the r.h.s. of (\ref{eq: tilde_K bound exp}) is exponentially small in $t$, i.e. that there exist  $\delta>0$ such that  inequality
\begin{equation}
	g(\omega, \psi, z_{c})- g(\omega,\psi,z_{-}(\omega, \psi))>\delta \label{diffrence}
\end{equation}
holds. The l.h.s. of the above inequality is difficult to analyze directly as it depends on two variables. We, however, can use the fact that it 
is monotonous in $\psi$ since the derivative
\begin{equation*}
	\frac{d}{d \psi } \left(g(\omega, \psi, z_{c})- g(\omega,\psi,z_{-}(\omega, \psi))\right)=\frac{1}{2}\ln\left(\frac{(1-z_-)z_-}{(1-\nu z_-)z_c^2}\right)<0
\end{equation*}
is negative, when  $z_-<z_c$.  Thus, we can bound the difference by that at   $\psi=-\omega$.
\begin{equation}
	g(\omega, \psi, z_{c})- g(\omega,\psi,z_{-}(\omega, \psi))>g(\omega, -\omega, z_{c})- g(\omega,-\omega,z_{-}(\omega, -\omega))\label{ineqg}
\end{equation}
To bound  the r.h.s. of this inequality we reproduce the arguments of lemma \ref{lem: uniform estimate phi psi}.   
We note that  the two quantities  $g(\omega,-\omega,z_\pm(\omega, -\omega))$   depend  non-analytically on $\omega$ at $\omega=0$ since $z_{\pm}(\omega, -\omega)$ does. Rather, they are analytic in the variable   $\tilde{\zeta}=\sqrt{-\omega}$. In particular,  
\begin{equation*}
\left.\frac{d}{d \tilde{\zeta}}\right|_{\tilde{\zeta}=0}g^{(0,0,2)}\left(-\tilde{\zeta}^2,\tilde{\zeta}^2, z_{\pm}(-\tilde{\zeta}^2, \tilde{\zeta}^2)\right)=\pm\tilde{\alpha} g^{(0,0,3)}(0,0,z_c),
\end{equation*}
where
\begin{equation*}
\tilde{\alpha}=\left|\frac{dz_-(\tilde{\zeta}^2,\tilde{\zeta}^2)}{d\tilde{\zeta}}\right|_{\tilde{\zeta}=0}=\sqrt{\frac{g^{(1,0,1)}(0,0,z_c)}{g^{(0,0,3)}(0,0,z_c)}}
\end{equation*}
and we used
\begin{eqnarray}
		g^{(1,0,1)}( 0, 0,z_c)&=&-\frac{1}{z_c},\quad	 g^{(0,1,1)}( 0,0 ,z_c)=0,\\
			g^{(0,0,3)}(0 ,0 ,z_c)&=&-4 \left( \frac{\kappa_{f}}{z_c}\right)^3.\label{g101,g011,g003}
\end{eqnarray}
As the second derivative $g^{(0,0,2)}\left(-\tilde{\zeta}^2,\tilde{\zeta}^2, z_{\pm}(-\tilde{\zeta}^2, \tilde{\zeta}^2)\right)$ is zero at $\tilde{\zeta}=0$ and its derivative is not, the inequalities
\begin{equation*}
g^{(0,0,2)}\left(-\tilde{\zeta}^2,\tilde{\zeta}^2, z_{\pm}(-\tilde{\zeta}^2, \tilde{\zeta}^2)\right)\lessgtr 0
\end{equation*}
must hold for small $\tilde{\zeta}>0$. They can also be extended to arbitrary $\tilde{\zeta}$ since no other double saddle points  appear in $(0,1)$.
Thus, we conclude that $z_{-}(-\tilde{\zeta}^2, \tilde{\zeta}^2)$ and $z_{+}(-\tilde{\zeta}^2, \tilde{\zeta}^2)$ are minimum and maximum the function  $g\left(-\tilde{\zeta}^2,\tilde{\zeta}^2, z\right)$ of the variable $z$ respectively and  this function is monotonously increasing in   $z_-(-\tilde{\zeta}^2,\tilde{\zeta}^2)<z< z_+(-\tilde{\zeta}^2,\tilde{\zeta}^2)$. In particular, there exists $\delta>0$ such that $\left(g(-\tilde{\zeta}^2, \tilde{\zeta}^2, z_{c})- g(-\tilde{\zeta}^2,\tilde{\zeta}^2,z_{-}(-\tilde{\zeta}^2, \tilde{\zeta}^2)\right)>\delta$, which  yields  (\ref{diffrence}).

Now suppose that $s_1+s_2<\epsilon t^{2/3}$. The critical points $z_{\pm}=z_{c}+u$  are to be found from 
	\begin{eqnarray}
		g^{(0,0,1)}(\omega,\psi ,z_c+u)&=&\omega g^{(1,0,1)}(0,0,z_c)+\psi g^{(0,1,1)}(0,0,z_c)\nonumber\\
		&+&u \omega g^{(1,0,2)}( 0, 0,z_c)+ u\psi g^{(0,1,2)}( 0, 0,z_c)\label{eq_u-}\\
		\nonumber &+&\frac{u^2}{2}g^{(0,0,3)}( 0, 0,z_c)+O(u^3, u^2\omega, u^2 \psi)=0.
	\end{eqnarray}
assuming that $u, \omega, \psi$ are small.	The involved derivatives are given explicitly by (\ref{g101,g011,g003}) and
	\begin{align*}
		g^{(1,0,2)}( 0, 0,z_c)=&\frac{1}{z_c^2}-\frac{1}{z_c^2\sqrt{1-\nu}} ,\ g^{(0,1,2)}( 0, 0,z_c)=\frac{1}{z_c^2\sqrt{1-\nu}}.\\
	\end{align*}
	Solving (\ref{eq_u-}) for $u$ we obtain  
	\begin{eqnarray}
		u_{\pm}=\frac{z_c(2t)^{-1/3}}{\kappa_{f}}\left(r_{21}\pm\sqrt{r_{21}^2+s_1+s_2}\right)+O\left(\frac{s_1+s_2}{t^{2/3}}\right), \label{u-}
	\end{eqnarray}
	where we use scaling form of $\omega, \psi $  (\ref{omega scaling}, \ref{psi scaling}) and the fact that $s_1+s_2\geq|s_1-s_2|$  to estimate the corrections.
	
The idea of the further analysis is to use the Taylor approximation  of $g(\omega,\psi,z)$  in the   vicinity $U_{\epsilon_1}(z_{-})$ of $z_-$ with some small $\epsilon_1>0$.	To this end, we limit the integration to  the part of the steepest descent contour  $\Gamma_{z_-\to\eta}$  within this domain,
 getting  the error of order of $O(e^{-\delta_{1}t})$ with some $\delta_1>0$. Then, 
 we  approximate the function $g(\omega, \psi, z)$ by its Taylor expansion near the critical point $z_-$.
The approximations  of  the function and its   derivatives at $z_-$ are obtained by substituting (\ref{u-}) into the exact expressions. 
	\begin{eqnarray}
		&&g(\omega,\psi,z_{-})= g(\omega,\psi,z_{c})	+ \frac{1}{t}\left(-\frac{2}{3}(r^2_{21}+s_1+s_2)^{3/2}+\frac{2}{3}r^{3}_{21}+r_{21}(s_2+s_1)\right)\nonumber\\
		&&\hspace{0.7\textwidth}+O\left(\frac{(s_1+s_2)^2}{t^{4/3}}\right),\nonumber\\
		&&g^{(0,0,1)}(\omega,\psi,z_{-})  =0,\nonumber\\
		&&g^{(0,0,2)}(\omega,\psi,z_{-})  =2^{5/3}\frac{ \kappa_f^{2}}{z_c^2}\sqrt{r_{21}^2+s_1+s_2}t^{-1/3} +O\left(\frac{s_1+s_2}{t^{2/3}}\right), \nonumber \\
		&&g^{(0,0,3)}(\omega,\psi,z_{-}) =-4 \left( \frac{\kappa_{f}}{z_c}\right)^3+O\left(\frac{\sqrt{s_1+s_2}}{t^{1/3}}\right). \label{gzc}
	\end{eqnarray}
As a result we obtain		
	\begin{eqnarray*}
		\tilde{K}_{t}^{alt}&(n_{1},x_{1};n_{2},x_{2}) = e^{tg(\omega,\psi,z_{c})-2/3 (r^2_{21}+s_1+s_2)^{3/2}+\frac{5}{3}r^{3}_{21}+r_{21}(s_2+s_1)+O\left(\frac{s_1+s_2}{t^{1/3}}\right)}\\
		&\times  \oint_{\Gamma_{z_{-}\to\eta}\bigcap U_{\epsilon}(z_{-})}e^{-t\left(z-z_{-}\right)^{3}\frac{2}{3}\left(\frac{\kappa_{f}}{z_c}\right)^3+\left(z-z_{-}\right)^{2}\left(\frac{\kappa_{f}}{z_c}\right)^22^{2/3}t^{2/3}\sqrt{r^2_{21}+s_1+s_2}}\\
		&\times  e^{ O\left((z-z_c)^2t^{1/3}(s_1+s_2) \right)+O\left(t(z-z_{-})^{3}t^{2/3}\sqrt{s_1+s_2}\right)+O\left(t\left(z-z_{-}\right)^{4}\right)}\frac{dz}{2\pi\mathrm{i}z}\\
		&+O\left(e^{t\left(g(\omega,\psi,z_{c})-\delta\right)}\right)+O(t^{-2/3}e^{-\delta_1(s_1+s_2)})
	\end{eqnarray*}
Finally, making  the variable change
	$$y=(2t)^{1/3}\left(z-z_{-}\right)\kappa_{f}/z_c-\sqrt{r_{21}^2+s_1+s_2},$$ transforming the contour accordingly as we did in lemma \ref{lem: uniform estimate phi psi} and estimating the corrections with the use of  inequality $|1-e^x|\le|x|e^{|x|}$ we arrive at  the statement of the lemma. 
%
\end{svmultproof}

\begin{lemma}
	\label{lem: Diffusive-part estimate2}(Uniform estimate of the diffusive
	part of the kernel)

	Under conditions of lemma \ref{lem: uniform estimate g2} the following estimate
	\begin{equation*}
		\phi^{*(n_{1},n_{2},)}(x_{1},x_{2})=\mathbbold{1}_{r_{2}>r_{1}} \frac{z_c^{x_2-x_1}(2t)^{-1/3}}{\kappa_{f}\sqrt{4\pi(r_{2}-r_{1})}}e^{-\frac{\left(s_{1}-s_{2}\right)^{2}}{4(r_{2}-r_{1})}}\times\left(1+O(t^{-1/3})\right)
	\end{equation*}
	holds uniformly  for $|s_1-s_2|< \epsilon t^{1/3}$ with some $\epsilon>0$.
\end{lemma}

\begin{svmultproof}
	The integral representation of the diffusive part of the kernel including the
	conjugation is given by
	\begin{equation}
		\phi^{*(n_{1},n_{2},)}(x_{1},x_{2})e^{-t g(\omega, \psi, z_c)}=\mathbbold{1}_{n_{2}>n_{1}}\oint_{\Gamma_{1}}\frac{dv}{2\pi\mathrm{i}}\frac{(\nu-1)e^{t^{2/3}\tilde{h}(v)}}{(1-\nu v)(1-v)},\label{eq: phi conj2}
	\end{equation}
	where
	\begin{align}
		\tilde{h}(v)=t^{1/3}\left(f(\chi_{2},\theta_{2},v)-f(\chi_{1},\theta_{1},v)-g(\omega, \psi, z_c)\right).\label{eq:h(v) alt}
	\end{align}
	We can see that the integrand is  similar to the one in lemma \ref{lem: Diffusive-part estimate} up to change of the function  $h(v)$ by $\tilde{h}(v)$, with all the difference in $v$-independent part $$g(\omega,\psi,z_c)=f(\chi_{2},\theta_{2},z_c)-f(\chi_{1},\theta_{1},z_c)$$ instead of $$f(\chi_{2},\theta_{2},z_c(\theta_2))-f(\chi_{1},\theta_{1},z_c(\theta_1)).$$
	Thus the proof  requires a simplified version of the analysis of that lemma. Specifically, the function 
		\begin{align*}
	\tilde{h}(z)=\tilde{h}_{1}(\Delta,z)  =2\Delta\ln\left(\left(\frac{1-z}{1-\nu z}\right)\frac{1}{z_c}\right)-t^{1/3}\delta\theta_{21}\ln\left(\frac{z}{z_c}\right)
	\end{align*}
	is a simplified version of  $h_1(\Delta,u)$, one of three parts of  the function $h(v)$ depending on the parameter $\Delta$ defined in (\ref{eq: X,Delta}), while the other two are absent. 
	The  estimate of interest for  $\Delta$ close to the critical value $\Delta_{c}  =-2^{2/3}\kappa_{c}r_{21}$, at which the integral is  maximal, 
	consists in evaluating the integral in the saddle point approximation, which is dominated by a single $\Delta$-dependent  saddle  point $z^*$ equal to $z_c$, when $\Delta=\Delta_c$. The analysis of the integral from lemma \ref{lem: Diffusive-part estimate} applied without any changes  then yields 
	\begin{align*}
		&(\nu-1)\int_{\Gamma_{1}} \frac{\exp\left(t^{2/3}\left(\tilde{h}(z)-\tilde{h}(z^*)\right)\right)}{(1-\nu z)(1-z)}\frac{dz}{2\pi\mathrm{i}}\\&=\frac{(2t)^{-1/3}}{2\kappa_{f}\sqrt{\pi(r_{2}-r_{1})}}\left(1+O\left(e^{-\delta t^{2/3}}\right)+O\left(t^{-1/3}(s_{1}-s_{2})\right)\right),
	\end{align*}
	while for  $h(z^*)$ we obtain 
	$$e^{t^{2/3}\tilde{h}(z^{*})}=e^{\left(-\frac{\left(s_{1}-s_{2}\right)^{2}}{4(r_{2}-r_{1})}\right)+O\left(\left(\Delta-\Delta_{c}\right)^{3}\right)},$$
	which is valid when $s_i <\epsilon t^{1/3}, i=1,2.$
	
	For larger deviations of $\Delta$ from $\Delta_c$ the integral can be bounded simply by the saddle point value  of the integrand. This estimate is slightly different 
	from that in lemma   \ref{lem: Diffusive-part estimate} since the second part of $h(v)$ proportional to $(s_1+s_2)$ is absent in the case of  the alternating IC.
	Instead we use the concavity of  $\tilde{h}_{1}(\Delta,z^{*})$   in $\Delta$ to show that 
	\begin{align*}
		\tilde{h}_{1}(\Delta,z^{*})<\tilde{h}_{1}(\Delta_c,z^{*})+\left.\frac{d^{2}\tilde{h}_{1}(\Delta,z^{*})}{d\Delta^{2}}\right|_{\Delta_c-2^{-2/3}\epsilon\kappa_f}2^{-2/3}\epsilon\kappa_f(\Delta-\Delta_c)\\=\tilde{h}_{1}(\Delta_c,z^{*})-\frac{(\epsilon+O(\epsilon^2)) t^{1/3}(s_{2}-s_{1})}{4(r_{2}-r_{1})},
	\end{align*}
	 when $\Delta<\Delta_c- 2^{-2/3}\epsilon\kappa_f$, i.e. $s_2-s_1>\epsilon t^{1/3}$. For further details we address the reader to proof lemma \ref{lem: Diffusive-part estimate}.
	 The last bound also proofs the next lemma.	
\end{svmultproof}
\begin{lemma}
	\label{lem: large deviation bound Diffusive-part }(Large deviation bound
	for the diffusive part of the kernel.)
	
	There exists a small $\epsilon>0$, such that for $|s_{1}-s_{2}|> \epsilon t^{1/3}$
	and $t$ large enough
	\[
	\phi^{*(n_{1},n_{2},)}(x_{1},x_{2})\leq \mathbbold{1}_{s_{2}>s_{1}} t^{-1/3}   e^ {t g(\xi,\psi, z_c)}e^{-\delta(s_{2}-s_{1})},
	\]
	for some $\delta>0$.
\end{lemma}

\begin{lemma}
	\label{lem: large deviation bound Kalttilda}(Large deviation bound
	for the main part of the kernel.)
	
	There exists small $\epsilon>0$, such that for $(s_{1},s_{2})\in[\underline{s},\infty)^{2}\backslash[\underline{s},\epsilon t^{2/3})^{2}$
	and $t$ large enough the inequality
	\[
	\widetilde{K}_t^{alt}(n_{1},x_{1};n_{2},x_{2})\leq t^{-1/3} e^ {t g(\omega,\psi, z_c)}e^{-\delta(s_{1}+s_{2})},
	\]
	holds for some $\delta>0$.
\end{lemma}

\begin{svmultproof}
	
	The integral can be majorized  by the maximum of the integrand times the total
	length of contour, which is $O(1).$  Therefore, we consider only  the exponential in
	 $t$ terms. Ignoring the power law factors we have.
	\[
	\widetilde{K}^{alt}(n_{1},x_{1};n_{2},x_{2})\sim e^{- tg(\omega,\psi, z_c)} \exp({t g(\omega, \psi , z_{-}(\omega, \psi))}).
	\]
	In this section, it is more convenient to use different variables 
	\[
	\omega_1=\frac{\omega+\psi}{2 }, \quad \psi_1=\frac{\omega-\psi}{2}.
	\]
	Suppose that $s_i t^{-2/3}>\epsilon,i=1,2$ for some $\epsilon>0$.
	Thus, we want to estimate the  function 
	$$g_1\left(\omega_1 ,\psi_1\right):= g\left(\omega,\psi, z_-(\omega, \psi)\right).$$
	Note that
	\begin{align}
		\frac{\partial}{\partial \omega_1}g_1(\omega_1, \psi_1)&=-\ln z_{-}>0,\label{der_omega}\\
		\frac{\partial}{\partial\psi_1}g_1(\omega_1, \psi_1)&=\ln\left(\frac{1-z_{-}}{1-\nu z_{-}}\right)<0,\label{der_psi}\\
		\frac{\partial^{2}}{\partial \omega_1^{2}}g_1(\omega_1, \psi_1)&=-\frac{1}{z^2_{-}g^{(0,0,2)}(\omega, \psi, z_- )}<0,\nonumber\\
		\frac{\partial^{2}}{\partial \psi_1^{2}}g_1(\omega_1, \psi_1)&=-\frac{(1-\nu)^2}{(1-z_{-})^2(1-\nu z_{-})^2g^{(0,0,2)}(\omega, \psi, z_-)}<0,\nonumber\\
		\frac{\partial^{2}}{\partial\psi_1 \partial \omega_1}g_1(\omega_1, \psi_1)&=-\frac{(1-\nu)}{(1-z_{-})(1-\nu z_{-})z_{-}g^{(0,0,2)}(\omega, \psi, z_-)}<0.\nonumber
	\end{align}
	To evaluate those derivatives we use the fact  that $g^{(0,0,1)}(\omega, \psi, z_-)=0$ and $g^{(0,0,2)}(\omega, \psi, z_-)>0$ for $0<z_-<z_c$ being minimum.
	Thus, $g_1(\omega_1, \psi_1)$ is a convex function of $\omega_1, \psi_1$. With 
	\begin{align*}
		\omega_0=2^{5/3}(r_{2}-r_1)\kappa_ct^{-1/3}-2^{1/3}\epsilon \kappa_{f},\\
		\psi_0=2^{5/3}(r_{1}-r_2)\kappa_ct^{-1/3}-2^{1/3}\epsilon \kappa_{f}.
	\end{align*}
	 the  following inequality follows from the convexity.
	\begin{align*}
		&g_1(\omega_1, \psi_1) \le \left. \frac{\partial g_1(\omega_1, \psi_0)}{\partial\omega_1}\right|_{\omega_1=\omega_0} (\omega_1-\omega_0)+\left.\frac{\partial g_1(\omega_0, \psi_1)}{\partial\psi_0}\right|_{\psi_1=\psi_0}(\psi_1-\psi_0)\\&+g_1(\omega_0, \psi_0) = {(\omega_0-\psi_0)}\ln z_c -\frac{2}{3}(2\epsilon)^{3/2}-\ln z_{-}2^{1/3}\kappa_{f} (\epsilon-s_{1}t^{-2/3})\\
		&+\ln\left(\frac{1- z_{-}}{1-\nu z_{-}}\right)2^{1/3}\kappa_{f}(\epsilon-s_{2}t^{-2/3})+O(\epsilon t^{-1/3}, \epsilon^{2})\\ &
		=\left(\sqrt{2\epsilon}+O(\epsilon)\right)\left(2\epsilon-(s_{1}+s_{2})t^{-2/3}\right)-\frac{2}{3}(2\epsilon)^{3/2}\\&+(\omega_1-\psi_1)\ln z_c+O( \epsilon t^{-1/3}, \epsilon^2)
	\end{align*}
	Here we use (\ref{der_omega},\ref{der_psi}) for the derivatives in the first line.  For  $z_-$  and  $g_1(\omega_0, \psi_0)$ we use the estimates  (\ref{gzc},\ref{u-}). The term $(\omega_1-\psi_1)\ln z_c$ multiplied by $t$ and exponentiated will produce the  conjugation (\ref{conjugation}). Hence
	\begin{align*}
		g_1(\omega_1, \psi_1)-(\omega_1-\psi_1)\ln z_c&\le -(s_{1}+s_{2})t^{-2/3}\sqrt{2\epsilon}\left(1+O(\sqrt{\epsilon})\right) 
	\end{align*}
	It follows then, that for any $\delta>0$ and large enough $t$
	\[
	\exp\left[t\left(g(\omega_1, \psi_1)-(\omega_1-\psi_1)\ln z_c\right)\right]\leq e^{-2\delta(s_{1}+s_{2})}<t^{-1/3}e^{-\delta(s_{1}+s_{2})}.
	\]
	The final estimate depends only on  $s_1+s_2$. Thus it remains correct even if  $s_i<\epsilon t^{2/3}$ for one of $i=1,2$. In this case we should replace $2\epsilon \to \epsilon+ s_i/t^{2/3}.$
\end{svmultproof}

Now we are  in a position to complete the proof of the second part of
Theorem \ref{thm: KPZ convergence}.
\begin{svmultproof}
	Using the above lemmas we can show that there exists $\delta>0,$ such that
	\[
	t^{1/3}e^ {-t g(\omega,\psi, z_c)}K_{t}^{alt}(n_{1},x_{1};n_{2},x_{2})\leq C(\,e^{-\delta\left(s_{1}+s_{2}\right)}+\mathbbold{1}_{r_{2}>r_{1}} e^{-\delta |s_2-s_1|})
	\]
	for constant $C $ independent of $t$.
	Following to \cite{Borodin2006} let us   consider a  kernel with conjugation
	\begin{align}\label{K_conj_Alt}
		K_t^{alt,conj} (n_{k},x_{k};n_{j},x_{j}) =e^ {-t g(\xi,\psi, z_c)}\frac{(1+s^2_k)^{k}}{(1+s^2_j)^{j}} K_t^{alt} (n_{k},x_{k};n_{j},x_{j}).
	\end{align}
	Assume that $r_k>r_j$ for $k>j$. Then,    we have the bounds 
	\begin{align}
		C&\left(\,e^{-\delta\left(s_{k}+s_{j}\right)}+e^{-\delta |s_{j}-s_{k}|}\right)\frac{(1+s^2_k)^{k}}{(1+s^2_j)^{j}}\le \frac{C_1}{1+s_j^2} ,\,\, j>k \\
		C&\left(\,e^{-\delta\left(s_{k}+s_{j}\right)}\right)\frac{(1+s^2_j)^{j}}{(1+s^2_k)^{k}}\le C_2 e^{-(s_k+s_j)} ,\,\, j\le k 
	\end{align}
	for some $C_1>0, C_2 >0$ independent of $t$.
	Thus 
	\begin{align*}
		t^{1/3}K_t^{alt,conj} (n_{k},x_{k};n_{j},x_{j})\le
		\begin{cases} 
			C_2\,e^{-\delta\left(s_{k}+s_{j}\right)}, \,\,  j\le k\\
			\frac{C_1}{1+s^2_j} ,\,\,\,\,\,\,\,\,\,\,\,\,\,\, \,\,\,\,\,\,\,\,\,j> k.
		\end{cases}
	\end{align*}
	By Hadamard inequality 
	\begin{align} 
		\det_{1\leq i,j\leq n} K^{alt}_t(n_{k},x_{k};n_{j},x_{j})\leq C_3 t^{-n/3}n^{n/2}\prod_{j=1}^{n}\max\left\{\frac{C_3}{1+s^2_j},\, C_1 e^{-\delta s_{j}}\right\},
	\end{align}
	and hence 
	\begin{align}\label{Hadamard3}
		&\sum_{x_{1}\leq a_{1}}\cdots\sum_{x_{n}\leq a_{n}}\det_{1\leq k,j\leq n}K^{step}_t(n_{k},x_{k};n_{j},x_{j})\sim\\
		&(C_4t^{1/3})^{n}\int_{s(a_1)}^{\infty}\dots \int_{s(a_n)}^{\infty} \det_{1\leq k,j\leq n}K^{step}_t(n(r_k),x(s_k,r_k);n(r_j),x(s_j,r_j))\times\\ &\times ds_1\dots d s_n\leq n^{n/2}C_5^{n}.
	\end{align}
	This ensures the absolute convergence of the Fredholm sum (\ref{eq: Fredholm series}).
	Interchanging the $t\to \infty$ limit with the   summation in $x_{1},\dots,x_{n}$  we obtain the 
	Fredholm determinant with the Airy$_{1}$ kernel in the form (\ref{eq:Airy1})
	plus  corrections vanishing in the limit $t\to\infty$.
\end{svmultproof}

\section{Asymptotic analysis: transitional regime}

\label{sec: Asymptotic analysis: transitional regime}
\subsection{Step initial configuration}

In this subsection we analyze  the asymptotics of the Fredholm determinants (\ref{eq: Fredholm series}) in the limit 
$t\to\infty$ and $\lambda\to\infty$ while  $\tau_{\beta}=\sqrt{t p(1-p)}\lambda^{\beta-1}$ fixed.
Though the general scheme of our analysis is similar to the one for KPZ case, the details turn out to be very different. 
As before, to study the convergence of the Fredholm determinant under the  simultaneous $t\to\infty$ and $\lambda\to\infty$ limit of the kernel we  prove the 
uniform convergence of the kernel on bounded sets and obtain the large deviation bounds for the whole summation region. A difficulty however appears with the diffusive part of the kernel, which   contains the Dirac delta function in the limit. Of course in this case we can not apply the Hadamard inequality straight away. 
Rather,  we  analyze the whole Fredholm sum evaluating  the sums converging to the integrals with  delta functions explicitly. 
The result is represented by a larger sum of determinants, which, however, converges, though much slower than in the KPZ case. 

First we obtain the estimate for the  kernel in $t \to \infty$ limit. We will see that  the naive steepest method is not applicable anymore. 
In one of the integrals  the entire integration contour contribute to the integral, unlike the KPZ regime.  
In the other, the saddle point coincides with the singularity of the integrand an the contour should avoid it. It can be done in such a way that
the dominant contribution still comes from a small part of the contour. Though the exponent of  integrand is not monotonous anymore on the whole contour, it can be shown 
to decrease beyond the small vicinity of the singular point.

For convenience let us represent the  kernel  as sum of products of  functions $\tilde{\Psi}_{\theta t-j}^{\theta t}(\chi t)$ and $\Phi_{\theta t-j}^{\theta t}(\chi t) $ defined in (\ref{K=phi+ K}-\ref{eq: Phi int}).
To simplify calculations we  make a variable change  $w=\frac{1-u}{1-\nu u}$ in the integrals from (\ref{eq: Psi int},\ref{eq: Phi int}) that  prevents the  poles $u=1/\nu$, $u=1$ and $u=1/\mu$ from sticking together in the limit $\lambda\to\infty$. The poles in new variables  are $w=\infty$ , $w=0$ , $w=\frac{p-1}{p}$. The integral representations  for the functions become
\begin{eqnarray}
	\tilde{\Psi}_{n-j}^{n}(x) &=&\oint_{\Gamma_{0}}e^{-tG(\chi,\theta,w)-j\ln\left(\frac{1-w}{1-\nu w}\right)}\frac{dw}{2\pi \mathrm{i} w}\label{Psi(w)}\\
	\Phi_{n-j}^{n}(x) & =&(\nu-1)\oint_{\Gamma_{1}}\frac{e^{tG(\chi,\theta,w)+j\ln\left(\frac{1-w}{1-\nu w}\right)}}{(1-w)(1-\nu w)}\frac{dw}{2\pi\mathrm{i}}\label{Phi(w)}\\
	\phi^{*(n_1,n_2)}(x_1,x_2)&=&\mathbbold{1}_{r_{2}>r_{1}}\oint_{\Gamma_{0}}e^{t(G(\chi_{2},\theta_{2},w)-G(\chi_{1},\theta_{1},w))}\frac{dw}{2\pi \mathrm{i} w}\label{phi(w)}
\end{eqnarray}
where we assigned 
\begin{eqnarray}
	n&=&t\theta , x=t\chi\\
	n_i&=&t\theta_i , x_i=t\chi_i,\quad i=1,2
	\end{eqnarray}
and
\begin{eqnarray}
G(\chi,\theta,w)=(\theta+\chi)\ln w+\theta\ln\left(\frac{1-w\nu}{1-w}\right)-\ln(1-p+ p w ).  \label{eq: f transition}\nonumber 
\end{eqnarray}
Note that  the contours $\Gamma_0$ and $\Gamma_1$ are interchanged here comparing to (\ref{eq: Psi int},\ref{eq: Phi int}).
 
In the next lemma we obtain the estimate for functions 
$\tilde{\Psi}_{n-j}^{n}(x)$ and $\Phi_{n-j}^{n}(x)$  in the limit    \begin{equation}
	\lambda\to \infty,\,t=\tau^2_{\beta}\lambda^{2-2\beta}/\left(p(1-p)\right).\label{scaling}
\end{equation}
under the scaling
\begin{eqnarray}
	n/t&=&\theta_r=r  p(1-p)\lambda^{3\beta-2} \tau_{\beta}^{-3} \label{theta_r} \\
	x/t&=&\chi_{r,s}=p-\theta_r- s p(1-p)\lambda^{\beta-1}\tau_{\beta}^{-1}, \label{chi_r,s}\\
	j&=&q  \lambda^{\beta}\tau^{-1}_{\beta}. \label{j}
\end{eqnarray}
where $r,s,q$ are supposed to be finite. 
\begin{lemma}
\label{lem: uniform estimate phi psi3}
Let $\theta_r,\chi_{r,s}$ and $j\le n$ be as in (\ref{theta_r}-\ref{j}). 
Then, given $r>0$ and  $\bar{s}>\underline{s}\in\mathbb{R}$
fixed,  estimates 
\begin{eqnarray}
\Phi_{t\theta_r-j}^{t\theta_r}(t\chi_{r,s})=  \,\, -\lambda^{-\beta}\tau_{\beta}\oint_{\Gamma_{0}}\frac{dx}{x^2}e^{  \left(\frac{r-q}{x}+s x+\frac{x^{2}}{2}\right)}+O(\lambda^{-\min(2\beta, 1)})\label{eq: Phi estimate}\\
\tilde{\Psi}_{t\theta_r-j}^{t\theta_r}(t\chi_{r,s})=  \,\,\frac{\lambda^{\beta-1}}{\tau_{\beta}}\int_{- 1-i \infty}^{-1+i \infty}dx
e^{\left(\frac{r-q}{x}+s x+\frac{x^2}{2}\right)}+O(\lambda^{2\beta-2})\label{eq: Psi estimate},
\end{eqnarray}
hold uniformly for $ \bar{s}> (s_1, s_2) >\underline{s}$ and $j>0$ for $\lambda$ large enough .
\end{lemma}

\begin{svmultproof} 
The first estimate is trivial. Let us consider  $\Phi_{\theta t-j}^{\theta t}(\chi t) $ and evaluate the integral using a small contour around $w=1$.
Making the variable change $w=1+x \frac{\lambda^{\beta-1}}{\tau_{\beta}}$, where $|x|=1$, and  the Taylor expansion of $G(\chi, \theta, w)$ near point $w=1$ we obtain
\begin{eqnarray}
G(\chi_{r,s}, \theta_r,w)=t^{-1}\left(\frac{x^2}{2}+s x+\frac{r}{x}+ O(\lambda^{\max(-\beta,\beta-1)})\right).\label{f approx}
\end{eqnarray}

Since the integration contour is compact and away from the singularities of the integrand we can exchange the limit and the integration to arrive at (\ref{eq: Phi estimate}).  

%
%

Now we consider   $\tilde{\Psi}_{\theta t-j}^{\theta t}(\chi t)$.
Let us show in the beginning that the  main contribution to the integral along  the contour $\Gamma_0=\{w: |w|=R\}$ and 
\begin{equation}R=1-\frac{\lambda^{\beta-1}}{\tau_{\beta}}\label{R},
\end{equation}  comes from a small region $\{w=Re^{i\phi}: \phi \in [-\epsilon , \epsilon]\} $.  Let us consider the behavior of   $|e^{-tG(\chi+j/t, \theta-j/t, w)}|$ by looking at the derivative of the  real part of the exponent. 
\begin{eqnarray*}
\frac{\partial}{\partial\phi}\Re G(\chi+j/t, \theta-j/t,R e^{i\phi})&=&\sin\phi\\\times\left(\frac{\theta-j/t}{(R\nu)^{-1}+R\nu-2\cos\phi}\right.&-&\left.\frac{\theta-j/t}{R^{-1}+R-2\cos\phi}+\frac{1}{\frac{1-p}{R p}+\frac{Rp}{1-p}+2\cos\phi}\right)
\end{eqnarray*}

The expression in parenthesis is positive, being a sum of a small negative and  a finite positive number. Indeed,  a sum of the first two  terms is a small negative number,  since
\begin{eqnarray*}
\frac{1}{\frac{1}{R\nu}+R\nu-2\cos\phi}-\frac{1}{\frac{1}{R}+R-2\cos\phi}\ge -\left|\frac{O(\lambda^{\beta-2})}{O(\epsilon^4)}\right|,
\end{eqnarray*}
and $(\theta-j/t)\geq 0 $ is finite. 
The third term is finite and positive, since  the  denominater is bounded by
\begin{eqnarray*}
\frac{p}{1-p}+\frac{1-p}{p}+2+O(\lambda^{\beta-1}) \ge\frac{1-p}{R p}+\frac{Rp}{1-p}+2\cos\phi>0.
\end{eqnarray*}
Then, the function $e^{-t G(\chi+j/t,\theta-j/t , R e^{i\phi})}$ is decreasing,  when $\phi$ goes away from  the $\epsilon$-vicinity  of $w=1$. 
Thus, we can restrict the integration to the small arc  within the  $\epsilon$-vicinity of $w=1$  for the price of an error 
\begin{eqnarray}\label{error}
O(2\pi e^{-t \delta_{\epsilon}}) |e^{- tG(\chi+j/t,\theta-j/t , R e^{i\phi})}|
\end{eqnarray}
 with $\delta_{\epsilon}$ being some  positive number. Note,  this fact does not depend of $s$.
\begin{figure}
\centerline{\includegraphics[clip,width=0.5\textwidth]{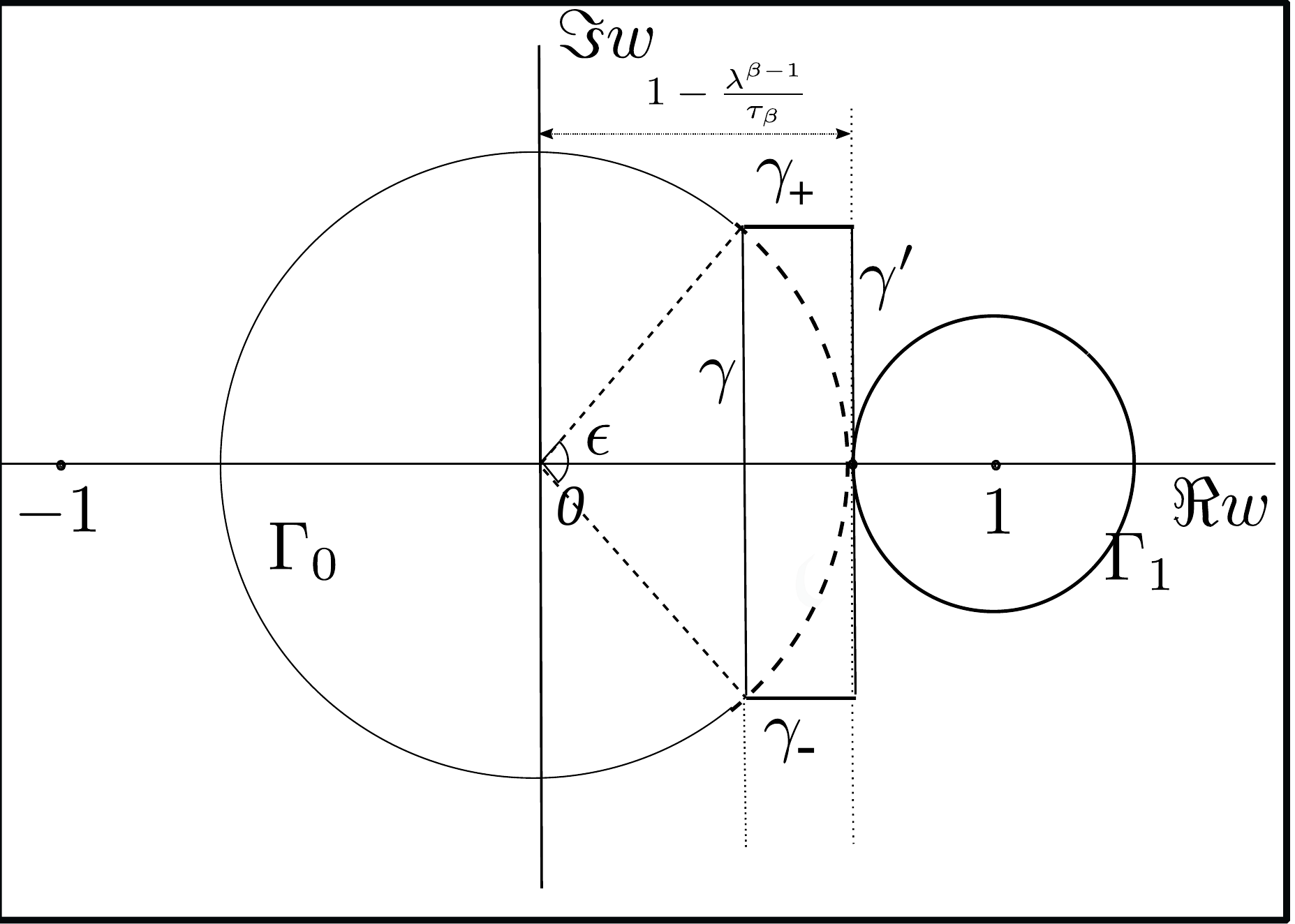}}
\caption{Integration contours $\Gamma_0$ and $\Gamma_1$  for  $\Phi_{t\theta-j}^{t\theta}(t\chi_{r,s})$ and $\tilde{\Psi}_{t\theta-j}^{t\theta}(t\chi_{r,s})$ and auxiliary contours $\gamma,\, \gamma' ,\, \gamma_{+} ,\,\gamma_{-}$  used for the estimates of  $\tilde{\Psi}_{t\theta-j}^{t\theta}(t\chi_{r,s})$ }
\label{fig: contours (double integral3)} 
\end{figure}

Let us  deform the small arc to the line segment $\gamma$ and parameterize it as $w=1+\frac{z}{\tau_{\beta}}$, where ${z} \in [Re^{i\epsilon}-1, Re^{-i\epsilon}-1]$, see fig.~\ref{fig: contours (double integral3)}. 
  The Taylor expansion of the function $G(\chi,\theta,w)$  in this parametrization gives
\begin{eqnarray}
G(\chi,\theta,w)=-\frac{p(1-p)}{\tau_{\beta}^2}\left(\frac{z^2}{2}+sz\label{exact f} \lambda^{\beta-1}+\frac{r\lambda^{3\beta-3}}{z}\right)+\\+O(z^3)+ O(z^2 \lambda^{\beta-1})+O(\lambda^{3\beta-3})+O\left(\lambda^{3\beta-4}z^{-2}\right).\notag
\end{eqnarray}
It is clear from the form of the expansion that the natural  integration variable $z$ would be $z\lambda^{1-\beta}$.  Under this scaling, however, the image of  $\gamma$ moves away to infinity, as its real 
part becomes $O(\epsilon^2\lambda^{1-\beta})$. To get rid of the infinities, we replace   
 $\gamma$ by another  vertical segment of the alike length
\begin{equation}
\gamma'=\{ z=-\lambda^{\beta-1}+iu: u\in [- \sin \epsilon, \sin \epsilon]\}
\end{equation} 
with a smaller real part and two horizontal segments
\begin{eqnarray*}
 \gamma_{\pm}= \{z=\pm iR\sin\epsilon+u: u\in [R\cos\epsilon-1, - \lambda^{\beta-1}]\} 
\end{eqnarray*}
 and use the fact that  the rectangle  with the sides $\gamma, \gamma',  \gamma_{+}, \gamma_{-} $ has no poles inside. The estimate of the integral along $ \gamma_{\pm}$ yields
\begin{eqnarray}\label{term1}
\Bigg|\int_{ \gamma_{\pm}}{\frac{dz}{\tau_{\beta}+z}}
\exp{ \left(s\lambda^{1-\beta}z+\frac{\lambda^{2-2\beta}z^2 }{2}+\frac{\lambda^{\beta-1}(r-q)}{ z}\right)}\\ \times
 e^{O(z^3\lambda^{2-2\beta})+O(z^{-2}\lambda^{\beta-2}) + O(\lambda^{\beta-1})+ O( z^2 \lambda^{1-\beta})}\Bigg| \nonumber\\
\le \frac{2 R}{\tau_{\beta}} \sin^2\frac{\epsilon}{2}O \left(e^{-\lambda^{2-2\beta} \sin^2 \epsilon /2}\right)\nonumber
\end{eqnarray}

 Now we can  make the variable change $x=z\lambda^{1-\beta}$ and integrate along $\gamma'$ to obtain the limiting result.  To estimate the error, we consider a difference between the integral with the exact  $G$ in the form (\ref{exact f}) and the integral with its approximation $G_{app}$  without corrections. 
\begin{eqnarray} \label{term2}
\left|\int_{\gamma'}dxe^{-t G(\chi ,\theta-j/t , 1+\frac{x \lambda^{\beta}-1}{\tau^{\beta}} )}-e^{-t G_{app}(\chi ,\theta-j/t , 1+\frac{x \lambda^{\beta}-1}{\tau^{\beta}} )}\right|\le\\\le\int_{\gamma'} \left| 1-e^{ O(x^3\lambda^{\beta-1})+ O(x^{-2}\lambda^{-\beta}) +O(\lambda^{\beta-1})+ O(x^2 \lambda^{\beta-1}))}\right|\times\\ \times \left|dxe^{\left(s x+\frac{x^2}{2}+\frac{r-q}{x}\right)}\right|=O(\lambda^{\beta-1})\nonumber
\end{eqnarray}
In the last step we use that $|1-e^{x}|\le |x|e^{|x|}$ and that $\int  x^n e^{-x^2}dx$ is finite.
Finally we extend $\gamma'$  to  the vertical  line $(-1-i\infty,-1+i\infty)$  obtaining an error 
\begin{eqnarray}\label{tail}
\left|\int_{-1-i\infty}^{-1-i \lambda^{1-\beta}\sin\epsilon} dx e^{-t G_{app}(\chi ,\theta-j/t , 1+\frac{x \lambda^{\beta}-1}{\tau^{\beta}} )}\right|=O\left(e^{-\lambda^{2-2\beta}\frac{\sin^2\epsilon}{2}}\right).\label{term3}
\end{eqnarray}
Collecting the error terms (\ref{term1}-\ref{term3}) we obtain  statement of  the lemma.
\end{svmultproof}
A similar statement about the uniform convergence on bounded sets of the diffusive part of the kernel also follows.
\begin{lemma}
\label{lem: uniform estimate phi }
Let  $n_i=t\theta_{r_i}$ and $x_i=t\chi_{r_i,s_i}, i=1,2$  with $\theta_{r_i}$ and $\chi_{r_i,s_i}$ related with  $r_i$ and $s_i$ by formulas (\ref{theta_r},\ref{chi_r,s})

Given $r_1, r_2 >0$ and $\bar{s}\in\mathbb{R}$,  estimate 
\begin{align*}
\phi^{*(n_{1},n_{2},)}(x_{1},x_{2})=\mathbbold{1}_{r_{2}>r_{1}}\big(\mathbbold{1}_{s_{2}=s_{1}}-\mathbbold{1}_{s_{2}>s_{1}}\lambda^{\beta-1}\sqrt{\frac{r_{2}-r_{1}}{s_{2}-s_{1}}}\times\label{phiStep}\\
\times I_{1}(2\sqrt{(r_{2}-r_{1})(s_{2}-s_{1})})+\mathbbold{1}_{s_{2}>s_{1}}O(\lambda^{-\max(1-\beta, \beta)+\beta-1})\big)\nonumber, 
\end{align*}
holds uniformly for $|s_1- s_2|<\bar{s}$  for $\lambda$ large enough, where  $I_{1}(\cdot)$ is the modified Bessel function of the first kind.
\end{lemma}
\begin{svmultproof}  The integral in (\ref{Phi}) for $\phi^{*(n_{1},n_{2},)}(x_{1},x_{2})$   is nonzero, when $$\theta_{2}-\theta_{1}+\chi_{2}-\chi_{1}=p(1-p)(s_{1}-s_{2})\frac{\lambda^{\beta-1}}{\tau_{\beta}}\le0.$$
 For $s_{2}>s_{1}$ the integrand has only 2 poles ($w=0$ and $w=1$) satisfying to
 $\mathrm{Res}_{0}(e^{G_{2}-G_{1}})=-\mathrm{Res}_{1}(e^{G_{2}-G_{1}})$, and  the change of the contour $\Gamma_0$ to $\Gamma_1$ does not affect the result. When  $s_1=s_2$, the integrand  acquires another simple pole at $w=\infty$. In this case the integral can be calculated explicitly.
\begin{eqnarray*}
\phi^{*(n_{1},n_{2},)}(x,x)=\frac{\mathbbold{1}_{n_{2}>n_{1}}}{2\pi \mathrm{i}}\oint_{\Gamma_{0}}\frac{e^{t(\theta_2-\theta_1)\ln\left(\frac{1-w\nu}{1-w}\right)}}{w}dw=\mathbbold{1}_{n_{2}>n_{1}}
\end{eqnarray*}
The case $s_2 > s_1$ is to be treated  similarly to that  for the function $\Phi$ in the lemma \ref{lem: uniform estimate phi psi3}. Using the Taylor approximation for   ${t(G(\chi_{2},\theta_{2},w)-G(\chi_{1},\theta_{1},w))}$ on the  contour  $\Gamma_1=1+x\frac{\lambda^{\beta-1}}{\tau_{\beta}}$, where $|x|=1$, 
\begin{eqnarray*}
{t(G(\chi_{2},\theta_{2},w)-G(\chi_{1},\theta_{1},w))}=- \left(s_{21}x+\frac{r_{21}}{x}\right)+O(\lambda^{-\max{(\beta,1-\beta)}}).
\end{eqnarray*}
we obtain 
\begin{eqnarray*}
\phi^{*(n_{1},n_{2},)}(x_1,x_2)=-\frac{\mathbbold{1}_{r_{2}>r_{1}}}{2\pi \mathrm{i}\lambda^{1-\beta}}(\oint_{\Gamma_{0}}e^{-(\frac{r_{2}-r_{1}}{x}-(s_{1}-s_{2})x)}dx+ O(\lambda^{-\max{(\beta,1-\beta)}}) )
\end{eqnarray*}
With the variable change $x=\sqrt{\frac{r_{2}-r_{1}}{s_{2}-s_{1}}}e^{i\phi}$ the integral in the last formula is nothing but the  integral representation of the modified Bessel function of the first kind.
\end{svmultproof} 
In the previous lemmas we obtained uniform estimates for bounded sets of $s_1, s_2$ for $\Psi$, $\Phi$ and $\phi$  . We will also  need uniform bounds for all values of $s_1 , s_2$.
\begin{lemma}
\label{lem: uniform estimate phi psi 2}Given $r>0$ and  $\underline{s}\in\mathbb{R}$
fixed, there exist $C_1, C_2,C_3>0$, such that  for any  $a,b>0$ 
\begin{eqnarray}
|\tilde{\Psi}_{t\theta-j}^{t\theta}(t\chi_{r,s})|& \le&  C_1 \lambda^{-\beta}   e^{-a s}\label{bound1}\\
|\Phi_{t\theta-j}^{t\theta}(t\chi_{r,s})|&\le& C_2 \lambda^{\beta-1}      e^{b s}\label{bound2}\\
|\phi^{*(t\theta_1,t\theta_2)}(x_{1},x_{2})|&\le& C_3(\lambda^{\beta-1}{\mathbbold{1}_{s_{2}> s_{1}}+    \mathbbold{1}_{s_{2}= s_{1}}       )\mathbbold{1}_{r_{2}>r_{1}}}e^{bs_{21}}\label{bound3}
\end{eqnarray}
hold uniformly for $s_1, s_2 >\underline{s}$ and $\lambda$ large enough .
\end{lemma}
\begin{svmultproof} 
In lemmas \ref{lem: uniform estimate phi psi3},\ref{lem: uniform estimate phi } we estimated the integrals (\ref{Psi(w)}-\ref{phi(w)}) under assumption that 
the value of the argument $\chi$ of $G(\chi,\theta,w)$ in the exponent of the integrands is $\chi=p-\theta-sp(1-p)\tau_{\beta}^{-1}\lambda^{\beta-1}$ with finite $s$. Here we would like to extend the estimates to arbitrary values of $s$. Note, however, that varying $s$ simply shifts the real part of  $G(\chi,\theta,w)$ as follows
\begin{eqnarray}
\Re t G(\chi,\theta,w)-\Re t G(p-\theta,\theta ,w)&=& tp(1-p)s\lambda^{\beta-1}\tau_\beta^{-1} \label{eq: G shift}\\
&=&s \tau_\beta\lambda^{1-\beta}\Re\ln w.\notag
\end{eqnarray}
We can estimate the effect of the shift by its maximal value on the integration contour, while for the remaining integrals with $\chi= p-\theta$ the estimates from  the lemmas \ref{lem: uniform estimate phi psi3},\ref{lem: uniform estimate phi } are applicable.
To this end, we choose the  contours $\Gamma_0$ and $\Gamma_1$ of lemmas \ref{lem: uniform estimate phi psi3},\ref{lem: uniform estimate phi }
being the following circles
\begin{eqnarray}
\Gamma_0=\{w: |w|=1- a \lambda^{\beta-1}\tau_{\beta}^{-1} \},\label{eq: Gamma_0}\\
\Gamma_1=\{w: |w-1|=b \lambda^{\beta-1}\tau_{\beta}^{-1} \}.\label{eq: Gamma_1}
\end{eqnarray}
Then, the integrands will be corrected  by the following shift-dependent factors
\begin{eqnarray}
\left|e^{s\lambda^{1-\beta}\tau_\beta\Re\ln w}\right|&\leq &c_1 e^{-as},\quad\quad w\in \Gamma_0,\quad s>0 \label{eq: diff1}\\
\left|e^{-s\lambda^{1-\beta}\tau_\beta\Re\ln w}\right|&\leq &c_2 e^{bs},\,\,\, \quad\quad w\in \Gamma_1, \quad s>0 \label{eq: diff2}\\
\left|e^{-(s_2-s_1)\lambda^{1-\beta}\tau_\beta\Re\ln w}\right|&\leq &c_2 e^{b(s_2-s_1)}, w\in \Gamma_1, \quad s_2>s_1, \label{eq: diff3}
\end{eqnarray}
with some $c_1,c_2>0$. 
Multiplying this by the estimates from  lemmas \ref{lem: uniform estimate phi psi3},\ref{lem: uniform estimate phi } for $\chi=p-\theta$, we arrive at (\ref{bound1}-\ref{bound3}).
\end{svmultproof}

We see that the estimate for $\Phi$ is exponentially increasing as $s_2$ grows. To make it summable we use  a conjugation of the kernel $K$, which does not affect the value of the determinants.
\begin{equation}
K^{conj}(n_1, x_1;n_2,x_2)=K(n_1, x_1;n_2,x_2)e^{-(a+b)s_{21}/2}\label{eq: conj}
\end{equation}
where we chose $a>b$. 
 Now we can make the necessary  estimate.
\begin{corollary}\label{cor: large deviation transition step}
Let the functions  $\tilde{K}^{step}(n_1,x_1;n_2,x_2)$	and $\phi^{*(n_{1},n_{2})}(x_{1},x_{2})$ be as in (\ref{Phi},\ref{eq:K_step}) and the functions $\left(\tilde{K}^{step}(n_1, x_1,n_2,x_2)\right)^{conj}$ and $\left(\phi^{*(n_{1},n_{2})}(x_{1},x_{2})\right)^{conj}$  be obtained from them under the conjugation (\ref{eq: conj}). Then, for any $a>b>0$ there exist constants $\tilde{C},C^*>0$ such that under the scaling  (\ref{scaling}-\ref{j}) the following inequalities hold, when $\lambda$ is large enough.
\begin{eqnarray}
&&\!\!\!\!\!\!\!\!\!\!\!\!\!\!\!\ \left|\left(\tilde{K}^{step}(n_1, x_1,n_2,x_2)\right)^{conj}\right|=\left|\sum_{j=0}^{n_2} \tilde{\Psi}_{t \theta_{r_1} -j}^{t \theta_{r_1} }(t\chi_{r_1,s_1} )\Phi_{t \theta_{r_2} -j}^{t\theta_{r_2} }(t \chi_{r_2,s_2} )\right|^{conj}\label{k_estimate} \\ \nonumber
&  \le&\sum_{j=0}^{t \theta_{r_2} }  C_1 \lambda^{-\beta} e^{-a s_1} C_2 \lambda^{\beta-1} e^{b s_2}e^{-(a+b)s_{21}/2} 
\le \tilde{C} \lambda^{\beta-1} e^{-(s_2 +s_1)(a-b)/2} \\
&&\!\!\!\!\!\!\!\!\!\!\!\!\!\!\!\left|\left(\phi^{*(n_{1},n_{2})}(x_{1},x_{2})\right)^{conj}\right|\le  C^*(\lambda^{\beta-1}{\mathbbold{1}_{s_{2}> s_{1}}+    \mathbbold{1}_{s_{2}= s_{1}}       )\mathbbold{1}_{r_{2}>r_{1}}}e^{-(a-b) s_{21}/2}\label{phi_estimate}
\end{eqnarray}
\end{corollary}
We see that the estimate for $\left(\tilde{K}^{step}(n_1, x_1,n_2,x_2)\right)^{conj}$  is  summable  in both $s_1$ and $s_2$ due to our choice of constants $a>b$. However,  one of the terms  in the estimate for $\left(\phi^{*(n_{1},n_{2})}(x_{1},x_{2})\right)^{conj}$ is not small, when $s_{21}$ is not large, even though both  $s_1$ and $s_2$ may grow  indefinitely. Furthermore, the other term  multiplied by the factor $\lambda^{1-\beta}$ diverges in the limit under consideration, effectively contributing the delta-function into the limiting expression. This makes problematic a straightforward  use of  Hadamard inequality.  We deal with this problem in the following subsection.

\subsection{Fredholm sum} 

Here, we are going to estimate the sum 
 \begin{equation}
 \sum_{x_1=a_1}^{\infty}\dots \sum_{x_n=a_n}^{\infty}\det\left[K(n_{{k}},x_{{k}};n_{{j}},x_{{j}})\right]_{1\leq k,j\leq n}\label{sum}
\end{equation}
to make sure that its growth with $n$ is slow enough to provide the convergence of the Fredholm sum.

Let us consider the determinant $\det\left[K(n_{{k}},x_{{k}};n_{{j}},x_{{j}})\right]_{1\leq k,j\leq n}$ from the sum (\ref{eq: Fredholm series}). We   fix the set  of $r_i$ such that $n_i>n_{j}$ if $i>j$. Let us denote $K(n_{{k}},x_{{k}};n_{{j}},x_{{j}})$,  $\tilde {K}(n_{{k}},x_{{k}};n_{{j}},x_{{j}})$  and  $\phi^{*(n_{1},n_{2},)}(x_{1},x_{2})$ as  $K_{kj}$,  $\tilde{K}_{kj}$ and $\phi_{kj}$  respectively and corresponding matrices as $K,\tilde{K}$, and $\phi$ for short. Then

\begin{eqnarray}
\det K = \begin{vmatrix}\label{matrix}
\tilde{K}_{11} &  \tilde{K}_{12}-\phi_{12} & \tilde{K}_{13}-\phi_{13}&\dots & \tilde{K}_{1n}-\phi_{1n} \\
\tilde{K}_{21}&  \tilde{K}_{22} & \tilde{K}_{23}-\phi_{23}& \dots & \tilde{K}_{2n}-\phi_{2n} \\
\dots & \dots & \dots & \dots & \dots \\
\tilde{K}_{n1} & \tilde{K}_{n2}&\tilde{K}_{n3}&\dots & \tilde{K}_{nn}
\end{vmatrix}.
\end{eqnarray}
 Note that  all  entries with $\phi_{i_k,j_k}$ are above the diagonal because of our choice of the set $\{n_i\}$. 

We can rewrite the determinant in the form of minor decomposition 
\begin{eqnarray}
\det K=\sum_{k=0}^{n-1} \sum_{I_k<J_k} \det \tilde{K}_{\bar{I}_k,\bar{J}_k}\prod_{l=1}^{k}\phi_{i_l,j_l},  \label{K_sum}
\end{eqnarray}
where the internal summation  is over all pairs of $k$-component  ordered sets 
\begin{eqnarray}
I_k&=&\{i_1<\dots<i_k\}\subset\{1,\dots,n\},\label{eq: I_k}\\
J_k&=&\{j_1,\dots,  j_k: j_i\neq j_l \quad \forall \quad 1\leq i\neq l\leq k \}\subset\{1,\dots,n\},\label{eq: J_k}
\end{eqnarray}
 such that $i_l<j_l$ for  $l=1,\dots,k$, and $\tilde{K}_{\bar{I}_k,\bar{J}_k}$
 is a sub-matrix of the  matrix $\tilde{K}$ with rows from $I_k$ and columns from $J_k$ removed.
We also use  notations
\begin{eqnarray*}
	\bar{I}_k&=&\{\bar{i}_1,\dots,\bar{i}_{n-k}\}=\{1,\dots,n\}\backslash I_k\\
	\bar{J}_k&=&\{\bar{j}_1,\dots,\bar{j}_{n-k}\}=\{1,\dots,n\}\backslash J_k
\end{eqnarray*}
for the complementary sets and their elements. 

Let us interchange the order of summations in (\ref{sum}) and the summations in (\ref{K_sum}). We will   show that   the sum  over $x_1,\dots,x_n$ of all the  summands  of the latter sum corresponding to a fixed $k$  can be given the same upper  bound.
\begin{lemma}\label{estimation_lemma}
With $0\leq k\leq n$ and the sets $\bar{I}_k,	\bar{J}_k$ fixed the following 	inequality holds
\begin{eqnarray} 
\sum_{x_1=a_1}^{\infty}\dots \sum_{x_n=a_n}^{\infty}\det  \tilde{K}_{\bar{I}_k,\bar{J}_k}\prod_{l=1}^{k}\phi_{i_l,j_l}  \le 3^n C^{n-k} (n-k)^{(n-k)/2}, \label{M_estimate1}
\end{eqnarray}
where $C>0$ is some  $n$-independent constant .
\end{lemma}

\begin{svmultproof} 
The determinant $\det \tilde{K}_{\bar{I}_k,\bar{J}_k}$ of the $(n-k)\times(n-k)$ matrix can be estimated using the Hadamard inequality and the bound (\ref{k_estimate})
\begin{equation}
\left|\det \tilde{K}_{\bar{I}_k,\bar{J}_k}\right|\leq \left(\lambda^{\beta-1}\tilde{C}\right)^{n-k}(n-k)^{(n-k)/2}e^{{-(a-b)(s_{\bar{i}_1}+\cdots+s_{\bar{i}_{n-k}})}},\label{eq: detK estimate}
\end{equation}
where $s_i$ is related to $x_i$ by (\ref{chi_r,s}), $\tilde{C}$ is a positive constant from the bound (\ref{k_estimate}), and without loss of generality we suppose that $(a-b)=1$. Using the bound (\ref{phi_estimate}) for the whole sum we have 
\begin{eqnarray}\label{M_estimate2}
\frac{\left(\mathrm{l.h.s.\,\, of\,\, (\ref{M_estimate1})}\right)}{\tilde{C}^{n-k}(n-k)^{\frac{n-k}{2}}} \label{eq: minor sums} &\leq& 
\lambda^{(n-k)(\beta-1)}(C^*)^k \\ &\times& \sum_{x_1=a_1}^{\infty}\dots \sum_{x_n=a_{n}}^{\infty} \prod_{l=1}^{k}(\lambda^{\beta-1}\mathbbold{1}_{s_{j_l}\ge s_{i_l}}+\mathbbm{1}_{s_{j_l}=s_{i_l}} )\prod_{l=1}^{n-k}e^{-\frac{s_{\bar{i}_l}}{2}} ,   \nonumber
\end{eqnarray}
where $C^*>0$ is the  constant from the bound (\ref{phi_estimate}).  The sum in the r.h.s. can be estimated by the integral  
\begin{eqnarray}
\int_{d_1}^{\infty}\dots \int_{d_{n}}^{\infty}\prod_{l=1}^{k}(\mathbbold{1}_{s_{j_l}\ge s_{i_l}}+\delta(s_{j_l}-s_{i_l})) \prod_{l=1}^{n-k}e^{-\frac{s_{\bar{i}_l}}{2}}\label{k=1}d s_n \dots ds_1   \nonumber
\end{eqnarray}
 where  going from  the summation in every  $x_i$ to an integration in  $s_i$ adds the factor of $\lambda^{1-\beta}$
  and changes the range of summation from $x_i\geq a_i$ to corresponding  $s_i\geq d_i$. All $d_i$  will be supposed to be positive for simplicity. The presence of the term  $\mathbbold{1}_{s_{j_l}= s_{i_l}}$, which is not accompanied by the vanishing factor of $\lambda^{\beta-1} $ 
  results in appearance of the delta function under the integral.

Let us consider the effect of factors $\mathbbold{1}_{s_{j_l}\ge s_{i_l}} $ on the integration. All the integrals in $s_{\bar{i}_l}$,  such that $\bar{i}_l\in \bar{I}_k \backslash J_k$  can be integrated explicitly, to obtain factor $2 e^{-d_{i_l}/2}\le2$.
 Every element  $s_{p_1}, p_1 \in \bar{I}_k\cap J_k$  is present in the integration in a combination $\mathbbold{1}_{s_{p_1}\ge s_{p_2}}e^{-s_{p1}/2}$, where $p_2 \in I_k$. There are two possibilities for $p_2$. It either belongs or not to the set $J_k$. In the first case we have the  integral with  only one  indicator function. In the second case there is another indicator function  $\mathbbold{1}_{s_{p_2}\ge s_{p_3}}$ where $p_3 \in I_k$. We can consider $p_3$ in the same way as $p_2$ and so on.  Thus for every $p_1$ we have the following chain of coupled integrals 
\begin{eqnarray*}
\int_{d_{p_m}}^{\infty}\dots\int_{d_{p_1}}^{\infty}\prod_{i=1}^{m-1}(\mathbbold{1}_{s_{p_i}\ge s_{p_{i+1}}}+\delta(s_{p_i}-s_{p_{i+1}})) e^{-s_{p_1}/2}d s_{p_1} \dots ds_{p_m}\\ \le 
\int_{d_{p_m}}^{\infty}\int_{s_{p_{m}}}^{\infty}\dots\int_{s_{p_2}}^{\infty}\prod_{i=1}^{m-1}(1+\delta(s_{p_i}-s_{p_{i+1}})) e^{-s_{p_1}/2}d s_{p_1} \dots ds_{p_m}\\=2\times 3^me^{-d_{p_{m}}/2}\le2\times 3^m,
\end{eqnarray*} 
where $\{p_i\}_{i \in \{2, \dots ,m \}} \subset I_k$. We used non-negativity  of the integrand to extend the lower integration limits.

 Collecting all the above estimates we conclude that the r.h.s of  (\ref{M_estimate2})  is bounded by $3^n$, which concludes the proof.
\end{svmultproof}

Now, we come back  to the summations of (\ref{K_sum}). Since the bound of the sum in r.h.s. of  (\ref{M_estimate1})  depends only on $k$ rather than on particular choice  of the pair of sets $I_k$ and $J_k$,  it is enough  to  enumerate the ways to choose this pair in order to estimate the sum over all possible choices. 
This number is given by  
\begin{eqnarray}
c_k&=&\#\{I_k<J_k\}\label{c_k}\\
&=&\sum_{j_k=k+1}^{n}\dots\sum_{j_1 = 2}^{j_2-1} (j_1-1)\times\dots\times(j_k- k)=S(n+k,n).\nonumber
\end{eqnarray}
where $S(n,k)$ are Stirling numbers of the second kind  $$S(n, k) = \frac{1}{k!}\sum\limits_{j=0}^k(-1)^{k+j}\binom{k}{j}j^n.$$

To estimate the large $n$ and $k$  asymptotics of $c_k$,  we change the summation variables in (\ref{c_k})  to  $b_l=j_l-l$ and estimate the sum by the integral 
\begin{eqnarray}
c_k&=&\sum_{b_k = 1}^{n-k}\dots \sum_{b_1 = 1}^{b_2}\prod_{l=1}^{k}b_l\notag \\
&\le&\int_{ 1}^{n-k}\dots \int_{1}^{b_2}  \prod_{l=1}^{k}b_l db_l=\frac{1}{k!}\int_{ 1}^{n-k}\dots \int_{1}^{n-k}  \prod_{l=1}^{k}b_l d \label{c_k estimate} b_l\\&=&\frac{1}{k!}\left(\int_{1}^{n-k} b d b\right)^k=\frac{(n-k)^{2k}}{2^kk!},\notag
\end{eqnarray}
where the inequality follows from the positivity and monotonicity of the function $f(x)=\int_1^xbdb$.

Thus, we are in position to estimate the finial sum in $k$. Using the estimates  (\ref{M_estimate1}) and (\ref{c_k estimate}) we find
\begin{eqnarray}
\sum_{s_1=d_1}^{\infty}\dots \sum_{s_n=d_n}^{\infty}\det K\asymp \sum_{k=0}^{n-1}\frac{(n-k)^{2k}}{ k!}(n-k)^{(n-k)/2} 
\end{eqnarray}
were by  $f(n)\asymp g(n)$ we mean that  $$\frac{\ln f(n)}{n}=O\left( \frac{\ln g(n)}{n}\right)$$, 
i.e. we compare only  the factors that grow faster than exponentially.
The sum in the r.h.s. can be approximated by $n$ times  the maximal term. To this end, using  the Stirling's approximation we find that  
\begin{eqnarray}
\frac{(n-k)^{2k}}{2^k k!}(n-k)^{(n-k)/2} \asymp e^{n g(x)} \label{intK},\end{eqnarray}
where $x=\frac{n-k}{n}$ and 
$$g(x)=-(1-x)\ln\left(n(1-x)\right)+\left(2-\frac{3}{2}x\right)\ln(nx).$$
The function $g(x)$ has a maximal value found from  $g'(x_0)=0$  at  $x_0 \simeq \frac{4}{\ln{n}}$, at which we have
$$e^{ng(x_0)}\asymp  n^n \left(\ln{n}\right)^{-n/2}.$$
To summarize we obtain
\begin{lemma}
	Given   $\{r_1<\cdots <r_n\}$ and $\{d_1,\dots,d_n\}$, let $n_i=[t\theta_{r_i}]$, $a_i=[\chi_{r_1,d_i}]$. Then, 
\begin{eqnarray}
\sum_{x_1=a_1}^{\infty}\dots \sum_{x_n=a_n}^{\infty}\det\left[K(n_{{k}},x_{{k}};n_{{j}},x_{{j}})\right]_{1\leq k,j\leq n}  \asymp n^n \left(\ln{n}\right)^{-n/2}
\end{eqnarray}
as $n\to \infty$ with $\lambda$ large enough.
\end{lemma}
The presence of $(\ln n)^{-n/2}$ ensures the absolute and uniform in $\lambda$ convergence of the Fredholm sum. Thus,  we are in a position to prove the first statement of  Theorem \ref{thm: Transitional regim}.
\begin{svmultproof}(Theorem \ref{thm: Transitional regim}, step IC)

\noindent Because of the uniform absolute convergence of the Fredholm sum we can interchange the order of taking the $\lambda\to \infty$ limit and the outer summation to analyse the convergence term by term.	The convergence of  remaining sums in $x_1,\dots, x_n$ in the $n$-th term to integrals follows from the uniform convergence of the  summands multiplied by the factor  $\lambda^{1-\beta}$  on  bounded sets, lemmas \ref{lem: uniform estimate phi psi3},\ref{lem: uniform estimate phi },  and the large deviation bounds, corollary \ref{cor: large deviation transition step}, except for the term of the diffusive part of the kernel with $\mathbbm{1}_{s_1=s_2}$, which would diverge if multiplied by the diverging factor $\lambda^{1-\beta}$. However its effect on the whole sum is finite. In fact, taken into account explicitly it simply reduces the multiplicity of some of multiple sums, so that the remaining sums with lower multiplicity contain only convergent  terms (see similar analysis in lemma \ref{estimation_lemma}). If we go back to the  $n$-fold  integrals,  the effect of the  term $\mathbbm{1}_{s_1=s_2}$  will be simply the effect of the  delta function under the integrals, for which we should accept convention that it contributes a unit mass being on the end of the integration domain.
\end{svmultproof}

\subsection{Alternating initial configuration}

The analysis for alternating IC is similar to that of step IC except for some  technical details. Below we briefly sketch the statements about the  uniform convergence of the kernel on bounded stets and the large deviation bounds specific for the the case of alternating IC. 
For the rest of the proof of the second statement of Theorem  \ref{thm: Transitional regim}  we refer the reader to the previous subsection. 

We start with  the  expression  (\ref{eq: K_t},\ref{Phi},\ref{K_alt}) of the kernel.
As before, we use the  notations (\ref{chi_r,s}, \ref{theta_r})  and \ref{f})  as before for $\chi_i , \theta_i , f(\chi , \theta, v) $.
For alternating IC $\beta=1/2$ thus  $\lambda\tau_{1/2}^2=p(1-p)t$.

\subsubsection*{Uniform convergence on bounded sets}
\begin{lemma}
\label{AltKestimate}

\noindent Given $r_1, r_2>0$ and  $\bar{s}$ , $\underline{s}\in\mathbb{R}$
fixed, let us  take define
\begin{eqnarray}
\theta=rp(1-p)\lambda^{-1/2}\tau_{1/2}^{-1} ,\, \chi=p-(2r+s)p(1-p)\lambda^{-1/2}\tau_{1/2}^{-1}\label{scaling Alt}
\end{eqnarray}
Then, estimates
\begin{align*}
&\widetilde{K}_{t}^{alt}(n_{1},x_{1};n_{2},x_{2}) \\
&=\lambda^{-1/2}\int_{-i \infty -1}^{+i\infty-1}\frac{dx}{2\pi \mathrm{i}} e^{\left(\tau^2_{1/2}\frac{x^2-x^{-2}}{2}+\tau_{1/2}\left(s_{1}x-\frac{s_{2}}{x}+r_{12}(x+x^{-1})\right)\right)}+O(\lambda^{-1})
\end{align*}
holds uniformly for $ \bar{s}>(s_1,\ s_2) >\underline{s}$  for $\lambda$ large enough.
\end{lemma}

\begin{svmultproof} 
 The integral in (\ref{K_alt})  is	
\begin{align}
\widetilde{K}_{t}^{alt}(n_{1},x_{1};n_{2},x_{2}) & =\oint_{\Gamma_{0}}\frac{dv}{2\pi\mathrm{i}v} e^{tF(v)}
\end{align}
where we  	introduce the function
\begin{eqnarray*}
		F(v)=f(\chi_{2},\theta_{2}, v )-f\left(\chi_1,\theta_1,\frac{1-v}{1-\nu v}\right)=\left( \theta_1+\theta_2+\chi_{2}\right)\ln\frac{1-v}{1-\nu v}-\\-( \theta_1+\theta_2+\chi_1)\ln v +\ln \left (\frac{(1-\nu v)(1-p + p v)}{(1-\mu v)}\right).
\end{eqnarray*}
It is nonzero, when  
\begin{eqnarray*}
\chi_{1}+\theta_{1}+\theta_{2}=p-(s_1+r_1-r_2)\frac{p(1-p)}{\lambda^{1/2}\tau_{1/2}}\ge0
\end{eqnarray*}
Let us show that the main contribution to the integral along the contour $\Gamma_0=\{v:|v|=R\}, $ where  
\begin{eqnarray*}
R=1-\frac{1}{\lambda^{1/2}}
\end{eqnarray*}
comes from a small region $v=Re^{i\phi},$ $\phi \in [-\epsilon , \epsilon ]$.
Let us consider the derivative of the real part of the exponent   $\frac{\partial}{\partial\phi}\Re F(Re^{i\phi})$. We  notice that 
\begin{eqnarray*}
F(v)=-G(\chi_1-\chi_2, \theta_1+\theta_2+\chi_2,   v)+\ln(1-\nu v)-\ln(1-\mu v).
\end{eqnarray*}
 The derivative $\partial  \Re G(\chi, \theta, Re^{i\phi})/ \partial\phi$ was investigated in lemma \ref{lem: uniform estimate phi psi3}. Since $ \theta_1+\theta_2+\chi_2\le C_r$, where $C_r$ is some  finite positive number independent of $s_2, \lambda$ then $$e^{-t G( \chi_1-\chi_2, \theta_1+\theta_2+\chi_2, v)}$$ is decreasing, when $\phi$ goes away from the $\epsilon$-vicinity of $v = 1$. We also should investigate the behavior of
\begin{eqnarray*}
\frac{\partial}{\partial\phi}\Re \ln\frac{(1-\nu v)}{(1-\mu v)}=\sin \phi \frac{\frac{1}{\mu R}-\frac{1}{\nu R}+(\mu-\nu) R }{\left(\frac{1}{\mu R}+\mu R -2 \cos\phi\right)\left(\frac{1}{\nu R}+\nu R -2 \cos\phi\right)}
\end{eqnarray*}
The denominator of this fraction is positive. Making a simple calculation one can show that
\begin{eqnarray*}
 \frac{1}{\mu R}-\frac{1}{\nu R}+(\mu-\nu) R <0.
\end{eqnarray*}
Then $\Re F(v)$ decreases outside the region $\phi \in [- \epsilon, \epsilon]$. We can consider only $\epsilon$-vicinity of $v=1$ for the price of an error $$O(2\pi e^{-t \delta_{\epsilon}})|e^{t F(Re^{i\epsilon})}|$$ with $\delta_{\epsilon}$ being some positive number.

The further analysis coincides with the one   function $\tilde{\Psi}$ in lemma \ref{lem: uniform estimate phi psi3}.  First we change integration contour  from the arc to the line segment and obtained error $O (e^{-t \sin^2 \epsilon/2 })$. Then we change the line segment to the whole line and  estimate function $F(v)$ by Taylor approximation $F(v)_{app}$ 
\begin{align*}
F\left(1+\frac{x}{ \lambda^{1/2}}\right)= \frac{p(1-p)}{\lambda}\left(\frac{x^2-x^{-2}}{2}+(s_{1}x-\frac{s_{2}}{x}+r_{12}(x+x^{-1}))\tau_{1/2}^{-1}\right)+\\+
\lambda^{-3/2}\left(O(x^3)+O(x^2 s_1)+O(x^{-2}s_2)+O(x^{-3})\right).
\end{align*}
Doing  that we obtain $O(\lambda^{-1})$ error term. Collecting all the error terms we derive the statement of the lemma.
\end{svmultproof}

\begin{lemma} \label{lem: uniform estimate phi2 }
Given $r_1, r_2 >0$ and  $\bar{s} \in\mathbb{R}$, the estimate 
\begin{align*}
\phi^{*(n_{1},n_{2},)}(x_{1},x_{2})=-\mathbbold{1}_{r_{2}>r_{1}}\mathbbold{1}_{s_{21}+r_{21}>0}\lambda^{-1/2}\sqrt{\frac{r_{12}}{r_{12}+s_{12}}}\times\\\nonumber
\times\big( I_{1}(2\tau_{1/2}\sqrt{(s_{12}+r_{12})r_{12}})+O(\lambda^{-1/2})\big)+\mathbbold{1}_{r_{2}>r_{1}}\mathbbold{1}_{s_{21}+r_{21}}\nonumber 
\end{align*}
holds uniformly for $|s_1- s_2|<\bar{s}$  for $\lambda$ large enough, where  $I_{1}(\cdot)$ is the modified Bessel function of the first kind.
\end{lemma}

\begin{svmultproof} 
 The statement   for $\phi^{*(n_{1},n_{2},)}(x_{1},x_{2})$  has already been  given in lemma \ref{lem: uniform estimate phi } for step IC. Here we can  use it with $\beta=1/2$,  changing     $s_{21}$ to $ s_{21}+r_{21}$. 
\end{svmultproof} 

\subsubsection*{Large deviation bounds}
\begin{lemma}
\label{lem: uniform estimate phi psi 3}Given $r>0$ and  $\underline{s}\in\mathbb{R}$
fixed, there exist $C_4, C_5>0 $, such that for any $a>0$  inequalities
\begin{align*}
|\widetilde{K}_{t}^{alt}(n_{1},x_{1};n_{2},x_{2})|&\le C_4   \lambda^{-1/2} e^{\tau_{1/2}(\frac{s_2}{a}-a s_1+ (a+1/a) r_{21})}\\
|\phi^{*(n_{1},n_{2},)}(x_{1},x_{2})|&\le C_5(\lambda^{-1/2}\mathbbold{1}_{s_{21}+r_{21}>0}+\mathbbold{1}_{s_{21}+r_{21}=0})\mathbbold{1}_{r_{2}>r_{1}}\times \\ \times e^{\tau_{1/2}(s_{21}+r_{21})/a}
\end{align*}
hold uniformly for $s_1, s_2 >\underline{s}$ and $\lambda$ large enough .
\end{lemma}

We again see that the estimate for $\widetilde{K}_{t}^{alt}$ is exponentially increasing as $s_2$ grows. The appropriate conjugation for the kernel will be  $$K^{conj}=K e^{-\frac{a^2+1}{2a}\tau_{1/2} s_{21}}$$. 
We can rewrite the estimate for the kernel using above conjugation as
\begin{align*}
|K_{t}^{conj}(n_{1},x_{1};n_{2},x_{2})|\le \lambda^{-1/2}C_4 e^{-\tau_{1/2} \frac{a^2-1}{2a}(s_1+s_2)}+\\+ (\lambda^{-1/2}\mathbbold{1}_{s_{21}+r_{21}>0}+\mathbbold{1}_{s_{21}+r_{21}=0})\mathbbold{1}_{r_{2}>r_{1}}e^{-\tau_{1/2} \frac{a^2-1}{2a}s_{21}}
\end{align*}
 We can choose $a$ in such a way that $\tau_{1/2} \frac{a^2-1}{2a}=1$  the fact of  convergence does no depend on the  value of $a$ as far as  $a>1$. From this we can obtain the  absolute convergence of the Fredholm sum, exactly the same way as for step IC case.

\subsection{Asymptotic tails of the transitional  kernels}

We have just obtained the results about  the crossover   between  the KPZ and DA regimes. The transitional distributions obtained are expected to have these two regimes as limiting cases. The crossover parameter is  $\tau_\beta$. The  limits $\tau_\beta\to \infty$ and $\tau_\beta\to 0$ are associated with the two regimes respectively. Below, without proofs, we  give an idea on  how   the known KPZ kernels and the kernels corresponding to the irreversible particle aggregation are obtained in these limits.

To state the result we introduce the notations $K'^{(\tau)}_{\mathcal{A}_{1}\to\mathcal{N}}(r_{1},s_{1};r_{2},s_{2})$ and $K'_{\mathcal{A}_{2}\to\mathcal{N}}(r_{1},s_{1}; r_{2},{s_{2}})$
for $K^{(\tau)}_{\mathcal{A}_{1}\to\mathcal{N}}(r_{1},s_{1};r_{2},s_{2})$ and $K_{\mathcal{A}_{2}\to\mathcal{N}}(r_{1},s_{1}; r_{2},{s_{2}})$ without the delta functions. 
Their limits can be proved using simple asymptotic analysis. We also give a simple explanation of what happens with the delta functions.

Before going to the statements yet another  remark has to be done. As we saw, for step IC  we obtain a single random process in the transitional regime, while the crossover parameter $\tau_\beta$ plays the role of the natural ``time''-unit, where by  ``time'' we mean the variable parametrizing  the process.  Therefore in the dimensionless framework the limits to be considered are 
actually  the   large and small  ``time''  limits.  On the contrary, in the case of the alternating IC the parameter $\tau$ can not be removed by rescaling, so that the whole one-parameter family of the random processes appear.  Thus, in this case the limit  taken involves both  the parameter $\tau$ and  the ``time''-scale which is also measured in the units depending on $\tau$. As we will see this leads to richer picture in the second case.   

\subsubsection*{KPZ tails}
In this limit the  integrands of the integrals  constituting the transitional kernels can be represented in the from of the exponential functions with the growing prefactor in the exponent. Similarly to how it worked in the  KPZ part, the  kernels attain their maximal values in the vicinity of the double saddle point, which finally  brings the dominant contribution to the Fredholm determinant. Then the standard saddle point analysis yields the following results.
\begin{itemize}
\item Step IC.\
\begin{eqnarray}
\lim_{r \to \infty}\left(\frac{3}{2}\right)^{1/3}{(2r)^{-1/9}}K'_{\mathcal{A}_{2}\to\mathcal{N}}(r_{1},s_{1};r_{2},s_{2}) e^{2^{1/3}(r_1^{1/3}s_1-r_2^{1/3}s_2)}\notag\\= K_{\mathcal{A}_{2}}(u_{1},v_{1}-u_{1}^2;u_{2},v_{2}-u_{2}^2) 
\end{eqnarray}
where $s_i=v_i (2r)^{-1/9}\left(2/3\right)^{-1/3}+(3/2)(2r_i)^{1/3}$ and $r_i=r+u_i(2r)^{7/9} \left(3/2\right)^{2/3}$.\\ \\

\item Alternating IC
\begin{eqnarray}
\lim_{{\tau} \to \infty}3^{1/3}\tau^{-1/3}K'^{(\tau)}_{\mathcal{A}_{1}\to\mathcal{N}}(r_{1},s_{1};r_{2},s_{2})e^{(s_{21}+2r_{21})} \\= K_{\mathcal{A}_{1}}(u_{1},v_{1};u_{2},v_{2}) \notag
\end{eqnarray}
where $s_i=\tau+v_i\tau^{-1/3}3^{1/3}$ and $r_i=u_i \tau^{1/3}3^{2/3}$ for $i=1,2$. 
\end{itemize}
To obtain the standard kernels of Airy processes we also had to choose appropriate conjugations. 
This justifies the Proposition \ref{Prop: KPZ tails} up to the fact that the original transitional kernels 
also contain delta-functions, which turn out to  fall off the kernel under  taking the limit despite the fact that they are invariant under simultaneous variable and kernel rescaling. To explain why the delta-functions are not present in the limiting kernel for step IC we  note that the  typical difference of values of $s_1$ and $s_2$ corresponding to fixed  $u_2>u_1$ is  of order of $r^{1/9}$,  which is much larger than the fluctuation scale $r^{-1/9}$. The fluctuation scale  defines the typical range of kernel arguments,     beyond which the kernel is negligibly small. In contrary the delta function contribution comes from the points $s_1=s_2$, so that   the values of $u_2$ and $u_1$ are the distance $r^{1/9}$ away from  their typical values and the rest of the kernel vanishes. However, being in the upper triangular  part of the kernel, $r_2>r_1$,  the delta-function within the determinants always appears  contracted with  the  main parts of the kernel, thus forcing them to be negligibly small. This is the reason, why the contribution from delta functions vanishes in this limit for the step IC. For the alternating IC the transitional process is stationary. Thus the typical values of the difference $s_{21}$ is zero. However in this case the difference $s_{21}$ appears in the delta-function together with   $r_{21}\sim \tau^{1/3}$, which is again much larger than the typical fluctuation scale $(v_2-v_1)\sim\tau^{-1/3}$, making the terms with delta functions vanish.

\subsubsection*{DA tails}
In this case the limits are even more simple, which is expectable, as in the DA case  the simple Gaussian behavior is anticipated. 
Technically,  in this limit  the  kernel integrals are reduced to simple Gaussian integrals, which are readily evaluated.  
The step IC transitional kernel  arrives straight away to the limit, in which all particles are described by a single normal random variable.   
The DA limit for the alternating IC kernel is taken in two steps. First, we obtain the nontrivial random process with Gaussian one-point distribution in the limit $\tau\to 0$. Second  we diminish the ``time'' scale, to arrive at the same fully correlated limit as in the case of step IC.  As a result we obtain.
\begin{itemize}	
\item Step IC
\begin{eqnarray}
\lim_{\epsilon \to 0} K'_{\mathcal{A}_{2}\to\mathcal{N}}(\epsilon r_{1},s_{1}; \epsilon r_{2},{s_{2}})  =\frac{\exp\left(-\frac{1}{2}\left(s_{1}\right)^{2}\right)}{\sqrt{2\pi}}\label{Gauss}
\end{eqnarray}

\item Alternating IC
\begin{equation}
\lim_{\tau \to 0} K'^{(\tau)}_{\mathcal{A}_{1}\to\mathcal{N}}({r_{1}},{s_{1}}; r_{2},s_{2}) =\frac{\exp\left(-\frac{1}{2}\left(s_{1}+r_{1}-r_{2}\right)^{2}\right)}{\sqrt{2\pi}}\label{Gauss_r21}
\end{equation}

\end{itemize}
The further limit of the  r.h.s. of   (\ref{Gauss_r21})   obtained by multiplying $r_1$ and $r_2$ by $\epsilon$ and sending $\epsilon\to0$ returns the same result (\ref{Gauss}) as for the step IC.      
To obtain the limiting kernels we should add back the terms with delta-functions from the corresponding transitional kernels without any change. Then we indeed obtain  the kernel (\ref{K_chi}) from (\ref{Gauss_r21}) and (\ref{Gauss full}) from (\ref{Gauss}).  Also after  a conjugation (\ref{Gauss full}) by factor   $\exp\left(\frac{1}{4}\left(s_{1}^{2}-s_{2}^{2}\right)\right)$  we obtain the following kernel 
\begin{eqnarray}
K_{\mathcal{N}}( r_{1},s_{1};  r_{2},{s_{2}})  =\frac{\exp\left(-\frac{1}{4}\left(s_{1}^{2}+s_{2}^{2}\right)\right)}{\sqrt{2\pi}}-\mathbbm{1}_{r_2> r_1}\delta(s_2-s_1)\label{Gausssymm}.
\end{eqnarray}
The Fredholm determinant of the operator with this kernel gives a joint distribution of random variables with standard Gaussian distribution, which are identical  almost surely.   This follows from the next proposition, which concludes the discussion justifying the Proposition \ref{Prop: DA tails}. 
\begin{proposition}\label{KN}
	Let $\xi_1=\cdots =\xi_n$ be a random variables with the same probability density $f_{\xi}(x)$, such that $ \xi_1=\cdots =\xi_n$ almost surely. 
	Then 
	\begin{align}\label{KN_statement}
		\mathbb{P}(\xi_1<a_{1},\dots,\xi_n<a_{n}) & =\mathbb{P}(\xi_1<\min(a_{1},\dots,a_{n}))\\\nonumber
		& =\det\left(\boldsymbol{1}-P_{a}K_{\xi}P_{a}\right)_{L^{2}((1,\dots,n)\times\mathbb{R})}\text{,}
	\end{align}
	where the operator $K_{\xi}$ has kernel
\begin{equation} \label{Kxi}
K_{\mathcal{\xi}}(r_{1},s_{1};r_{2},s_{2})=\sqrt{f_{\xi}(s_{1})f_{\xi}(s_{2})}-\mathbbold{1}_{r_{2}>r_{1}}\delta(s_{2}-s_{1}).
\end{equation}
\end{proposition}

To prove this proposition  we use the following fact.

\begin{lemma}\label{min}
	If $r_{m}>r_{m-1}>\dots>r_{1}$ then for kernel (\ref{Kxi})
	\begin{eqnarray*}
		\int^{\infty}_{a_{1}}ds_{1}\int^{\infty}_{a_{2}}ds_{2}\dots\int^{\infty}_{a_{m}}ds_{m}\det(K_{\mathcal{\xi}}(r_{k},s_{k};r_{l},s_{l}))_{1\le k,l\le m}=\\
		=\int^{\infty}_{\max\{a_{1},a_{2},\dots,a_{m}\}}f(s) ds
	\end{eqnarray*}
\end{lemma}

\begin{svmultproof}
Let us rewrite the determinant in the form of  minor decomposition  (\ref{K_sum}) with 
\[
\tilde{K}_{i,j}=\sqrt{f_{\xi}(s_{i})f_{\xi}(s_{j})} ,\, \phi_{i,j}= \mathbbold{1}_{r_{j}>r_{i}}\delta(s_{j}-s_{i})
\]
As $\mathrm{rank}(\tilde{K})=1$, the only non-zero minors are  order  one. On the other hand, since $\phi_{i,j}$ is upper triangular, the only product $\phi_{i_1j_1}\cdots \phi_{i_{m-1} j_{m-1}}$ that survives corresponds to $i_k=k,j_k=k+1,k=1,\dots m-1$, which yields.
\begin{eqnarray}
&&\det(K_{\mathcal{\xi}}(r_{k},s_{k};r_{l},s_{l}))_{1\le k,l\le m}\\
&&\,\,\,\,\,\,\,\,=\sqrt{f_{\xi}(s_{m})f_{\xi}(s_{1})}\delta(s_{1}-s_{2})\delta(s_{2}-s_{3})\dots\delta(s_{m-1}-s_{m}).
\notag
\end{eqnarray}

Integrating it in $s_1, \dots, s_n$ obtain the desired statement.
\end{svmultproof}

\begin{svmultproof}(Proposition \ref{KN})
Rewrite r.h.s. of (\ref{KN_statement})  using the lemma \ref{min} as 
\begin{eqnarray*} 
	&1&-\sum_{i=1}^{n}\int^{\infty}_{a_{i}}K(r_{i},s_{i};r_{i},s_{i})ds_{i}+\\ \nonumber
	&+&\sum_{1 \le i_1<i_2 \le n}\int^{\infty}_{a_{i_1}}\int^{\infty}_{a_{i_2}}\det(K(r_{i_k},s_{i_k};r_{i_l},s_{i_l}))_{1\le k,l\le2}ds_{i_1}ds_{i_2}-\dots=\\ \nonumber
	&=&1-\sum_{i=1}^{n}\int^{\infty}_{a_i}f(s)ds+\sum_{1 \le i<j\le n}\int^{\infty}_{\max(a_{i},a_{j})}f(s)ds\\ \nonumber
	&-&\sum_{1 \le i<j<k \le n}\int^{\infty}_{\max(a_{i},a_{j},a_{k})}f(s)ds+\dots=\\ \nonumber
	&=&1-\int^{\infty}_{\min(a_1,\dots,a_n)}f(s)ds=\int_{-\infty}^{\min\{a_1,\dots, a_n\}}f(s)ds
\end{eqnarray*}
To do the last step we use the inclusion-exclusion principle.
\end{svmultproof}

\bibliography{DerbyshevPovolotsky}

\begin{thebibliography}{10}
\providecommand{\url}[1]{{#1}}
\providecommand{\urlprefix}{URL }
\expandafter\ifx\csname urlstyle\endcsname\relax
  \providecommand{\doi}[1]{DOI~\discretionary{}{}{}#1}\else
  \providecommand{\doi}{DOI~\discretionary{}{}{}\begingroup
  \urlstyle{rm}\Url}\fi

\bibitem{KPZ}
Kardar, K., Parisi, G., Zhang, Y.: Dynamic scaling ofgrowing interfaces.
\newblock Phys. Rev. Lett.56: 889-892  (1982)

\bibitem{Corwin2012}
Corwin, I.: The {Kardar-Parisi-Zhang} equation and universality class.
\newblock Random Matrices: Theory and Applications 01(01)  (2012)

\bibitem{KorepinBogoliubovIzergin}
Korepin, V., Bogoliubov, N., Izergin, A.: Quantum inverse scattering method and
  correlation functions.
\newblock Cambridge university press (1997)

\bibitem{Derrida1998}
Derrida, B.: An exactly soluble non-equilibrium system: the asymmetric simple
  exclusion process.
\newblock Physics Reports \textbf{301}, 65--83 (1998)

\bibitem{dhar1987exactly}
Dhar, D.: An exactly solved model for interfacial growth.
\newblock In: Phase transitions, vol.~9, pp. 51--51. Gordon Breach Sci Publ Ltd
  C/O STBS Ltd PO Box 90, Reading, Berks, England (1987)

\bibitem{gwa1992six}
Gwa, L.H., Spohn, H.: Six-vertex model, roughened surfaces, and an asymmetric
  spin hamiltonian.
\newblock Physical review letters \textbf{68}(6), 725 (1992)

\bibitem{gwa1992bethe}
Gwa, L.H., Spohn, H.: Bethe solution for the dynamical-scaling exponent of the
  noisy burgers equation.
\newblock Physical Review A \textbf{46}(2), 844 (1992)

\bibitem{Derrida_Lebowitz}
Derrida, B., Lebowitz, J.: Exact large deviation function in the asymmetric
  exclusion process.
\newblock Phys. Rev.Lett. 80, 209  (1998)

\bibitem{kim1995bethe}
Kim, D.: Bethe ansatz solution for crossover scaling functions of the
  asymmetric xxz chain and the kardar-parisi-zhang-type growth model.
\newblock Physical Review E \textbf{52}(4), 3512 (1995)

\bibitem{lee1999large}
Lee, D.S., Kim, D.: Large deviation function of the partially asymmetric
  exclusion process.
\newblock Physical Review E \textbf{59}(6), 6476 (1999)

\bibitem{de2005bethe}
De~Gier, J., Essler, F.H.: Bethe ansatz solution of the asymmetric exclusion
  process with open boundaries.
\newblock Physical review letters \textbf{95}(24), 240601 (2005)

\bibitem{de2006exact}
De~Gier, J., Essler, F.H.: Exact spectral gaps of the asymmetric exclusion
  process with open boundaries.
\newblock Journal of Statistical Mechanics: Theory and Experiment
  \textbf{2006}(12), P12011 (2006)

\bibitem{de2011large}
de~Gier, J., Essler, F.H.: Large deviation function for the current in the open
  asymmetric simple exclusion process.
\newblock Physical review letters \textbf{107}(1), 010602 (2011)

\bibitem{derrida1993exact}
Derrida, B., Evans, M.R., Hakim, V., Pasquier, V.: Exact solution of a 1d
  asymmetric exclusion model using a matrix formulation.
\newblock Journal of Physics A: Mathematical and General \textbf{26}(7), 1493
  (1993)

\bibitem{blythe2007nonequilibrium}
Blythe, R.A., Evans, M.R.: Nonequilibrium steady states of matrix-product form:
  a solver's guide.
\newblock Journal of Physics A: Mathematical and Theoretical \textbf{40}(46),
  R333 (2007)

\bibitem{SchutzG.1997}
Sch\"{u}tz, G.M.: Exact solution of the master equation for the asymmetric
  exclusion process.
\newblock J. Stat. Phys.88, 427  (1997)

\bibitem{Johansson}
Johansson, K.: Shape fluctuations and random matrices.
\newblock Comm. Math. Phys.,13: 1380-1418  (2000)

\bibitem{nagao2004asymmetric}
Nagao, T., Sasamoto, T.: Asymmetric simple exclusion process and modified
  random matrix ensembles.
\newblock Nuclear Physics B \textbf{699}(3), 487--502 (2004)

\bibitem{rakos2005current}
R{\'a}kos, A., Sch{\"u}tz, G.M.: Current distribution and random matrix
  ensembles for an integrable asymmetric fragmentation process.
\newblock Journal of statistical physics \textbf{118}(3), 511--530 (2005)

\bibitem{tracy2008integral}
Tracy, C.A., Widom, H.: Integral formulas for the asymmetric simple exclusion
  process.
\newblock Communications in Mathematical Physics \textbf{279}(3), 815--844
  (2008)

\bibitem{tracy2008fredholm}
Tracy, C.A., Widom, H.: A fredholm determinant representation in {ASEP}.
\newblock Journal of Statistical Physics \textbf{132}(2), 291--300 (2008)

\bibitem{BorodinOkounkovOlshaski}
Borodin, A., Okounkov, A., Olshanski, G.: Asymptotics of {Plancherel} measures
  for symmetric groups.
\newblock Journal of American Mathematical Society \textbf{13}, 491--515 (2000)

\bibitem{Borodin2015}
Borodin, A.: The Oxford Handbook of Random Matrix Theory, chap. Determinantal
  point processes.
\newblock Oxford University Press (2015)

\bibitem{Sasamoto2005}
Sasamoto, T.: Spatial correlations of the 1{D} {KPZ} surface on a flat
  substrate.
\newblock Journal of Pysics A: Mathematical and General \textbf{38}, L549--L556
  (2005)

\bibitem{borodin2007fluctuation}
Borodin, A., Ferrari, P.L., Pr{\"a}hofer, M., Sasamoto, T.: Fluctuation
  properties of the {TASEP} with periodic initial configuration.
\newblock Journal of Statistical Physics \textbf{129}(5-6), 1055--1080 (2007)

\bibitem{Borodin2}
Borodin, A., Ferrari, P.: Large time asymptotics of growth models on space-like
  paths {I}: Push{ASEP}.
\newblock Electron. J. Probab., 13:1380-1418  (2008)

\bibitem{BFS}
Borodin, A., Ferrari, P., Sasamoto, T.: Large time asymptotics of growth models
  on space-like paths {II}: {PNG} and parallel {TASEP}.
\newblock Comm. Math. Phys. 283, 417-449  (2008)

\bibitem{imamura2007dynamics}
Imamura, T., Sasamoto, T.: Dynamics of a tagged particle in the asymmetric
  exclusion process with the step initial condition.
\newblock Journal of Statistical Physics \textbf{128}(4), 799--846 (2007)

\bibitem{PoghosyanS.2010}
Poghosyan, S.S., Priezzhev, V.B., Sch\"{u}tz, G.M.: Green functions for the
  {TASEP} with sublattice parallel update.
\newblock J. Stat. Mech. P04022  (2010)

\bibitem{exit}
Poghosyan, S.S., Povolotsky, A.M., Priezzhev, V.B.: Universal exit
  probabilities in the {TASEP}.
\newblock J. Stat. Mech. P08013  (2012)

\bibitem{Ferrari2008}
Ferrari, P.L.: Integrable systems and random matrices, \emph{Contemporary
  Mathematics}, vol. 458, chap. The universal {Airy}$_1$ and {Airy}$_2$
  processes in the totally asymmetric simple exclusion process, pp. 321--332.
\newblock American Mathematical Society, Providence, R.I. (2008)

\bibitem{MQR}
Matetski, K., Quastel, J., Remenik, D.: The {KPZ} fixed point.
\newblock arXiv preprint arXiv:1701.00018. Dec 30.  (2016)

\bibitem{BL_1}
Baik, J., Liu, Z.: Fluctuations of {TASEP} on a ring in relaxation time scale.
\newblock Communications on Pure and Applied Mathematics  (2016)

\bibitem{BL_2}
Baik, J., Liu, Z.: Multi-point distribution of periodic {TASEP}.
\newblock Journal of the American Mathematical Society 32(3)  (2017)

\bibitem{Baik2000}
Baik, J., Rains, E.: Limiting distributions for a polynuclear growth model with
  external sources.
\newblock J. Stat. Phys., 100:523-542  (2000)

\bibitem{BFPS_2007}
Borodin, A., Ferrari, P., Pr\"{a}hofer, M., Sasamoto, T.: Fluctuations in the
  discrete {TASEP} with periodic initial configurations and the {Airy}$_1$
  process.
\newblock Int. Math. Res. Papers  (2007)

\bibitem{Spohn2002}
Pr\"{a}hofer, M., Spohn, H.: Scale invariance of the {PNG} droplet and the
  {Airy} process.
\newblock J. Stat. Phys. 108, 1071-1106  (2002)

\bibitem{Baik2010}
Baik, J., Ferrari, P., Peche, S.: Limit process of stationary {TASEP} near the
  characteristic line.
\newblock Comm. Pure Appl. Math 63 (2010), 1017-1070  (2010)

\bibitem{Mehta}
Mehta, M.: Random matrices.
\newblock Elsevier (2004)

\bibitem{WeissFerrariSpohn2017}
Weiss, T., Ferrari, P., Spohn, H.: Reflected {Brownian} motions in the {KPZ}
  universality class.
\newblock Springer International Publishing (2017)

\bibitem{Johansson2005}
Johansson, K.: The arctic circle boundary and the {Airy} process.
\newblock Ann. Probab. Volume 33, Number 1 (2005), 1-30.  (2005)

\bibitem{Okounkov2003}
Okounkov, A., Reshetikhin, N.: Correlation function of {Schur} process with
  application to local geometry of a random 3-dimensional {Young} diagram.
\newblock J. Amer. Math. Soc.16:581-603  (2003)

\bibitem{tracy2009asymptotics}
Tracy, C.A., Widom, H.: Asymptotics in {ASEP} with step initial condition.
\newblock Communications in Mathematical Physics \textbf{290}(1), 129--154
  (2009)

\bibitem{ferrari2015tracy}
Ferrari, P.L., Vet{\H{o}}, B.: Tracy-{Widom} asymptotics for $ q $-{TASEP}.
\newblock In: Annales de l'IHP Probabilit{\'e}s et statistiques, vol.~51, pp.
  1465--1485 (2015)

\bibitem{vetHo2015tracy}
Vet{\H{o}}, B., et~al.: Tracy-{Widom} limit of $ q $-{Hahn TASEP}.
\newblock Electronic Journal of Probability \textbf{20} (2015)

\bibitem{imamura2019fluctuations}
Imamura, T., Sasamoto, T.: Fluctuations for stationary q-{TASEP}.
\newblock Probability Theory and Related Fields \textbf{174}(1-2), 647--730
  (2019)

\bibitem{Borodin2008}
Borodin, A., Ferrari, P.L., Sasamoto, T.: Transition between {Airy}$_1$ and
  {Airy}$_2$ processes and {TASEP} fluctuations.
\newblock Comm. Pure Appl. Math., 61(11):1603-1629  (2008)

\bibitem{Imamura2004}
Imamura, T., Sasamoto, T.: Fluctuations of the one-dimensional polynuclear
  growth model with external sources.
\newblock Nuclear Phys. B, 699(3):503-544  (2004)

\bibitem{krug1997origins}
Krug, J.: Origins of scale invariance in growth processes.
\newblock Advances in Physics \textbf{46}(2), 139--282 (1997)

\bibitem{sasamoto2010one}
Sasamoto, T., Spohn, H.: One-dimensional {Kardar-Parisi-Zhang} equation: an
  exact solution and its universality.
\newblock Physical review letters \textbf{104}(23), 230602 (2010)

\bibitem{dotsenko2010bethe}
Dotsenko, V.: Bethe ansatz derivation of the {Tracy-Widom} distribution for
  one-dimensional directed polymers.
\newblock EPL (Europhysics Letters) \textbf{90}(2), 20003 (2010)

\bibitem{Amir2011ProbabilityDO}
Amir, G., Corwin, I., Quastel, J.: Probability distribution of the free energy
  of the continuum directed random polymer in 1 + 1 dimensions.
\newblock Communications on Pure and Applied Mathematics \textbf{64}, 466--537
  (2011)

\bibitem{calabrese2010free}
Calabrese, P., Le~Doussal, P., Rosso, A.: Free-energy distribution of the
  directed polymer at high temperature.
\newblock EPL (Europhysics Letters) \textbf{90}(2), 20002 (2010)

\bibitem{Woelki}
Woelki, M.: Master's thesis, University of Duisburg-Essen (2005)

\bibitem{gtasep}
Derbyshev, A., Poghosyan, S.S., Povolotsky, A., Priezzhev, V.: The totally
  asymmetric exclusion process with generalized update.
\newblock J. Stat. Mech. P05014  (2012)

\bibitem{P2013}
Povolotsky, A.: On integrability of zero-range chipping models with factorized
  steady state.
\newblock J. Phys. A: Math. Theor. 46 (2013) 465205  (2013)

\bibitem{BCPS2013}
Borodin, A., Corwin, I., Petrov, L., Sasamoto, T.: Spectral theory for
  interacting particle systems solvable by coordinate {Bethe} ansatz.
\newblock Comm. Math. Phys. 339 (2015), no. 3, 1167--1245  (2015)

\bibitem{KnizelPetrovSaenz}
Knizel, A., Petrov, L., Saenz, A.: Generalizations of {TASEP} in discrete and
  continuous inhomogeneous space.
\newblock Communications in Mathematical Physics. Dec 1;372(3):797-864.  (2019)

\bibitem{DPP2015}
Derbyshev, A., Povolotsky, A., Priezzhev, V.: Emergence of jams in the
  generalized totally asymmetric simple exclusion process.
\newblock Phys. Rev. E. Feb;91(2):022125.  (2015)

\bibitem{aneva2016matrix}
Aneva, B., Brankov, J.: Matrix-product ansatz for the totally asymmetric simple
  exclusion process with a generalized update on a ring.
\newblock Physical Review E \textbf{94}(2), 022138 (2016)

\bibitem{bunzarova2017one}
Bunzarova, N.Z., Pesheva, N.C.: One-dimensional irreversible aggregation with
  dynamics of a totally asymmetric simple exclusion process.
\newblock Physical Review E \textbf{95}(5), 052105 (2017)

\bibitem{brankov2018model}
Brankov, J., Bunzarova, N.Z., Pesheva, N., Priezzhev, V.: A model of
  irreversible jam formation in dense traffic.
\newblock Physica A: Statistical Mechanics and its Applications \textbf{494},
  340--350 (2018)

\bibitem{bunzarova2019one}
Bunzarova, N.Z., Pesheva, N., Brankov, J.: One-dimensional discrete
  aggregation-fragmentation model.
\newblock Physical Review E \textbf{100}(2), 022145 (2019)

\bibitem{bunzarova2021aggregation}
Bunzarova, N.Z., Pesheva, N.: Aggregation-fragmentation of clusters in the
  framework of gtasep with attraction interaction.
\newblock Physics of Particles and Nuclei \textbf{52}(2), 169--184 (2021)

\bibitem{baik2005phase}
Baik, J., Arous, G.B., P{\'e}ch{\'e}, S., et~al.: Phase transition of the
  largest eigenvalue for nonnull complex sample covariance matrices.
\newblock Annals of Probability \textbf{33}(5), 1643--1697 (2005)

\bibitem{baik2006painleve}
Baik, J., et~al.: Painlev{\'e} formulas of the limiting distributions for
  nonnull complex sample covariance matrices.
\newblock Duke Mathematical Journal \textbf{133}(2), 205--235 (2006)

\bibitem{barraquand2015phase}
Barraquand, G.: A phase transition for q-tasep with a few slower particles.
\newblock Stochastic Processes and their Applications \textbf{125}(7),
  2674--2699 (2015)

\bibitem{rajewsky1998asymmetric}
Rajewsky, N., Santen, L., Schadschneider, A., Schreckenberg, M.: The asymmetric
  exclusion process: Comparison of update procedures.
\newblock Journal of statistical physics \textbf{92}(1), 151--194 (1998)

\bibitem{Rost}
Rost, H.: Lectures in Probability and Statistics, chap. On the behavior of the
  hydrodynamical limit for stochastic particle systems, pp. 129--164.
\newblock Springer, Berlin, Heidelberg (1986)

\bibitem{Pablo_Ferrari}
Ferrari, P.: {TASEP} hydrodynamics using microscopic characteristics.
\newblock Probab. Surveys Volume 15, 1-27  (2018)

\bibitem{KM1990}
Krug, J., Meakin, P.: Universal finite-size effects in the rate of growth
  processes.
\newblock Journal of Physics A: Mathematicaland General, 23(18), p.L987  (1990)

\bibitem{amar1992universal}
Amar, J.G., Family, F.: Universal scaling function and amplitude ratios in
  surface growth.
\newblock Physical Review A \textbf{45}(6), R3373 (1992)

\bibitem{KMH1992}
Krug, J., Meakin, P., Halpin-Healy, T.: Amplitude universality for driven
  interfaces and directed polymersin random media.
\newblock Physical Review A, 45(2), p.638.  (1992)

\bibitem{imamura2004fluctuations}
Imamura, T., Sasamoto, T.: Fluctuations of the one-dimensional polynuclear
  growth model with external sources.
\newblock Nuclear Physics B \textbf{699}(3), 503--544 (2004)

\bibitem{imamura2005polynuclear}
Imamura, T., Sasamoto, T.: Polynuclear growth model with external source and
  random matrix model with deterministic source.
\newblock Physical Review E \textbf{71}(4), 041606 (2005)

\bibitem{baik2000limiting}
Baik, J., Rains, E.M.: Limiting distributions for a polynuclear growth model
  with external sources.
\newblock Journal of Statistical Physics \textbf{100}(3), 523--541 (2000)

\bibitem{forrester2000painlev}
Forrester, P.: Painlev$\backslash$'e transcendent evaluation of the scaled
  distribution of the smallest eigenvalue in the laguerre orthogonal and
  symplectic ensembles.
\newblock arXiv preprint nlin/0005064  (2000)

\bibitem{kuijlaars2011non}
Kuijlaars, A., Mart{\'\i}nez-Finkelshtein, A., Wielonsky, F.: Non-intersecting
  squared bessel paths: critical time and double scaling limit.
\newblock Communications in mathematical physics \textbf{308}(1), 227--279
  (2011)

\bibitem{Nagao_Sasamoto_2004}
Nagao, T., Sasamoto, T.: Asymmetric simpleexclusion process and modified random
  matrix ensembles.
\newblock Nuclear PhysicsB. Nov 8;699(3):487-502.  (2004)

\bibitem{Borodin_Rains}
Borodin, A., Rains, E.M.: {Eynard-Mehta} theorem, {Schur} process, and their
  {Pfaffian} analogs.
\newblock Journal of Statistical Physics. 121(3):291-317.  (2005)

\bibitem{GTW}
Gravner, J., Tracy, C.A., Widom, H.: Limit theorems for height fluctuations in
  a class of discrete space and time growth models.
\newblock Journal of Statistical Physics \textbf{102}(5), 1085--1132 (2001)

\bibitem{Johansson1}
Johansson, K.: Discrete polynuclear growth and determinantal processes.
\newblock Commun. Math. Phys. 242, 277-329  (2003)

\bibitem{Borodin2006}
Borodin, A., Ferrari, P., Prahofer, M.: Fluctuations in the discrete {TASEP}
  with periodic initial configurations and the {Airy}$_1$ process.
\newblock Int. Math. Res. Papers 2007, rpm002  (2006)

\bibitem{LeV92}
LeVeque, R.J.: Numerical methods for conservation laws, Lectures in Mathematics
  ETH Zurich, vol. 132.
\newblock Basel: Birkh\"{a}user,

\bibitem{KRUG_1997_rev}
Krug, J.: Origins ofscale invariance in growth processes.
\newblock Advances in Physics46.2 : 139-282.  (1997)

\end{thebibliography}

\appendix
\section{Hydrodynamic, KPZ and transitional heuristics from the stationary
	state\label{sec:Hydrodynamic,-KPZ-and}}
\subsection{Hydrodynamics \label{app: Hydrodynamics}}

In the nonuniform setting of step IC the particle density $c(x,t)$  and current $j(x,t)$ are related by the continuity equation that expresses 
the particle conservation law,
\[
\partial_{t}c(x,t)+\partial_{x}j(x,t)=0.
\]
Then, the local quasi-equilibrium suggests the same stationary state  current-density relation $j(x,t)=j_{\infty}(c(x,t))$ holding locally in the
varying density landscape, yielding  the hyperbolic PDE for the function $c(x,t)$
\begin{equation}
\partial_{t}c+j'_{\infty}(c)\partial_{x}c=0.
\end{equation}
This equation can be solved by the method of characteristics, which  
in general  can be non-trivial due to presence of shocks, see e.g. Ex. 3.6 in \cite{LeV92}.
However,  for step IC, $c(x,0)=\mathbbold1_{x\leq0}$, with  $j_{\infty}(c)$ being  differentiable,
convex function vanishing at the ends of density range 
\begin{equation}
j_{\infty}(0)=j_{\infty}(1)=0,\label{eq:j endpoints}
\end{equation}
the solution  corresponding to the rarefaction fan  is straightforward,  Ex.  3.7 in \cite{LeV92}.
In this case, all nontrivial characteristics of this equation are outgoing from the origin $(x,t)=(0,0)$.
Hence the solution $c(x,t)=c(\chi)$ is a function of $\chi=x/t$
given by an inversion of relation
\begin{align}
j'_{\infty}(c) & =\chi,\label{eq: j'=00003Dchi}
\end{align}
in the range $j'(1) \leq \chi< j'_{\infty}(0) $ and  $c(\chi)  =0$ or $1$ when $\chi\geq j'_{\infty}(0)$ or  $\chi< j'_{\infty}(1)$
respectively otherwise. 

To relate this solution to the position of a particle $x_{n}(t)$
we note that the number $n=\theta t$ is exactly the number of particles
to the right of $x_{n}(t)\simeq t\chi(\theta)$, i.e.
\begin{align}
\theta & =\int_{\chi}^{j'_{\infty}(0)}c(y)dy  =-\chi c-\int_{\chi}^{j'_{\infty}(0)}yc'(y)dy\label{eq: thetalegandre} =-\chi c+j_{\infty}(c), 
\end{align}
where the second equality  is an integration by parts and for  the third
one we used the variable change $y\to c(y)$ together with (\ref{eq: j'=00003Dchi})
and (\ref{eq:j endpoints}). According to (\ref{eq: j'=00003Dchi})
and (\ref{eq: thetalegandre}) the function $\theta(\chi)$ is nothing
but minus the Legendre transform of $j_{\infty}(c)$, which is stated in eq.(\ref{eq: thetavar}).
Being strictly monotonous, when  $c\in(0,1)$, it can obviously be inverted.

\subsection{Stationary state and deterministic relations \label{app: Stationary state and deterministic relations}}

To prepare a translationally invariant steady state we consider first
the model on a finite periodic lattice of $L$ sites, $\mathbb{Z}/L\mathbb{Z}$,
implying that $L$ will be sent to infinity in the end. There are
two complementary approaches that were shown to be effective in studies
of the simplest stationary state of GTASEP. The first one is based
on the so called ZRP-ASEP mapping and canonical partition function
formalism, while the second works directly with GTASEP though within
the framework of grand-canonical ensemble. 

Within the first approach we overcome the difficulty connected with
the non-locality of dynamical rules of GTASEP by mapping the ASEP-like
system, where particles obey the exclusion interaction, to an equivalent
zero-range process (ZRP)-like system, where many particles in a site
are allowed. To this end, we replace occupied sites of an $n$-particle
cluster plus one empty site ahead with a site occupied by $n$ particles.
As a result from the ASEP like system of $M$ particles on the lattice
of size $L$ we obtain the ZRP-like system with the same number of
particles on the lattice of size $N=L-M$. The dynamical rules of
GTASEP prescribe probability $\varphi(m|n)$ form (\ref{eq: phi(m|n)})
to jumps of $m$ particles out of sites with $n$ particles, all sites
being updated simultaneously and independently of the others at every
time step. 

A crucial observation is that the jumping probabilities (\ref{eq: phi(m|n)})
can be written in the product form 
\[
\varphi(m|n)=\frac{v(m)w(n-m)}{f(n)},
\]
where 
\begin{eqnarray}
	v(k) & = & \mu^{k}(\delta_{k,0}+(1-\delta_{k,0})(1-\nu/\mu)),\label{eq:v(k)}\\
	w(k) & = & (\delta_{k,0}+(1-\delta_{k,0})(1-\mu)),\label{eq:w(k)}
\end{eqnarray}
and
\begin{equation}
	f(n)=\sum_{k=0}^{n}v(n)w(n-k)=(\delta_{n,0}+(1-\delta_{n,0})(1-\nu)).\label{eq: f(n) form}
\end{equation}
This fact is responsible for  the stationary measure of the ZRP obtained from  GTASEP
on the ring having a factorized form. This is to say that the stationary
state probability for $n_{1},\dots,n_{N}$ particles to occupy sites
$1,\dots,N$ respectively is given by product 
\[
P_{st}(n_{1},\dots,n_{N})=\frac{1}{Z(M,N)}\prod_{i=1}^{N}f(n_{i}),
\]
where $Z(M,N)=\sum_{||n||=M}\prod_{i=1}^{N}f(n_{i})$ is the partition
function, aka sum of the stationary weights over particle configurations
$\boldsymbol{n}$ constrained by $||\boldsymbol{n}||:=n_{1}+\cdots+n_{N}=M$
. The partition function can be represented in the form of a contour integral
\begin{equation}
	Z\left(M,N\right)=\oint_{\Gamma_{0}}\frac{\left[F(z)\right]^{N}}{z^{M+1}}\frac{dz}{2\pi \mathrm{i}},\label{eq: Z(N,M)}
\end{equation}
where $F(z)=\sum_{n\geq0}f(n)z^{n}$ is the generating function of
stationary weights, and the contour of integration encircles the origin.
An explicit form of $F(z)$ is given by a product $F(z)=V(z)W(z)$
of generating functions of $v(k)$ and $w(k)$, 
\[
V(z)=\frac{1-\nu z}{1-\mu z},\,\,W(z)=\frac{1-\mu z}{1-z},\,\,F(z)=\frac{1-\nu z}{1-z},
\]
for the stationary weight $f(n)$ being the convolution of the $v(k)$
and $w(k)$. The integral representation suits ideally for the asymptotic
analysis we perform in the thermodynamic (large lattice, fixed density)
limit 
\[
L\to\infty,M\to\infty,M/L=c.
\]
In this limit we evaluate integral $(\ref{eq: Z(N,M)})$ in the saddle
point approximation. The equation for the critical point 
\begin{equation}
	\frac{c}{(1-c)}\frac{1}{z}+\frac{\nu}{1-\nu z}-\frac{1}{1-z}=0\label{eq: saddle point eq.}
\end{equation}
has two solutions 
\[
z_{c}^{\pm}=1+\frac{(1-\nu)}{2c\nu}\left(1\pm\sqrt{1+\frac{4(1-c)c\nu}{1-\nu}}\right).
\]
from which $z_{c}^{-}$ is the one that brings dominating contribution
into the integral. As, the second saddle point does not play any role,
we will omit the minus sign for brevity of notations implying $z_{c}\equiv z_{c}^{-}.$

The value of $z_{c}$ being function of the density $c$ increases
from zero to one as $c$ decreases from one to zero. Many observables
of the stationary state can be represented in the form of similar contour
integrals. Then, in the thermodynamic limit they will be the functions
of fugacity $z_{c}$. 

Now we can change the point of view and consider the stationary state
observables as functions of the parameter  $z_{c}$, which takes values
in the range $z_{c}\in(0,1)$\footnote{This is equivalent to going to the grand-canonical ensemble from the
	canonical one, which is simpler and suits well for description of
	genuinely infinite systems. We could start directly from the grand-canonical
	description having defined fugacity a priori and fixing the density
	as an average occupation number. We, however, started from the canonical
	partition function for the finite system keeping in mind that we are
	interested not only in the thermodynamic quantities, but also in finite
	size corrections to them.}. In particular the total number of particle jumps from a site per
one time step (translated into the language of ASEP-like system) having
the exact integral representation \cite{DPP2015},
\begin{equation}
	j_{L}=\frac{N/L}{Z\left(M,N\right)}\oint_{\Gamma_{0}}\frac{\left[F\left(z\right)\right]^{N}}{z^{M}}\frac{V'(z)}{V(z)}\frac{dz}{2\pi \mathrm{i}},\label{eq: j_L}
\end{equation}
converges to 

\begin{equation}
	j_{\infty}=\big(\frac{1}{1-\mu z_{c}}-\frac{1}{1-\nu z_{c}}\big)(1-c)
\end{equation}
in the thermodynamic limit. In this way we obtain parametric dependence
of $j_{\infty}$ on $c$. Though it can be explicitly resolved to
give the current-density relation obtained in \cite{Woelki,DPP2015},
the further derivation of $\chi(\theta)$ is possible only in parametric
form. Using the relations (\ref{eq: j'=00003Dchi},\ref{eq: thetalegandre})
we obtain the  functional dependence (\ref{eq: c(z)}-\ref{eq: theta}) between $\chi,\theta$
and $c$, which are expressed as functions of the parameter $z_{c}$
varying in the range $0<z_{c}<1$. 
These three functions determine the behavior of GTASEP particles
in the deterministic scale. Formally, to express one of them as the
function of another, one has to eliminate $z_{c}$ between them by
solving the large degree polynomial equation, which is hardly suitable
for further calculations. In contrast, the parametric form is the
one which is obtained from the asymptotic analysis of the exact distributions
and also is enough to proceed with scaling constants defining the fluctuation
and correlation scales. 

\subsection{KPZ dimensional invariants and model-dependent scaling constants \label{app: KPZ dimensional invariants }}

The next step is to understand the meaning of scaling constants $\kappa_{f}$
and $\kappa_{c}$ defining correlation and fluctuation scales respectively.
To this end we refer to the papers \cite{KM1990},\cite{KMH1992} and review
\cite{KRUG_1997_rev}, where predictions for the scaling form of cumulants
of the interface height were made on the basis of the analysis of KPZ
equation and conjectured to be universal for the large class of models
belonging to KPZ class. 

To summarize, the large and small-time scaling behavior of height
$h(x,t)$ of an interface governed by the KPZ equation 
\begin{equation}
	\frac{\partial h}{\partial t}=\widetilde{\nu}\Delta h+\widetilde{\lambda}\left(\nabla h\right)^{2}+\eta,\label{eq:KPZ}
\end{equation}
with the Gaussian white noise $\eta$  defined by  zero mean $\langle\eta(x,t) \rangle=0$ and covariance 
\[
\left\langle \eta(x,t)\eta(x',t')\right\rangle =D\delta(x-x')\delta(t-t'),
\]
depends on two dimensional invariants $\widetilde{\lambda}$ and $A=D/2\widetilde{\nu}$.\footnote{The notations for parameters $\tilde{\lambda}$ and $\tilde{\nu}$
	bring the tilde sign to keep the custom KPZ notations and to distinguish
	them from the $\nu$ and $\lambda$ of our paper.} Here and further within this section we use the accepted in physical literature  notations $\langle \xi^n \rangle$ and $\langle \xi^n \rangle_c$ for $n$-th moment and cumulant of the random variable $\xi$ respectively.

In the transient (short time) regime, which can be thought of either
as the large time evolution of an infinite system or that of the finite
system with the large time limit $t\to\infty$ taken after the large
size limit $L\to\infty$, the fluctuations of the interface height
are scaled as $t^{1/3}$ with time units being inverse of $\left|\widetilde{\lambda}\right|A^{2}$,
\begin{equation}
	h-\left\langle h\right\rangle \sim\mathrm{const}\left(\left|\widetilde{\lambda}\right|A^{2}t\right)^{1/3}\mathcal{X}.\label{eq: short time}
\end{equation}
Here $\left\langle h\right\rangle $ is the mean value obtained from
averaging over the noise realization and $\mathcal{X}$ is some universal
(parameterless) random variable, which still depends on IC . On the
other hand, in the late time regime, where the large time limit is
taken in a finite system, which is then supposed to be large, the height
deviation from its spacial average $\bar{h}=L^{-1}\int hdx$ is purely
Gaussian, 
\begin{equation}
	h-\bar{h}\sim\mathrm{const}\left(AL\right)^{1/2}\mathcal{N}\label{eq: late time}
\end{equation}
with the variance proportional to the distance measured in the units
inverse to $A$. The notation $\mathcal{N}$ is used for the standard
normal random variable. These two regimes can be sewed together within
the so called Family-Vicek scaling of interface width
\begin{equation}
	w=\sqrt{\left\langle \left(h-\bar{h}\right)^{2}\right\rangle }=\left(AL\right)^{1/2}\mathcal{F}_{FV}(L/\xi(t)),\label{eq:Family Vicek}
\end{equation}
where the asymptotic behavior of the scaling function, $\mathcal{F}_{FV}(0)=\mathrm{const}$
and $\mathcal{F}_{FV}(x)=O(x^{-1/2})$ as $x\to\infty$, is dictated
by the requirement of attaining both limits (\ref{eq: late time},\ref{eq: short time}).
The correlation length that would reproduce (\ref{eq: short time})
must have the form 
\begin{equation}
	\xi(t)=\frac{\left(\left|\widetilde{\lambda}\right|A^{2}t\right)^{2/3}}{A}.\label{eq: xi(t)}
\end{equation}
Note that all the dimensional scaling constants are defined up to
dimensionless numbers, which being the universal normalization should
be chosen consistently with the definitions of limiting random variables
and processes, depending on initial and boundary conditions. Qualitatively
$\xi(t)$ defines the correlation scale, within which the fluctuations
are supposed to be non-trivially correlated, and, in particular, the
fluctuations become stationary, when the correlation length is comparable
to the system size. In other words we expect that $\xi(t)$ gives
the natural spacial correlation scale, i.e. the units  of spacial coordinate
within the limiting multipoint correlation functions.

The key hypothesis from \cite{KMH1992} states that the scaling arguments
of the same type are applicable to the wide range systems within the
whole KPZ class far beyond the KPZ equation itself, up to the only
difference that the a priori unknown parameters $\tilde{\lambda}$ and
$A$ should be read off from the properties of the stationary state
of the models. The recipe of finding them was also proposed in \cite{KM1990},\cite{KMH1992}.
For growing KPZ interfaces of arbitrary origin the lateral growth
parameter $\widetilde{\lambda}$, related to the response of the interface
velocity to a small tilt $h(x,t)\to h(x,t)+\kappa x$, can be determined
from 
\begin{equation}
	\widetilde{\mbox{\ensuremath{\lambda}}}=\frac{\partial^{2}v_{\infty}}{\partial\kappa^{2}},\label{eq:lambda-def}
\end{equation}
where $v_{\infty}$ is the $L\to\infty$ limit of stationary interface
velocity $v_{L}=\lim_{t\to\infty}\left\langle \partial h/\partial t\right\rangle ,$
and leading finite size correction to the interface velocity 
\[
b_{v}=\lim_{L\to\infty}\lim_{t\to\infty}L\left(\left\langle \partial h/\partial t\right\rangle -v_{\infty}\right)
\]
is expected to be given by 
\[
b_{v}=-\frac{A\widetilde{\lambda}}{2}.
\]
Finding these two quantities is enough for defining the two necessary
dimensional constants. An independent consistency check can be done
with calculation of spacial correlation function 
\begin{equation}
	\lim_{t\to\infty}\left\langle \left(h(x,t)-h(y,t)\right)^{2}\right\rangle _{c}=A\left|x-y\right|,\label{eq:A-def}
\end{equation}
which amplitude is nothing but $A.$

All these quantities are accessible for our GTASEP system by identifying
it with an interface using mapping 
\[
h_{i+1}-h_{i}=1-2\eta_{i},
\]
where $\eta_{i}=0,1$ is the occupation number of the $i-$th site,
$h_{i}$ is the interface height above the bond connecting sites $i-1$
and $i$ of the lattice, $i=1,\dots,L,$ satisfying helicoidal boundary
conditions 
\[
h_{i+L}=h_{i}-(L-2M),
\]
which gives a tilt $\kappa=1-2c$ to the interface. Observing that
the interface velocity is twice the particle current $v_{L}=2j_{L}(c)$
we obtain 
\begin{align}
	\widetilde{\lambda} & =\frac{1}{2}\frac{d^{2}j_{\infty}}{dc^{2}}=\frac{1}{2}\left(\frac{1}{dc/dz_{c}}\frac{d}{dz_{c}}\right)^{2}j_{\infty}\\
	& =-\frac{(1-\mu)(\mu-\nu)}{(1-\nu)^{2}}\frac{\left(1-\nu\left(2-z_{c}\right)z_{c}\right){}^{3}\left(1-\mu\nu z_{c}^{3}\right)}{\left(1-\mu z_{c}\right){}^{3}\left(1-\nu z_{c}^{2}\right){}^{3}},\nonumber 
\end{align}
The first order finite size correction to the current obtained from
(\ref{eq: j_L}) is 
\[
b_{v}=2\frac{(1-\mu)(\mu-\nu)}{(1-\nu)}\times\frac{\left(1-z_{c}\right)z_{c}\left(1-\nu z_{c}\right)\left(1-\mu\nu z_{c}^{3}\right)}{\left(1-\mu z_{c}\right){}^{3}\left(1-\nu z_{c}^{2}\right){}^{2}}.
\]

From these formulas $\widetilde{\lambda}$ and $A$ can be found in
terms of the fugacity $z_{c}.$ Alternatively we could obtain $A$
from knowing correlation function (\ref{eq:A-def}), which was derived
in \cite{DPP2015} within the grand canonical formalism applied directly
to the ASEP-like system and was shown to be consistent with the other
formulas. We refer the reader to \cite{DPP2015} for further details.

Finally to translate  the fluctuation and correlation scales, (\ref{eq: short time})
and (\ref{eq: xi(t)}), obtained for the interface to
the language of particles, we note that the height increase is twice
the number of particles having traversed a particular bond. The ratio
of the distance scale and the $h-$scale is the particle density $c$,
i.e. the ratio of the number of jumps per particle to the number of
jumps per bond. Hence we define
\begin{align}
	\kappa_{f} & =\frac{1}{2c}\left(\frac{|\widetilde{\lambda}|A^{2}}{2}\right)^{1/3} \label{eq: kappa_f KPZ}
\end{align}
Conversely, to go from $x-$scale to $n$-scale we have to multiply
$\xi(t)$ by $c$:

\begin{align}
	\kappa_{c} & =\frac{c}{A}\left(\frac{|\widetilde{\lambda}|A^{2}}{2}\right)^{2/3} \label{eq: kappa_c KPZ}
\end{align}
After substitution the explicit expressions for $A$ and $\tilde{\lambda}$ in terms of  $z_c$ into (\ref{eq: kappa_f KPZ},\ref{eq: kappa_c KPZ}) we arrive at formulas  (\ref{eq: kappa_f},\ref{eq: kappa_c}). We used the coefficient $1/2$ with the dimensional constant $|\widetilde{\lambda}|A^{2}$
in these definitions just for aesthetic reasons. As was mentioned
above, all the dimensional scales can be defined up to dimensionless
numbers, which then can be consistently taken into account in the
statements of convergence and definition of limiting processes.

\end{document}